\documentclass[a4paper,11pt]{article}
\pdfoutput=1 

\usepackage{pub}

\usepackage[T1]{fontenc} 
\usepackage[dvipsnames]{xcolor}
\usepackage{amssymb, bm, graphicx, graphics, color, mathrsfs, multirow,amsmath}
\usepackage{subfigure} \subfiguretopcaptrue

\newcommand{\be}{\begin{equation}}
\newcommand{\ee}{\end{equation}}
\newcommand{\ben}{\begin{eqnarray}}
\newcommand{\een}{\end{eqnarray}}

\newcommand{\la}{{\lambda}}

\newcommand{\cL}{{\cal L}}

\newcommand{\na}{\nabla}

\newcommand{\tpe}{{\tilde p}}

\newcommand{\ah}{\tpe_h}
\newcommand{\aX}{\tpe_X}
\newcommand{\POv}{_{,v}}
\newcommand{\POu}{_{,u}}
\newcommand{\POuv}{_{,uv}}

\makeatletter
\newcommand{\raisemath}[1]{\mathpalette{\raisem@th{#1}}}
\newcommand{\raisem@th}[3]{\raisebox{#1}{$#2#3$}}
\makeatother

\title{\boldmath Gravitational dynamics in the toy model of the Higgs--dark matter sector --- the field theoretic perspective}

\author[1]{Anna Nakonieczna \note{Corresponding author.}}
\author{and {\L}ukasz Nakonieczny}

\affiliation{Institute of~Theoretical Physics, Faculty of~Physics, University of~Warsaw, \\
Pasteura 5, 02-093 Warszawa, Poland}

\emailAdd{Anna.Nakonieczna@fuw.edu.pl}
\emailAdd{Lukasz.Nakonieczny@fuw.edu.pl}

\abstract{The objective of the paper was to examine gravitational evolutions in the Higgs--dark matter sector toy model. The~real part of the Higgs doublet was modelled by a~neutral scalar. Two dark matter candidates introduced were the dark photon and a~charged complex scalar. Non-minimal couplings of both scalars to gravity were included. The~coupling channels between the ordinary and dark matter sectors were kinetic mixing between the electromagnetic and dark $U(1)$ fields and the Higgs portal coupling among the scalars. The~structures of emerging singular spacetimes were either of Schwarzschild or Reissner-Nordstr\"{o}m types. The~non-minimal scalar--gravity couplings led to an appearance of timelike portions of apparent horizons where they transform from spacelike to null. The~features of dynamical black holes were described as functions of the model parameters. The~black holes formed later and their radii and masses were smaller as the mass parameter of the complex scalar increased. The~dependencies on the coupling of the Higgs field to gravity exhibited extrema, which were a~maximum for the time of the black holes formation and minima in the cases of their radii and masses. A~set of quantities associated with an observer moving with the evolving matter was proposed. The~energy density, radial pressure and pressure anisotropy within dynamical spacetimes get bigger as the singularity is approached. The~increase is more considerable in the Reissner-Nordstr\"{o}m spacetimes. The~apparent horizon local temperature changes monotonically in the minimally coupled case and non-monotonically when non-minimal scalar--gravity couplings are involved.
}

\begin{document} 
\maketitle
\flushbottom

\section{Introduction}
\label{sec:intro}

Considering dynamically formed spacetimes containing complex matter systems, their geometric structures and features of objects existing within them, such as black holes, white holes or wormholes, are in most cases impossible to describe analytically. Moreover, even for relatively simple gravity--matter systems, such as for example a~gravitationally self-interacting electrically charged scalar field, structures of dynamically formed spacetimes differ significantly from their non-dynamical counterparts. The~dynamical Reissner-Nordstr\"{o}m spacetime is considerably different from the static one~\cite{PoissonIsrael1990-1796}. There exists a~central spacelike singularity and a~null Cauchy horizon, which is also singular due to the the mass inflation phenomenon. In the static case, the central singularity is timelike and the Cauchy horizon is non-singular. For the above reasons, numerical investigations are usually performed to obtain a~description of spacetimes, which were formed or evolving during dynamical processes, such as gravitational collapse or matter accretion.

A set of various matter-geometry systems were hitherto considered for establishing outcomes of dynamical processes within them. The~simplest ones are self-interacting scalar fields, neutral~\cite{HamadeStewart1996-497} and electrically charged~\cite{OrenPiran2003-044013,HodPiran1998-1554,HodPiran1998-1555,HongHwangStewartYeom2010-045014,HwangYeom2011-064020}. The~gravitational evolution of the inflaton field was tracked in~\cite{HansenHwangYeom2009-016,HwangYeom2011-ChenYeom2015} and within the $f(R)$ gravity in~\cite{HwangLeeYeom2011-006,ChenYeom2015-022}. The~role of dilatonic and phantom couplings during the collapse within the Einstein-Maxwell theory was elaborated in~\cite{BorkowskaRogatkoModerski2011-084007,NakoniecznaRogatko2012-3175,NakoniecznaRogatkoModerski2012-044043,NakoniecznaRogatko-AIP}. The~gravitational evolution of scalar fields in the Brans--Dicke theory was inspected in~\cite{HwangYeom2010-205002,HansenYeom2014-040,HansenYeom2015-019}. The~influence of the dark sector, i.e., the presence of dark matter and dark energy in spacetime, on the collapse of an electrically charged scalar field was examined in~\cite{NakoniecznaRogatkoNakonieczny2015-012}.

Investigating gravitational dynamics within various physical models is often accompanied by inspecting additional issues, related to the dynamics. Pair creation during collapse and subsequent evaporation of black holes, i.e., quantum effects accompanying the collapse, were described in~\cite{SorkinPiran2001-084006,SorkinPiran2001-124024,HongHwangStewartYeom2010-045014}. The~issues of the cosmic censorship conjecture and the information loss problem were raised in~\cite{HongHwangStewartYeom2010-045014,HwangYeom2011-064020,HansenHwangYeom2009-016,HwangYeom2011-155003}. The~impact of changing the number of dimensions on the course and outcomes of the collapse was described in~\cite{HwangKimYeom2012-055003}, while the role of various topologies in~\cite{HwangLeeLeeYeom2012-003,HwangLeeParkYeom2012-003}. The~problem of measuring time using a~collapsing scalar field was discussed in~\cite{NakoniecznaLewandowski2015-064031,NakoniecznaYeom2016-049,NakoniecznaYeom2016-155}.

The existence and features of dark matter and the Higgs field have been a~viable topic of both experimental and theoretical research recently. How these two matter components influence dynamical gravitational processes is an intriguing issue, which we decided to raise in the current paper. To describe the matter sector we opted to use the model containing two scalar fields and two $U(1)$ gauge fields.
One of the gauge fields describes electromagnetism and the second one can be associated with the dark photon 
\cite{1983Natur.306..765G,HOLDOM1986196,BaekKoPark2015-255} (see also~\cite{Fabbrichesi:2020wbt} for a~modern review). Among the scalars, one is real and not charged under either of the $U(1)$ gauge fields. This field may represent a~real part of the 
Higgs doublet written in the unitary gauge. The~second one, which is complex scalar charged under a~$U(1)$ field may 
also represent a~stable dark matter candidate~\cite{BaekKoPark2015-255,PhysRevD.96.103540,ARCADI20201}.
Additionally, we also considered a~possibility that both scalars possess non-minimal couplings to gravity. Moreover, there are two coupling channels among the ordinary matter sector, which consists of the real scalar and electromagnetic field, and the dark sector, composed of the complex scalar and an additional $U(1)$ field.
These channels are given by a~kinetic mixing between the $U(1)$ gauge fields and the Higgs portal coupling among the scalars.

The double null formalism in investigating dynamical gravitational processes is beneficial as it allows to track their course from an asymptotically flat region situated close to past null infinity up to the forming singularity. It means that both the external and internal geometrical structures of the forming objects can be resolved. Formation of additional spacetime special regions, such as event and Cauchy horizons, dynamical horizons, wormhole throats, vacuum bubbles, etc., can be followed within this formalism. A~comprehensive review on employing the double null formalism in investigations of gravitational collapse of matter can be found in~\cite{NakoniecznaNakoniecznyYeom2019-1930006}.

The aim of the current paper is investigating the course and results of gravitational dynamics within the toy model of the Higgs--dark matter sector using the double null formalism. The~structures of emerging spacetimes will be analyzed via locations of dynamically formed horizons and singularities. The~dependence of the characteristics of forming objects, precisely dynamical black holes, on the parameters of the model of interest, will be presented. Additionally, a~set of observables will be proposed to describe the outcomes of the processes.

The paper is organized as follows. In section~\ref{sec:model} the theoretical model of the matter--gravity system of interest is presented. Section~\ref{sec:particulars} contains basic information on solving the derived evolution equations and particulars of the results interpretation. In sections~\ref{sec:noncoupl} and~\ref{sec:coupl} the course and outcomes of the investigated process are discussed. The~summary of the obtained results is placed in section~\ref{sec:conclusions}. A~presentation of the numerical setup for solving the equations of motion within the examined model and tests of the code constitute appendix~\ref{sec:appendix}.

\section{Theoretical model of~the~evolution}
\label{sec:model}

The action functional for our theory consists of three additive parts and is given by
\ben
S = \int d^{4} x \sqrt{-g} \left( \cL_G + \cL_M + \cL_{GM} \right),
\label{a1}
\een
where $\mathcal{L}_{G} = \frac{1}{16 \pi G} R$ is the Einstein-Hilbert action and the matter part is
\ben
\mathcal{L}_{M} &=& -\frac{1}{4} B_{\mu \nu} B^{\mu \nu} - \frac{1}{4}F_{\mu \nu}F^{\mu \nu} - \frac{\alpha}{2}B_{\mu \nu}F^{\mu \nu} -d_{\mu}X d^{\mu}X^{*} - m_{X}^2 |X|^2 - \frac{\lambda_{X}}{4}|X|^4 +\nonumber \\
&&- \frac{1}{2} \nabla_{\mu} h \nabla^{\mu}h + \frac{1}{2}m_{h}^2 h^2 - \frac{\lambda_h}{4}h^4 - \lambda_{hX}h^2 |X|^2, \\
\mathcal{L}_{GM} &=& - \frac{1}{2} \xi_h R h^2 - \xi_X R |X|^2.
\een
In the above formulas $B_{\mu \nu} \equiv 2 \nabla_{[\mu} B_{\nu]}$ is the strength tensor for the dark photon field~$B_{\mu}$,
$F_{\mu \nu} \equiv 2 \nabla_{[\mu}A_{\nu ]}$ is the Maxwell tensor for an ordinary electromagnetic field $A_{\mu}$. In~the employed representation the gauge fields are coupled through the kinetic mixing term $\frac{\alpha}{2}F_{\mu \nu}B^{\mu \nu}$. The~strength of the mixing is controlled by the coupling constant $\alpha$, whose upper value is experimentally constrained to be less than $10^{-3}$~\cite{GuptaPrimulandoSaraswat2015-079}.

The remaining matter is represented by two scalar fields. The~real neutral scalar field~$h$ may be regarded as the neutral part of the Higgs field (for an appropriate choice of the 
$\lambda_h$ and $m_{h}^2$ parameters, roughly speaking $\lambda_h \sim 0.13$ and $m_{h}^2 \gtrsim 0$). The~complex scalar field $X$ is charged under a~dark $U(1)$ field (the covariant derivative is $d_\mu=\na_\mu+ieB_\mu$, where $e$~is its charge) and may be a~model of a~dark matter candidate, see for example 
\cite{PhysRevD.43.2314,PhysRevD.100.063515,PhysRevD.83.063509,PhysRevD.99.075001} 
and references therein. 
The fields $h$ and $X$ interact through the Higgs portal term $\lambda_{hX}h^2 |X|^2$.
An influence of non-minimal couplings of the scalars to gravity is described by the $\mathcal{L}_{GM}$
part of the action.

Before we proceed further it is convenient to diagonalize the gauge sector. For this purpose we define $C_{\mu} \equiv B_{\mu} + \alpha A_{\mu}$ and write  
\ben
\cL_M &=& -\frac{1}{4} C_{\mu\nu} C^{\mu\nu} - \frac{\bar\alpha^2}{4} F_{\mu\nu} F^{\mu\nu} - d_\mu X d^\mu X^\ast - m_{X}^2|X|^2 
- \frac{\lambda_X}{4} \Big(|X|^2\Big)^2 +\nonumber\\
&&-\frac{1}{2}\na_\mu h \na^\mu h + \frac{1}{2}m_{h}^2 h^2 - \frac{\lambda_h}{4}h^4 - \lambda_{hX}h^2|X|^2,
\een
where $d_\mu=\na_\mu+ie \left (C_\mu-\alpha A_\mu \right )$ and $\bar\alpha^2=1-\alpha^2$. In this representation the $X$ field 
is also charged under the ordinary electromagnetic $U(1)$ group, but its charge is suppressed by the small $\alpha$ parameter
(in the dark matter literature this type of field is called millicharged dark matter, provided that $m_{X}^2 > 0$). On the other 
hand, if $m_{X}^2 <0$ the appropriate dark matter candidate is the dark photon which also becomes massive due to the existence 
of the non-zero vacuum expectation value for the $X$ field. 

To write the equations of motion in a~fully explicit form we need to specify the metric ansatz. We will be using the double null
$(u,v,\Theta,\Phi)$ spherically symmetric line element~\cite{MisnerThorneWheeler}
\ben
ds^2 = - a(u, v)^2 du dv + r^2(u, v) d \Omega^2,
\label{m}
\een
where~$u$ and~$v$ null coordinates are called retarded and~advanced time, respectively, and~$d \Omega^2 = d\Theta^2 + \sin^2\Theta d\Phi^2$ is~the~line element of~the~unit sphere, where $\Theta$ and~$\Phi$ are angular coordinates. Such a~coordinate choice determines the double null spacetime foliation~\cite{InvernoSmallwood1980-1223}. From now on, partial derivatives with respect to~the~null coordinates will be denoted by~$\POu $ and~$\POv$. Vector and tensor elements indexes will be marked by adequate sub- and superscripts.

Before we present the equations of motion it is convenient to introduce a~set of dimensionless quantities. 
For this purpose we make the following rescaling:
\begin{align}
&\overline{M}_{P} u \rightarrow \tilde{u}, \quad \overline{M}_{P} v \rightarrow \tilde{v}, \quad
\overline{M}_{P} r(u,v) \rightarrow \tilde{r}(\tilde{u},\tilde{v}), \quad \overline{M}^{-2}_{P} R \rightarrow \tilde{R}, \nonumber\\
&\overline{M}_{P}^{-1} h \rightarrow \tilde{h}, \quad \overline{M}_{P}^{-1} X \rightarrow \tilde{X}, \quad
\overline{M}_{P}^{-2} m_{h}^2 \rightarrow \alpha_{m}, \quad \overline{M}_{P}^{-2} m_{X}^2 \rightarrow m^2, \\
&\overline{M}_{P}^{-1} A_{u} \rightarrow \tilde{A}_{u}, \quad \overline{M}_{P}^{-1} A_{v} \rightarrow \tilde{A}_{v}, \quad
\overline{M}_{P}^{-1} C_{u} \rightarrow \tilde{C}_{u}, \quad \overline{M}_{P}^{-1} C_{v} \rightarrow \tilde{C}_{v}, \nonumber
\end{align} 
where $\overline{M}_{P}^2 = \frac{1}{8 \pi G}$ is the reduced Planck mass squared. In what follows we will be using the rescaled dimensionless 
fields and coordinates, and moreover for the purpose of making the notation more readable we will drop the tilde above them. 
To recapitulate, the Lagrangian densities of the action~\eqref{a1} written in terms of the rescaled variables are
\ben
\mathcal{L}_{G} &=& \frac{1}{2}R, \\
\cL_M &=& -\frac{1}{4} C_{\mu\nu} C^{\mu\nu} - \frac{\bar\alpha^2}{4} F_{\mu\nu} F^{\mu\nu} - d_\mu X d^\mu X^\ast - m^2|X|^2 
- \frac{\lambda_X}{4} \Big(|X|^2\Big)^2 +\nonumber\\
&&-\frac{1}{2}\na_\mu h \na^\mu h + \frac{1}{2}\alpha_{m} h^2 - \frac{\lambda_h}{4}h^4 - \lambda_{hX}h^2|X|^2, \\
\mathcal{L}_{GM} &=& - \frac{1}{2} \xi_h R h^2 - \xi_X R |X|^2.
\een

The equations of motion for the evolving fields were obtained via the~variation of~the~action~\eqref{a1}. The~equations for the Higgs field $h$, the complex scalar field $X$ and gauge fields $A_\mu$ and $C_\mu$ are
{\allowdisplaybreaks
\ben
\na^2 h + \alpha_m h - \lambda_h h^3 - \xi_h hR - 2\lambda_{hX} h |X|^2 = 0,
\label{eqn:h-1} \\
\na^2 X - e^2 \left(C_\mu-\alpha A_\mu\right) \left(C^\mu-\alpha A^\mu\right) X 
+ ie \na_\mu \left(C^\mu-\alpha A^\mu\right) X +\nonumber\\
+ 2ie \left(C_\mu-\alpha A_\mu\right) \na^\mu X
- m^2 X - \frac{\lambda_X}{2} X |X|^2 - \xi_X XR -\lambda_{hX} h^2 X = 0,
\label{eqn:X-1} \\
\na^2 X^\ast - e^2 \left(C_\mu-\alpha A_\mu\right) \left(C^\mu-\alpha A^\mu\right) X^\ast 
- ie \na_\mu \left(C^\mu-\alpha A^\mu\right) X^\ast +\nonumber\\
- 2ie \left(C_\mu-\alpha A_\mu\right) \na^\mu X^\ast
- m^2 X^\ast - \frac{\lambda_X}{2} X^\ast |X|^2 - \xi_X X^\ast R -\lambda_{hX} h^2 X^\ast = 0,
\label{eqn:Xcc-1} \\
\na_\mu C^{\mu\nu} - 2e^2 C^\nu |X|^2 + 2\alpha e^2 A^\nu |X|^2 
- ie \left( X \na^\nu X^\ast - X^\ast \na^\nu X \right) = 0,
\label{eqn:C-1} \\
\na_\mu F^{\mu\nu} - \frac{2\alpha^2 e^2}{\bar\alpha^2} A^\nu |X|^2 + \frac{2\alpha e^2}{\bar\alpha^2} C^\nu |X|^2 
+ ie \frac{\alpha}{\bar\alpha^2} \left( X \na^\nu X^\ast - X^\ast \na^\nu X \right) = 0.
\label{eqn:F-1}
\een}
The~Einstein equations derived by~varying the~action~\eqref{a1} with respect to~gravitational field complement the~above set of~equations, which describes the dynamics of the~inspected dynamical system.

Regarding spherical symmetry, in~the~chosen coordinate system, the~only non-vanishing components of~the~field tensors are $F_{uv}$, $F_{vu}$, $C_{uv}$ and~$C_{vu}$. Due~to~the~gauge freedom $A_{u} \to A_{u} + \na_{u} \theta^\prime$ and~$C_{u} \to C_{u} + \na_{u} \theta^{\prime\prime}$, where $\theta^\prime = \int A_{v}dv$ and~$\theta^{\prime\prime} = \int C_{v}dv$, the~only non-zero four-vector components are~$A_{u}$ and~$C_{u}$. They are functions of~retarded and~advanced time.

The evolution equations of the fields from within the examined theoretical setup expressed in double null coordinates, i.e., using the~assumed line element~\eqref{m}, are the following. The~dynamics of the~Higgs field described covariantly by~\eqref{eqn:h-1} is governed by the equation
\ben
r h\POuv + r\POu h\POv + r\POv h\POu + \frac{a^2r}{2} \lambda_{hX} h \left(X_1^2 + X_2^2\right) +\nonumber\\
- \frac{a^2r}{4} \left\{ \alpha_m h - \lambda_h h^3 - \xi_h h\left[ \frac{2}{r^2} 
+ \frac{8}{a^2} \left( \frac{a\POuv}{a} - \frac{a\POu a\POv}{a^2} + \frac{2r\POuv}{r} + \frac{r\POu r\POv}{r^2} \right) \right] \right\} = 0.
\label{eqn:h-2}
\een
The~equations of motion of~the~scalar field $X$~\eqref{eqn:X-1} and its complex conjugate $X^{*}$~\eqref{eqn:Xcc-1} are given by
\ben
rX\POuv + r\POu X\POv + r\POv X\POu 
+ \frac{1}{2} ierX \left(C_{u,v}-\alpha A_{u,v}\right) + ier \left(C_u-\alpha A_u\right) X\POv +\nonumber\\
+ ie r\POv \left(C_u-\alpha A_u\right) X 
+ \frac{a^2r}{4} X \left\{ m^2 + \frac{\lambda_X}{2} |X|^2 + \lambda_{hX} h^2 \right. +\nonumber\\
\left. + \xi_X \left[ \frac{2}{r^2} 
+ \frac{8}{a^2} \left( \frac{a\POuv}{a} - \frac{a\POu a\POv}{a^2} + \frac{2r\POuv}{r} + \frac{r\POu r\POv}{r^2} \right) \right] \right\} = 0,
\label{eqn:X-2} \\
rX^\ast\POuv + r\POu X^\ast\POv + r\POv X^\ast\POu 
- \frac{1}{2} ierX^\ast \left(C_{u,v}-\alpha A_{u,v}\right) - ier \left(C_u-\alpha A_u\right) X^\ast\POv +\nonumber\\
- ie r\POv \left(C_u-\alpha A_u\right) X^\ast 
+ \frac{a^2r}{4} X^\ast \left\{ m^2 + \frac{\lambda_X}{2} |X|^2 + \lambda_{hX} h^2 \right. +\nonumber\\
\left. + \xi_X \left[ \frac{2}{r^2} 
+ \frac{8}{a^2} \left( \frac{a\POuv}{a} - \frac{a\POu a\POv}{a^2} + \frac{2r\POuv}{r} + \frac{r\POu r\POv}{r^2} \right) \right] \right\} = 0.
\label{eqn:Xxx-2}
\een
The equations of motion of~the~only non-zero components of~the~respective four-vectors of~the~$U(1)$ gauge fields
\ben
C_{u, v} - \frac{Ta^2}{2r^2} &=& 0,
\label{eqn:C-2}\\
A_{u, v} - \frac{Qa^2}{2r^2} &=& 0
\label{eqn:A-2}
\een
define the gauge fields related charges $T$ and $Q$, whose evolution equations are
\ben
T\POv - ier^2 \left(X X^\ast\POv - X^\ast X\POv\right) &=& 0,
\label{eqn:T-2} \\
Q\POv + \frac{ier^2}{\bar\alpha^2} \left( X X^\ast\POv - X^\ast X\POv \right) &=& 0.
\label{eqn:Q-2}
\een
The above pairs~\eqref{eqn:C-2},~\eqref{eqn:T-2} and~\eqref{eqn:A-2},~\eqref{eqn:Q-2} stem from a~separation of the~$v$-components of~the gauge fields equations~\eqref{eqn:C-1} and~\eqref{eqn:F-1}, respectively, into two first-order differential equations. Both $T$ and $Q$ depend on retarded and~advanced time. They correspond to~charges associated with the~$C_\mu$ and $A_\mu$ fields contained within a~sphere of~a~radius $r(u,v)$, on~a~spacelike hypersurface containing the~point~$(u,v)$. In the latter case, the charge is simply the electric charge.

The~stress-energy tensor for~the~considered theory is
\ben
\label{ogolnyTmunu}
T_{\mu \nu} &=& \frac{1}{M\left(\xi\right)^2} \left( T_{\mu\nu}^M + T_{\mu\nu}^{GM} \right),
\label{eqn:T}
\een
where $M\left(\xi\right)^2=1-\xi_h h^2-2\xi_X|X|^2$ and
\ben
T_{\mu\nu}^M &=& g_{\mu\nu}\cL_M + C_{\mu\alpha}C_\nu^\alpha + \bar\alpha^2 F_{\mu\alpha}F_\nu^\alpha + d_\mu X d_\nu X^\ast + d_\nu X d_\mu X^\ast + \na_\mu h \na_\nu h, \\
\label{eqn:TM}
T_{\mu\nu}^{GM} &=& \xi_h\na^2 h^2 g_{\mu\nu} - \xi_h \na_\mu\na_\nu h^2 + 2 \xi_X\na^2|X|^2 g_{\mu\nu} - 2 \xi_X\na_\mu\na_\nu|X|^2.
\label{eqn:Txi}
\een
Its non-vanishing components expressed in~double null coordinates are composed of
{\allowdisplaybreaks
\ben
\label{eqn:TMuu}
T_{uu}^M &=& h\POu ^2 + 2\Big[X\POu X^\ast\POu - ie\left(C_u-\alpha A_u\right)\left(X\POu X^\ast-XX^\ast\POu \right) + e^2\left(C_u-\alpha A_u\right)^2XX^\ast\Big],\hspace{1cm} \\
T_{vv}^M &=& h\POv^2+2X\POv X^\ast\POv, \\
T_{uv}^M &=& \frac{T^2a^2}{4r^4} + \bar\alpha^2\frac{Q^2a^2}{4r^4} +\nonumber\\
&&+ \frac{a^2}{2} \left[ m^2|X|^2 
+ \frac{\lambda_X}{4} \Big(|X|^2\Big)^2 - \frac{1}{2}\alpha_m h^2 + \frac{\lambda_h}{4}h^4 + \lambda_{hX}h^2|X|^2 \right], \\
T_{\theta\theta}^M &=& \frac{T^2}{2r^2} + \bar\alpha^2\frac{Q^2}{2r^2} - r^2 \left[ m^2|X|^2 + \frac{\lambda_X}{4} \Big(|X|^2\Big)^2 - \frac{1}{2}\alpha_m h^2 + \frac{\lambda_h}{4}h^4 + \lambda_{hX}h^2|X|^2 \right] +\nonumber\\
&& + \frac{2r^2}{a^2} \Big[h\POu h\POv + X\POu X^\ast\POv+X\POv X^\ast\POu +ie\left(C_u-\alpha A_u\right)\left(XX^\ast\POv-X^\ast X\POv\right)\Big], \\
\label{eqn:TMtt}
\label{eqn:Txiuu}
T_{uu}^{GM} &=& -2\xi_h\left(h\POu ^2+hh_{,uu}-2h\frac{a\POu }{a}h\POu \right) 
 - 4\xi_X\left(|X|\POu ^2+|X||X|_{,uu}-2h\frac{a\POu }{a}|X|\POu \right), \\
T_{vv}^{GM} &=& -2\xi_h\left(h\POv^2+hh_{,vv}-2h\frac{a\POv}{a}h\POv\right) 
 - 4\xi_X\left(|X|\POv^2+|X||X|_{,vv}-2h\frac{a\POv}{a}|X|\POv\right), \\
T_{uv}^{GM} &=& 2\xi_h\left(h\POu h\POv+hh\POuv\right) 
 + \frac{4}{r}\xi_hh\left(r\POu h\POv+r\POv h\POu \right) +\nonumber\\
&&+4\xi_X\left(|X|\POu |X|\POv+|X||X|\POuv\right) 
 + \frac{8}{r}\xi_X|X|\left(r\POu |X|\POv+r\POv |X|\POu \right), \\
T_{\theta\theta}^{GM} &=& -\frac{8r^2}{a^2}\xi_h \left(h\POu h\POv + hh\POuv\right) - \frac{4r}{a^2}\xi_h h \left(r\POu h\POv+r\POv h\POu \right) +\nonumber\\
&&-\frac{16r^2}{a^2}\xi_X \left(|X|\POu |X|\POv + |X||X|\POuv\right) - \frac{8r}{a^2}\xi_X |X| \left(r\POu |X|\POv+r\POv |X|\POu \right),
\label{eqn:Txitt}
\een}
with $|X|=\sqrt{XX^\ast}$. Finally, the~gravitational field equations are obtained using the~adequate components of~the~Einstein tensor resulting from the~line element~\eqref{m} and~the~stress-energy tensor~\eqref{eqn:T} components
{\allowdisplaybreaks
\ben
\frac{2 a\POu r\POu}{a} - r_{,u u} &=&
\bigg\{ \frac{r}{2} h\POu ^2 + r\Big[X\POu X^\ast\POu - ie\left(C_u-\alpha A_u\right)\left(X\POu X^\ast-XX^\ast\POu \right) +\nonumber\\
&&+ e^2\left(C_u-\alpha A_u\right)^2XX^\ast\Big] - r\xi_h\left(h\POu ^2+hh_{,uu}-2h\frac{a\POu }{a}h\POu \right) +\nonumber\\
&&- 2r\xi_X\left(|X|\POu ^2+|X||X|_{,uu}-2|X|\frac{a\POu }{a}|X|\POu \right) \bigg\} \cdot\nonumber\\
&&\cdot \Big(1-\xi_h h^2-2\xi_X|X|^2\Big)^{-1},
\label{eqn:e1} \\
\frac{2 a\POv r\POv}{a} - r_{,vv} &=&
\bigg\{ \frac{r}{2}h\POv^2+rX\POv X^\ast\POv - r\xi_h\left(h\POv^2+hh_{,vv}-2h\frac{a\POv}{a}h\POv\right) +\nonumber\\
&&- 2r\xi_X\left(|X|\POv^2+|X||X|_{,vv}-2|X|\frac{a\POv}{a}|X|\POv\right) \bigg\} \cdot\nonumber\\
&&\cdot \Big(1-\xi_h h^2-2\xi_X|X|^2\Big)^{-1}, \\
\frac{a^2}{4r} + \frac{r\POu r_{v}}{r} + r\POuv &=&
\bigg\{ \frac{T^2a^2}{8r^3} + \bar\alpha^2\frac{Q^2a^2}{8r^3} + \frac{a^2r}{4} \bigg[ m^2|X|^2 
+ \frac{\lambda_X}{4} \Big(|X|^2\Big)^2 +\nonumber\\
&&- \frac{1}{2}\alpha_m h^2 + \frac{\lambda_h}{4}h^4 + \lambda_{hX}h^2|X|^2 \bigg] +\nonumber\\
&&+r\xi_h\left(h\POu h\POv+hh\POuv\right) + 2\xi_hh\left(r\POu h\POv+r\POv h\POu \right) +\nonumber\\
&&+2r\xi_X\left(|X|\POu |X|\POv+|X||X|\POuv\right) +\nonumber\\
&& +4\xi_X|X|\left(r\POu |X|\POv+r\POv |X|\POu \right) \bigg\} \Big(1-\xi_h h^2-2\xi_X|X|^2\Big)^{-1}, \\
\frac{a\POu a\POv}{a^2} - \frac{a\POuv}{a} - \frac{r\POuv}{r} &=&
\bigg\{ \frac{T^2a^2}{8r^4} + \bar\alpha^2\frac{Q^2a^2}{8r^4} - \frac{a^2}{4} \bigg[ m^2|X|^2 + \frac{\lambda_X}{4} \Big(|X|^2\Big)^2 +\nonumber\\
&&- \frac{1}{2}\alpha_m h^2 + \frac{\lambda_h}{4}h^4 + \lambda_{hX}h^2|X|^2 \bigg] + \frac{1}{2} \Big[h\POu h\POv + X\POu X^\ast\POv+X\POv X^\ast\POu +\nonumber\\
&&+ie\left(C_u-\alpha A_u\right)\left(XX^\ast\POv-X^\ast X\POv\right)\Big] +\nonumber\\
&&-2\xi_h \left(h\POu h\POv + hh\POuv\right) - \frac{1}{r}\xi_h h \left(r\POu h\POv+r\POv h\POu \right) +\nonumber\\
&&-4\xi_X \left(|X|\POu |X|\POv + |X||X|\POuv\right) - \frac{2}{r}\xi_X |X| \left(r\POu |X|\POv+r\POv |X|\POu \right) \bigg\} \cdot\nonumber\\
&&\cdot \Big(1-\xi_h h^2-2\xi_X|X|^2\Big)^{-1}.
\label{eqn:e4}
\een}
The set~\eqref{eqn:h-2}--\eqref{eqn:Q-2},~\eqref{eqn:e1}--\eqref{eqn:e4} forms a~complete set of equations of motion which describe the dynamics of the system of interest.

In~order to~solve the~obtained system of equations of~motion, a~set of~auxiliary variables
\ben\label{eqn:substitution}
\begin{split}
c &= \frac{a\POu }{a}, \qquad &d&= \frac{a\POv}{a},
\qquad &f&= r\POu , \qquad &g&= r\POv, \\
x &= h\POu , \qquad &y&= h\POv, \\
w &= X\POu , \qquad &z&= X\POv, \qquad &\beta &= A_u, \qquad &\gamma &= C_u,
\end{split}
\een
and quantities defined as
\ben
\la\equiv \frac{a^2}{4}+fg, \qquad \eta\equiv fy+gx, \qquad \kappa\equiv gw+fz,
\een
was introduced. The~substitutions enabled us to~rewrite the~second-order differential equations from the set~\eqref{eqn:h-2}--\eqref{eqn:Q-2},~\eqref{eqn:e1}--\eqref{eqn:e4} as~first-order ones. Additionally, the~real fields $X_1$ and $X_2$ were introduced instead of~conjugate fields $X$ and $X^\ast$ simply according to~$X = X_{1} + i X_{2}$ and $X^\ast = X_{1} - i X_{2}$. These relations result in
\ben
\begin{split}
X &= X_1 + i X_2, \qquad &w& = w_1 + i w_2, \qquad &z& = z_1 + i z_2, \\
\kappa &= \kappa_1 + i \kappa_2, \qquad &\kappa_1& = f z_1 + g w_1, \qquad &\kappa_2& = fz_2 + gw_2.
\label{defdef}
\end{split}
\een

The~final system of~equations of~motion, which governs the~investigated evolution yields
{\allowdisplaybreaks
\ben \label{eqn:P1-2}
P1: && a\POu = ac,\\
P2: && a\POv = ad,\\
P3: && r\POu = f,\\
P4: && r\POv = g,\\
P5: && X_{1(2),u} = w_{1(2)},\\
P6: && X_{1(2),v} = z_{1(2)},\\
P7: && h_{,v} = y,\\
E1: && f\POu = 2cf - \Big[1-\xi_h h^2-2\xi_X\left(X_1^2+X_2^2\right)^2\Big]^{-1} \cdot r \bigg\{ \frac{x^2}{2} + w_1^2+w_2^2+\nonumber\\
&&+2e\left(\gamma-\alpha\beta\right)\left(X_1w_2-X_2w_1\right) 
+ e^2\left(\gamma-\alpha\beta\right)^2\left(X_1^2+X_2^2\right) +\nonumber\\
&&- \xi_h\left(x^2+hx\POu -2chx\right) +\nonumber\\
&&- 2\xi_X\left[w_1^2+w_2^2+X_1w_{1,u}+X_2w_{2,u}-2c\left(X_1w_1+X_2w_2\right)\right] \bigg\},
\label{eqn:E1} \\
\label{eqn:E2}
E2: && g\POv = 2dg - \Big[1-\xi_h h^2-2\xi_X\left(X_1^2+X_2^2\right)^2\Big]^{-1} \cdot r \bigg\{ \frac{y^2}{2} + z_1^2+z_2^2  +\nonumber\\
&&- \xi_h\left(y^2+hy\POv-2dhy\right) +\nonumber\\
&&- 2\xi_X\left[z_1^2+z_2^2+X_1z_{1,v}+X_2z_{2,v}-2d\left(X_1z_1+X_2z_2\right)\right] \bigg\}, \\
E3: && g\POu = f\POv = - \frac{\la}{r} + \Big[1-\xi_h h^2-2\xi_X\left(X_1^2+X_2^2\right)^2\Big]^{-1} \cdot \bigg\{ \frac{T^2a^2}{8r^3} + \bar\alpha^2\frac{Q^2a^2}{8r^3} +\nonumber\\
&&+ \frac{a^2r}{4} \bigg[m^2\left(X_1^2+X_2^2\right)+\frac{\la_X}{4}\left(X_1^2+X_2^2\right)^2 -\frac{1}{2}\alpha_m h^2+\frac{\la_h}{4}h^4 +\nonumber\\
&&+\la_{hX}h^2\left(X_1^2+X_2^2\right)\bigg] + r\xi_h\left(xy+hx\POv\right) +2\xi_hh\eta +\nonumber\\
&&+2r\xi_X\left(w_1z_1+w_2z_2+X_1w_{1,v}+X_2w_{2,v}\right) +4\xi_X\left(X_1\kappa_1+X_2\kappa_2\right) \bigg\}, \hspace{1cm} \\
E4: && d\POu = c\POv = \frac{\lambda}{r^2} - \Big[1-\xi_h h^2-2\xi_X\left(X_1^2+X_2^2\right)^2\Big]^{-1} \cdot \bigg[ \frac{1}{2}xy + \frac{T^2 a^2}{4 r^4} + \bar\alpha^2\frac{Q^2 a^2}{4 r^4} +\nonumber\\
&&+ w_1 z_1 + w_2 z_2 -e\left(\gamma-\alpha\beta\right)\left(X_2z_1 - X_1z_2\right) -\xi_h\left(xy+hx\POv\right) +\frac{1}{r}\xi_hh\eta +\nonumber\\
&&-2\xi_X \left(w_1z_1+w_2z_2 + X_1w_{1,v}+X_2w_{2,v}\right) +\frac{2}{r}\xi_X\left(X_1\kappa_1+X_2\kappa_2\right) \bigg],
\label{eqn:E4} \\
H: && y_{,u} = x_{,v} = - \frac{\eta}{r} - \frac{a^2}{2}\lambda_{hX}h\left(X_1^2+X_2^2\right) +\nonumber\\
 && + \frac{a^2}{4}h \left\{\alpha_m -\lambda_h h^2-\xi_h\left[\frac{2}{r^2}+\frac{8}{a^2}\left(d\POu +\frac{fg}{r^2}+2\frac{f\POv}{r}\right)\right]\right\}, \\
X_{_{\left(Re\right)}}: && z_{1,u} = w_{1,v} = - \frac{\kappa_1}{r} + eX_2\frac{\left(T-\alpha Q\right)a^2}{4r^2} 
+ e \left(z_2+\frac{g}{r}X_2\right)\left(\gamma-\alpha\beta\right) +\nonumber\\
 && - \frac{a^2}{4}X_1\left\{m^2+\frac{\lambda_X}{2}\left(X_1^2+X_2^2\right)+\lambda_{hX}h^2 +\right. \nonumber\\
 && \left. +\xi_X\left[\frac{2}{r^2}+\frac{8}{a^2}\left(d\POu +\frac{fg}{r^2}+2\frac{f\POv}{r}\right)\right]\right\},
\label{eqn:HR} \\
X_{_{\left(Im\right)}}: && z_{2,u} = w_{2,v} = - \frac{\kappa_2}{r} - eX_1\frac{\left(T-\alpha Q\right)a^2}{4r^2} 
- e \left(z_1+\frac{g}{r}X_1\right)\left(\gamma-\alpha\beta\right) +\nonumber\\
 && - \frac{a^2}{4}X_2\left\{m^2+\frac{\lambda_X}{2}\left(X_1^2+X_2^2\right)+\lambda_{hX}h^2 +\right. \nonumber\\
 && \left. +\xi_X\left[\frac{2}{r^2}+\frac{8}{a^2}\left(d\POu +\frac{fg}{r^2}+2\frac{f\POv}{r}\right)\right]\right\}, \\
D1: && \beta\POv = \frac{Qa^2}{2 r^2},
\label{eqn:D1} \\
D2: && Q\POv = -\frac{2}{\bar\alpha^2} e r^2 \left( X_1 z_2 - X_2 z_1 \right),
\label{eqn:D2} \\
G1: && \gamma\POv = \frac{Ta^2}{2 r^2}, \\
G2: && T\POv = 2e r^2 \left( X_1 z_2 - X_2 z_1 \right).
\label{eqn:C2}
\een}

\section{Details of computer simulations and results analysis}
\label{sec:particulars}

The final system of evolution equations~\eqref{eqn:P1-2}--\eqref{eqn:C2} was solved numerically. Appendix~\ref{sec:appendix} presents the details of the numerical code and its tests.

\begin{figure}[tbp]
\begin{minipage}{0.35\textwidth}
\centering
\includegraphics[width=0.75\textwidth]{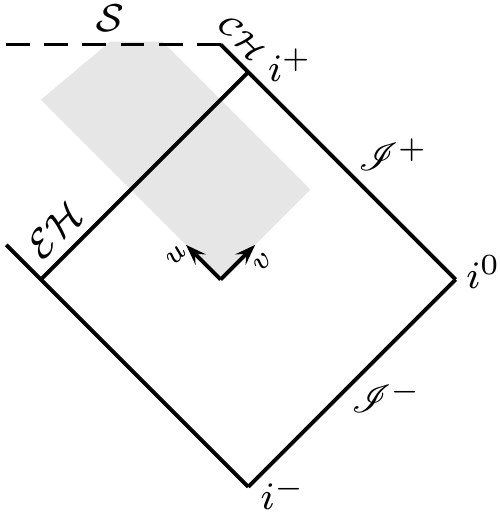}
\end{minipage}
\hfill
\begin{minipage}{0.625\textwidth}
\caption{The domain of computations (marked gray) on the background of the Carter-Penrose diagram of the dynamical Reissner-Nordstr\"{o}m spacetime. $\mathcal{S}$, $\mathcal{EH}$ and $\mathcal{CH}$ are the central singularity along $r=0$, the event and Cauchy horizons, respectively. The~infinities $\mathscr{I}^\pm$, $i^\pm$ and $i^0$ are null, timelike and spacelike, respectively.}
\label{fig:domain}
\end{minipage}
\end{figure}

The equations were solved in the region of the $\left(vu\right)$-plane, which is shown on a~Carter-Penrose diagram in figure~\ref{fig:domain}. The~background spacetime is the one containing a~dynamical Reissner-Nordstr\"{o}m black hole with a~spacelike central singularity and a~Cauchy horizon, both surrounded by an event horizon~\cite{OrenPiran2003-044013}. In all current simulations the computational domain spread from $v=0$ to $v=7.5$ and from $u=0$ to $u=22.5$. The~only arbitrary input data were initial profiles of the evolving fields, posed on the null $u=0$ hypersurface. The~initial profile of the $h$ field was Gaussian
\ben
h = \ah\cdot v^2\cdot \left(v-v_f\right)^3\cdot e^{-\left(\frac{v-v_0}{D}\right)^2}
\label{phi-prof}
\een
and the complex field was represented by the trigonometric profile
\ben
X = \aX \cdot \sin^4\left(\pi\frac{v-v_i}{v_f-v_i}\right)
\cdot\Bigg[\cos\left(2\pi\frac{v-v_i}{v_f-v_i}\right)+i\cos\left(2\pi\frac{v-v_i}{v_f-v_i}+\delta_{ph}\right)\Bigg].
\label{psichi-prof}
\een
The profiles were one-parameter families with free family parameters being the amplitudes $\ah$ and $\aX$, which regulate the strength of the particular field gravitational self-interaction~\cite{Choptuik1993-9}. The~values of the remaining constants $v_0=1.3$, $D=0.21$, $v_i=0$ and $v_f=2.5$ were chosen arbitrarily and were the same in all simulations. The~parameter~$\delta_{ph}$ determining the amount of initial charge was equal to $\frac{\pi}{2}$. The~employed profiles~\eqref{phi-prof} and~\eqref{psichi-prof} ensure that the spacetime slice at the initial null hypersurface is regular and hence their choice is representative for the examined evolutions as the outcomes of computations are independent of the profiles types~\cite{Choptuik1993-9,NakoniecznaLewandowski2015-064031}.

The non-zero value of the electric coupling constant does not influence the results of the gravitational collapse~\cite{BorkowskaRogatkoModerski2011-084007}. Hence, it was invariable in all evolutions and equal to $1$. Due to the argument relating the field $h$ with the neutral part of the Higgs field raised in section~\ref{sec:model}, the parameter $\lambda_h$ was set as equal to $0.13$ during computations. The~value of the parameter $\alpha$ was set as equal to $10^{-5}$ in all simulations, which is in agreement with its experimental constraint~\cite{PhysRevLett.118.011802,Fabbrichesi:2020wbt,GuptaPrimulandoSaraswat2015-079}.

From equations~\eqref{eqn:T-2} and~\eqref{eqn:Q-2} it can be seen that the difference between the absolute values of the $U(1)$ charges $T$ and $Q$ is of the order of $\alpha^2$. Taking into account the value of~$\alpha$ adopted in the computations, the difference is of the order of $10^{-10}$. It was confirmed in the course of numerical simulations and for this reason the text of subsequent sections of the paper will refer only to one of the charges.

There is a~significant number of quantities to be set to completely define a~particular collapse case (the parameters of the model $\lambda_X$, $\lambda_{hX}$, $m^2$, $\xi_X$, $\xi_h$, $\alpha_m$ and field amplitudes $\ah$,~$\aX$). For this reason, a~vast collection of simulations has been run, which involved possible configurations of the above parameters from within their admissible ranges. It should be emphasized that the spacetime structures as well as observables and fields spacetime distributions presented in sections~\ref{sec:noncoupl} and~\ref{sec:coupl} are representative for the particular case. Moreover, to make the presentation of the results as clear as possible, the field amplitudes $\ah$ and $\aX$ were set as equal to each other, with no loss of generality.

Spacetime structures stemming from the studied dynamical processes will be shown on Penrose diagrams. The~diagrams present $r=const.$ contour lines in the $\left(vu\right)$-plane. The~outermost thick line refers to $r=0$. The~spacetime regions depicted on Penrose diagrams will be those essential for the performed analyses. The~structures of spacetimes will be described qualitatively at first and then analyzed via the behavior of selected black hole features listed in the next section when the parameters of the model change.

\subsection{Horizons}

Since the spacetimes obtained as outcomes of the gravitational evolutions of interest are dynamical, global characteristics are of limited use in their structures interpretation. Instead, local notions ought to be employed in order to interpret the obtained structures. One of the most straightforward tools to describe causal relations within dynamical spacetimes are local horizons, which are lines of expansion
\ben
\theta_i\equiv\frac{2}{r}r_{,i}
\label{eqn:expansion}
\een
vanishing in null directions $i=u,v$. The~line $r\POv=0$ will be marked on the spacetime diagrams as a~green solid line. The~horizon corresponding to the $r\POu =0$ line is not present in neither of the obtained spacetimes.

As null coordinates tend to infinity, the dynamics of the spacetime-matter system diminishes and hence in these regions local horizons coincide with the global ones. In particular, considering the placement of the numerical domain within spacetime, the location of the event horizon of the black hole forming during the gravitational collapse can be determined by inspecting the behaviour of a~local horizon as $v$ tends to infinity. For large values of advanced time, the local horizon settles along a~hypersurface $u=const.$ indicating the location of the null event horizon in spacetime. 

For purposes of interpretation of the results and numerical checks of the code, a~set of quantities characterizing the emerging black hole will be inspected. These are the charges $Q$ and $T$ and mass contained within the event horizon. The~mass will be calculated as the quasi-local Hawking mass for two coupled gauge fields $A_\mu$ and $C_\mu$~\cite{NakoniecznaRogatkoNakonieczny2015-012}
\ben
m_H\left(u,v\right) = \frac{r}{2}\left(1+\frac{4fg}{a^2}+\frac{\bar\alpha^2 Q^2+T^2}{r^2}\right).
\label{eqn:mH}
\een
It describes mass contained in a~sphere of a~radius $r\left(u,v\right)$ on a~spacelike hypersurface containing the point $\left(u,v\right)$. The~values of $Q$, $T$ and $m_H$ will be obtained at a~local horizon in the area where it becomes null, i.e., on the boundary of the numerical domain, where $v=v_f$. Additionally, $u$-locations of the black hole event horizons and their radii will be also read out at the same point and will be used in the results interpretation.

The above quantities characterizing emerging black holes will be investigated within the model parameters ranges which, apart from the scalar--gravity couplings, are limited by the model determinants. In the case of the constants $\xi_X$ and $\xi_h$, their ranges were established such that the apparent horizons of the nascent black holes are separated from the initial data hypersurface. This is equivalent to ensuring that the calculations begin in the region outer with respect to the black hole horizon which can be regarded as nearly~flat.

\subsection{Observables}
\label{ssec:obs}

Another way of interpreting the results of gravitational collapse within the model of interest is inspecting a~set of local spacetime quantities, whose derivation is the following.

The tetrad connected to the observer moving with the medium, associated with the double null metric~\eqref{m} is~\cite{NielsenVisser2006-4637,Nielsen_2008}
\ben
{\bf{e}^0} &=& \frac{a}{2} \left( du + dv \right), \\
{\bf{e}^1} &=& \frac{a}{2} \left( du - dv \right), \\
{\bf{e}^2} &=& r d \theta, \\
{\bf{e}^3} &=& r \sin\theta d \phi,
\een
or, alternatively, may be written as
\ben
e^{ \mu}_{i} &=& \left( V^{\mu}, S^{\mu}, \hat{\theta}^{\mu}, \hat{\phi}^{\mu} \right)
\een
with
\ben
V^{\mu} &=& \frac{1}{a} (1,1,0,0), \\
S^{\mu} &=& \frac{1}{a} (1,-1,0,0), \\
\hat{\theta}^{\mu} &=& \frac{1}{r} (0,0,1,0), \\
\hat{\phi}^{\mu} &=& \frac{1}{r \sin\theta} (0,0,0,1),
\een
where
\ben
V^\mu V_\mu &=& -1, \\
S^\mu S_\mu &=& \hat{\theta}^\mu \hat{\theta}_\mu = \hat{\phi}^\mu \hat{\phi}_\mu = +1.
\een
The energy density $\hat{\rho}$, energy fluxes $\hat{f}$, radial $\hat{p}_{r}$ and tangential pressures $\hat{p}_{t}$
as seen by the observer can be defined as
\ben
\hat{\rho} &=& T_{00} = e^{ \mu}_{0} e^{ \nu}_{0} T_{\mu \nu} = V^{\mu} V^{\nu} T_{\mu \nu}, \\
\hat{f}_{j} &=& T_{0j} = e^{ \mu}_{0} e^{ \nu}_{j} T_{\mu \nu}, \qquad j \neq 0, \\
\hat{p}_{r} &=& T_{11} = e^{ \mu}_{1} e^{ \nu}_{1} T_{\mu \nu} = S^{\mu} S^{\nu} T_{\mu \nu}, \\
\hat{p}_{t} &=& T_{22} = T_{33} = e^{ \mu}_{2} e^{ \nu}_{2} T_{\mu \nu}.
\een
Moreover, defining two null vectors
\ben
l^{\mu} &=& V^{\mu} + S^{\mu}, \\
n^{\mu} &=& V^{\mu} - S^{\mu},
\een
the surface gravity for a~dynamic black hole is~\cite{NielsenVisser2006-4637,Nielsen_2008}
\ben
\kappa_{l} &=& l^{\mu} l^{\nu} \nabla_{\nu} n_{\mu}.
\een

The quantities used for the results interpretation were the energy density $\hat{\rho}$, radial pressure $\hat{p}_{r}$, pressure anisotropy $\hat{p}_{a} \equiv \hat{p}_{t} - \hat{p}_{r}$ and local temperature $T_l = \frac{\kappa_l}{2 \pi}$ with the surface gravity 
\ben
\kappa_{l} = \frac{a\POu }{a^2}
\label{eqn:surfgrav}
\een
as defined in~\cite{NielsenVisser2006-4637,Nielsen_2008}. It is worth emphasizing that there is no unique definition of surface gravity for dynamical black holes, as dynamical spacetimes within which they emerge do not admit a~Killing vector field~\cite{Wald}.

The first three of the above quantities are related with the stress-energy tensor components via
\ben
\hat{\rho} &=& \frac{1}{a^2} \left( T_{uu} + 2 T_{uv} + T_{vv} \right), \\
\hat{p}_{r} &=& \frac{1}{a^2} \left( T_{uu} - 2 T_{uv} + T_{vv} \right), \\
\hat{p}_{a} &=& \frac{1}{r^2}T_{\theta \theta} - \frac{1}{a^2} \left( T_{uu} - 2 T_{uv} + T_{vv} \right)
\een
and, taking into account \eqref{ogolnyTmunu}, for the inspected gravity--matter model considered in double null coordinates are given by the following relations:
{\allowdisplaybreaks
\ben
\hat{\rho} &=& 2\left\{a^2 r\Big[1-\xi_h h^2-2\xi_X\left(X_1^2+X_2^2\right)^2\Big]\right\}^{-1} 
\Bigg\{\frac{rx^2}{2} + \frac{ry^2}{2} + \frac{T^2a^2}{4r^3} + \bar\alpha^2\frac{Q^2a^2}{4r^3} + \nonumber\\
&& + r \left[w_1^2 + w_2^2 + 2 e \left(\gamma-\alpha\beta\right) \left(X_1 w_2 - X_2 w_1\right) + 
e^2 \left(\gamma-\alpha\beta\right)^2 \left(X_1^2 + X_2^2\right)\right] + \nonumber\\
&& + \frac{a^2r}{2} \bigg[m^2\left(X_1^2+X_2^2\right)+\frac{\la_X}{4}\left(X_1^2+X_2^2\right)^2 -\frac{1}{2}\alpha_m h^2+\frac{\la_h}{4}h^4+\la_{hX}h^2\left(X_1^2+X_2^2\right)\bigg] + \nonumber\\
&& + 4\xi_h h \eta + r \left(z_1^2 + z_2^2\right) - r \xi_h \Big[x^2 + y^2 + h\left(x\POu + y\POv\right) - 2\left(chx + dhy + h x\POv + xy\right)\Big] + \nonumber\\
&& - 2\xi_X r \Big[w_1^2 + w_2^2 + X_1 w_{1,u} + X_2 w_{2,u} - 2c \left(X_1 w_1 + X_2 w_2\right) + \nonumber\\
&& + z_1^2 + z_2^2 + X_1 z_{1,v} + X_2 z_{2,v} - 2d \left(X_1 z_1 + X_2 z_2\right) + \nonumber\\
&& - 2\left(X_1 w_{1,v} + X_2 w_{2,v} + w_1 z_1 + w_2 z_2\right) \Big] + 8\xi_X \left(X_1 \kappa_1 + X_2 \kappa_2\right) \Bigg\},
\label{eqn:rho}\\
\hat{p}_{r} &=& 2\left\{a^2 r\Big[1-\xi_h h^2-2\xi_X\left(X_1^2+X_2^2\right)^2\Big]\right\}^{-1} 
\Bigg\{\frac{rx^2}{2} + \frac{ry^2}{2} - \frac{T^2a^2}{4r^3} - \bar\alpha^2\frac{Q^2a^2}{4r^3} + \nonumber\\
&& + r \left[w_1^2 + w_2^2 + 2 e \left(\gamma-\alpha\beta\right) \left(X_1 w_2 - X_2 w_1\right) + 
e^2 \left(\gamma-\alpha\beta\right)^2 \left(X_1^2 + X_2^2\right)\right] + \nonumber\\
&& - \frac{a^2r}{2} \bigg[m^2\left(X_1^2+X_2^2\right)+\frac{\la_X}{4}\left(X_1^2+X_2^2\right)^2 -\frac{1}{2}\alpha_m h^2+\frac{\la_h}{4}h^4+\la_{hX}h^2\left(X_1^2+X_2^2\right)\bigg] + \nonumber\\
&& - 4\xi_h h \eta + r \left(z_1^2 + z_2^2\right) - r \xi_h \Big[x^2 + y^2 + h\left(x\POu + y\POv\right) - 2\left(chx + dhy + h x\POv + xy\right)\Big] + \nonumber\\
&& - 2\xi_X r \Big[w_1^2 + w_2^2 + X_1 w_{1,u} + X_2 w_{2,u} - 2c \left(X_1 w_1 + X_2 w_2\right) + \nonumber\\
&& + z_1^2 + z_2^2 + X_1 z_{1,v} + X_2 z_{2,v} - 2d \left(X_1 z_1 + X_2 z_2\right) + \nonumber\\
&& + 2\left(X_1 w_{1,v} + X_2 w_{2,v} + w_1 z_1 + w_2 z_2\right) \Big] - 8\xi_X \left(X_1 \kappa_1 + X_2 \kappa_2\right) \Bigg\},
\label{eqn:pr}\\
\hat{p}_{a} &=& \left\{a^3 r^4 \Big[1-\xi_h h^2-2\xi_X\left(X_1^2+X_2^2\right)^2\Big]\right\}^{-1} 
\Big\{a^3 \left(T^2 + \bar\alpha^2 Q^2\right) + \nonumber\\
&& + 4 ac r^4 \big[-\xi_h hx - 2\xi_X\left(X_1w_1 + X_2w_2\right)\big] 
- a~r^3 \Big[-4\xi_h h \eta - 8\xi_X \left(X_1 \kappa_1 + X_2 \kappa_2 \right) + \nonumber\\
&& + r \Big(\left(-2 \xi_h + 1\right) \left(x - y\right)^2 + 2\left(-2\xi_X + 1\right) \left(w_1^2 + w_2^2 + z_1^2 + z_2^2\right) + \nonumber\\
&& + 2e^2\left(\gamma-\alpha\beta\right)^2\left(X_1^2+X_2^2\right) +4e\left(\gamma-\alpha\beta\right)\left(X_2z_1 - X_1z_2\right) + \nonumber\\
&& + 4 w_1 (\alpha \beta e X_2 - e \gamma X_2 - z_1 + 2 \xi_X z_1) - 4 w_2 (\alpha \beta e X_1 - e \gamma X_1 + z_2 - 2 \xi_X z_2) + \nonumber\\
&& + 8\xi_X d \left(X_1 z_1 - X_2 z_2\right) - 2\xi_h h \left(x\POu - 2x\POv + y\POv - 2dy\right) + \nonumber\\
&& - 4\xi_X \left(X_1w_{1,u} + X_2w_{2,u} - 2X_1 w_{1,v} - 2X_2 w_{2,v} + X_1 z_{1,v} + X_2 z_{2,v} \right) \Big) \Big] \Big\}.
\label{eqn:pa}
\een}



\section{Collapse dynamics within a~model involving scalars minimally coupled to gravity}
\label{sec:noncoupl}

To begin with, a~set of results of gravitational collapse within a~simplified version of the investigated model will be presented. The~non-minimal couplings of scalars to gravity will be omitted, that is $\xi_X=\xi_h=0$.

\subsection{Spacetime structures}
\label{sec:noncoupl-struc}

\begin{figure}[tbp]
\subfigure[][]{\includegraphics[width=0.4\textwidth]{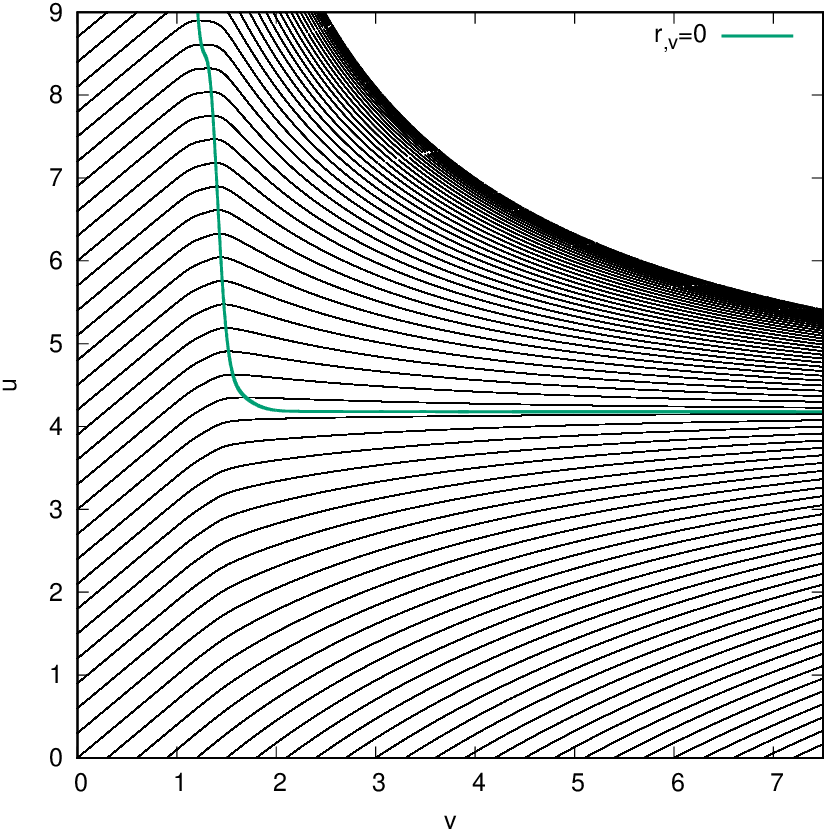}\label{fig:noncoupl-str-a}}
\hfill
\subfigure[][]{\includegraphics[width=0.4\textwidth]{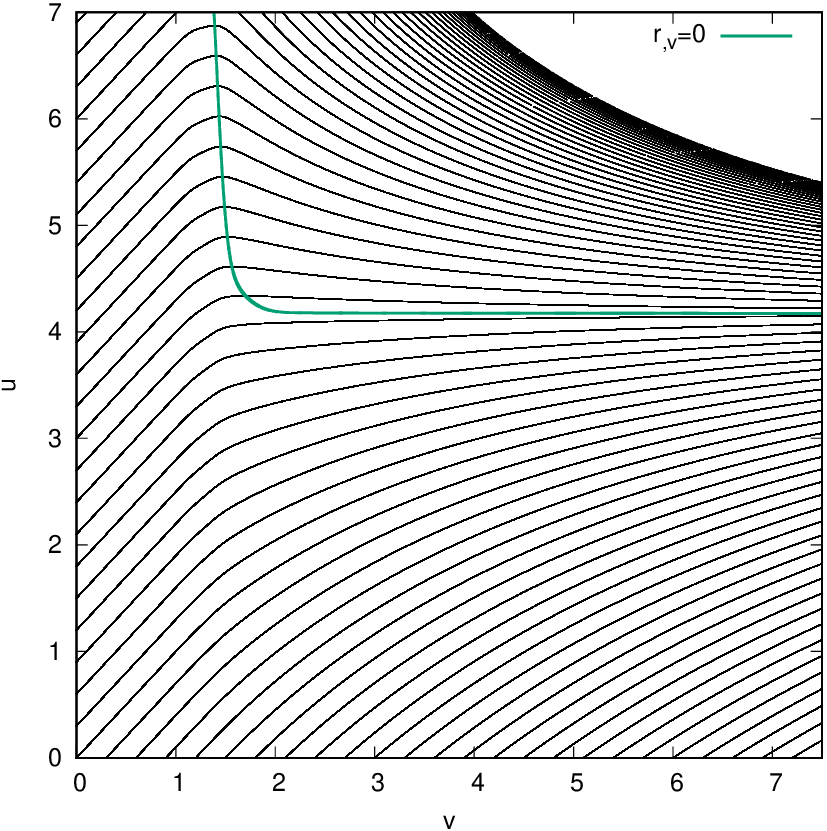}\label{fig:noncoupl-str-b}}\\
\subfigure[][]{\includegraphics[width=0.4\textwidth]{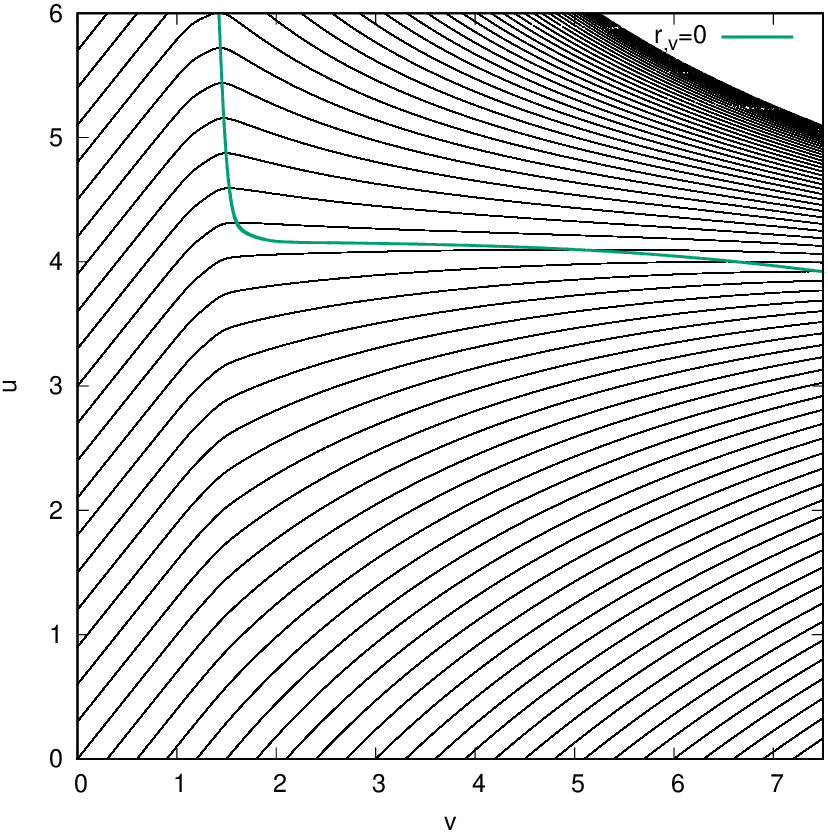}}
\hfill
\subfigure[][]{\includegraphics[width=0.4\textwidth]{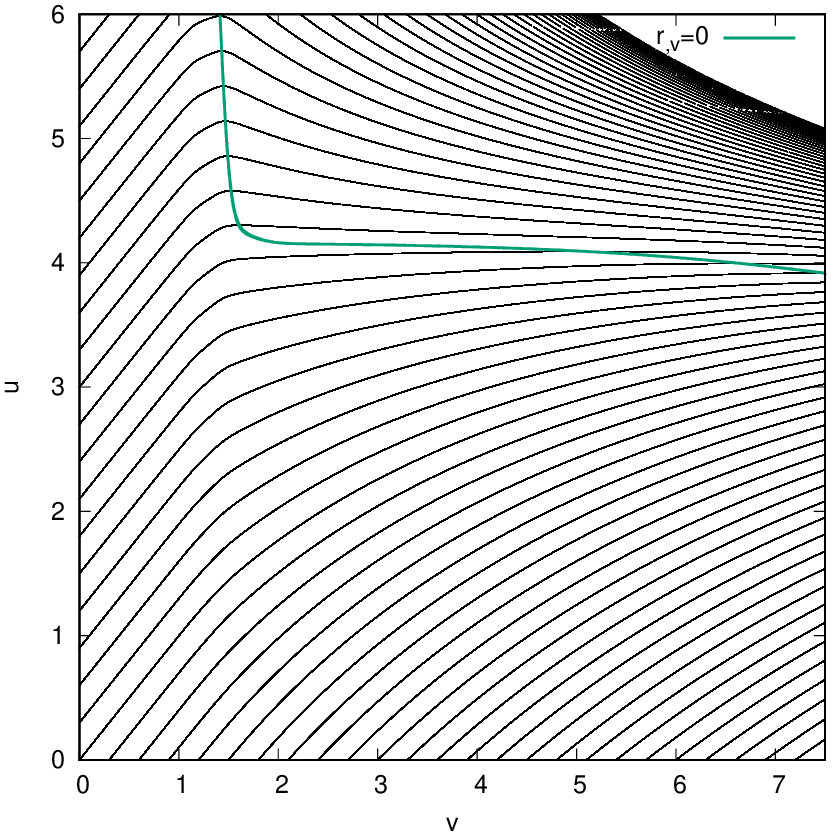}}
\caption{(color online) Penrose diagrams of spacetimes emerging from evolutions with scalars minimally coupled to gravity. The~field amplitudes $\aX$ and $\ah$ were equal to $0.05$, while the parameters $\lambda_X=\lambda_{hX}=0.1$ and $\xi_X=\xi_h=0$. The~remaining parameters were (a)~$m^2=0.25$, $\alpha_m=0$, (b)~$m^2=0.25$, $\alpha_m=0.1$, (c)~$m^2=-0.25$, $\alpha_m=0$ and (d)~$m^2=-0.25$, $\alpha_m=0.1$. }
\label{fig:noncoupl-str}
\end{figure}

The structures of singular spacetimes resulting from the dynamical evolution of fields are presented in figure~\ref{fig:noncoupl-str}. All the spacetimes contain a~spacelike central singularity along $r=0$ surrounded by an apparent horizon that is either spacelike or null, when situated along a~$u=const.$ hypersurface. For positive values of $m^2$ the resulting spacetime is a~typical dynamical Schwarzschild spacetime. The~apparent horizon which is spacelike for small values of advanced time, settles along a~null hypersurface of constant retarded time as $v$ tends to infinity thus indicating the location of the event horizon in spacetime. When $m^2$ is negative, the apparent horizon has two spacelike sections, for small and large values of advanced time, separated by a~null section. Such a~behavior has been already observed during collapses within models containing more than one scalar~\cite{NakoniecznaRogatkoModerski2012-044043,NakoniecznaRogatko2012-3175} and can be interpreted as an accretion of one of the scalars on a~resulting black hole.

\subsection{Black hole characteristics}
\label{sec:noncoupl-char}

The characteristics of emerging black holes were investigated for the case with the following parameters: $\lambda_X=\lambda_{hX}=0.1$, $m^2=0.25$, $\alpha_m=0$ and $\aX=\ah=0.05$. The~spacetime structure for this collection is shown in figure~\ref{fig:noncoupl-str-a}. While the dependence on the particular parameter of the model is presented, the remaining ones are as above.

\begin{figure}[tbp]
\subfigure[][]{\includegraphics[width=0.45\textwidth]{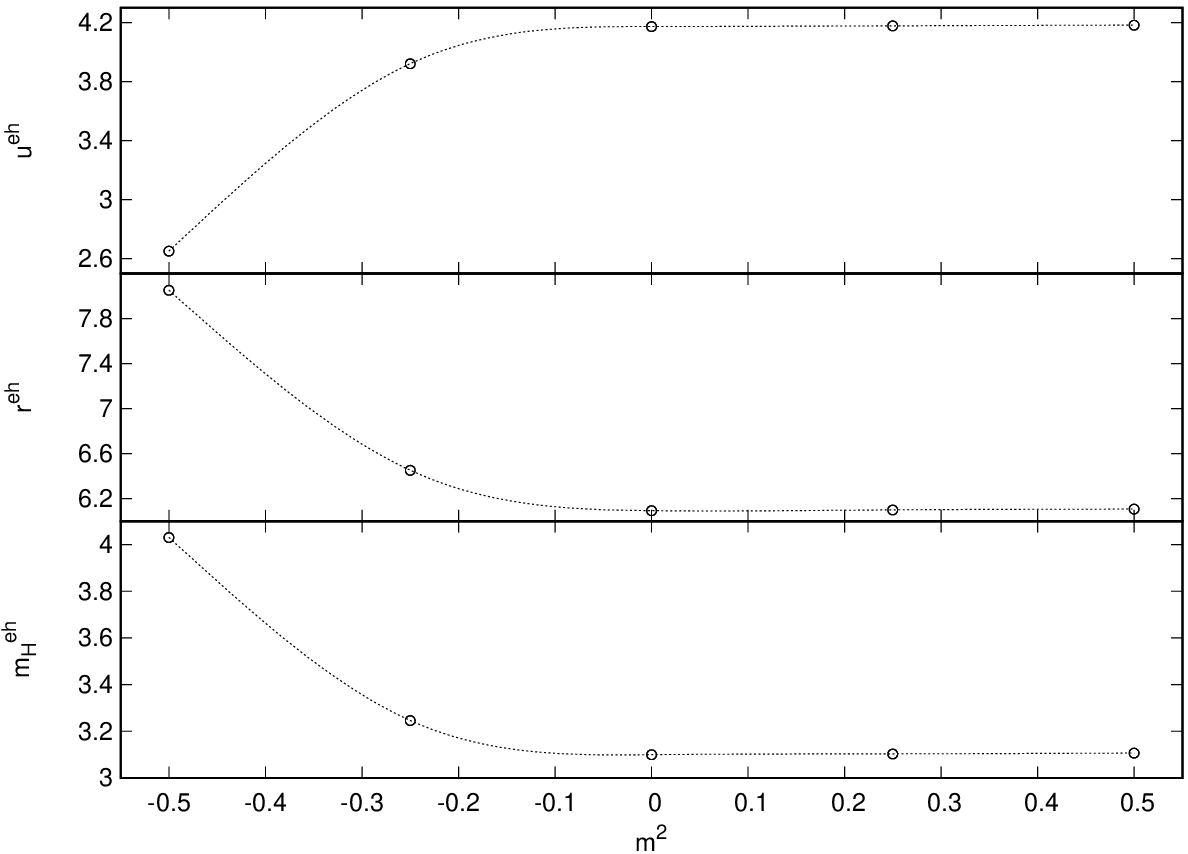}}
\hfill
\subfigure[][]{\includegraphics[width=0.45\textwidth]{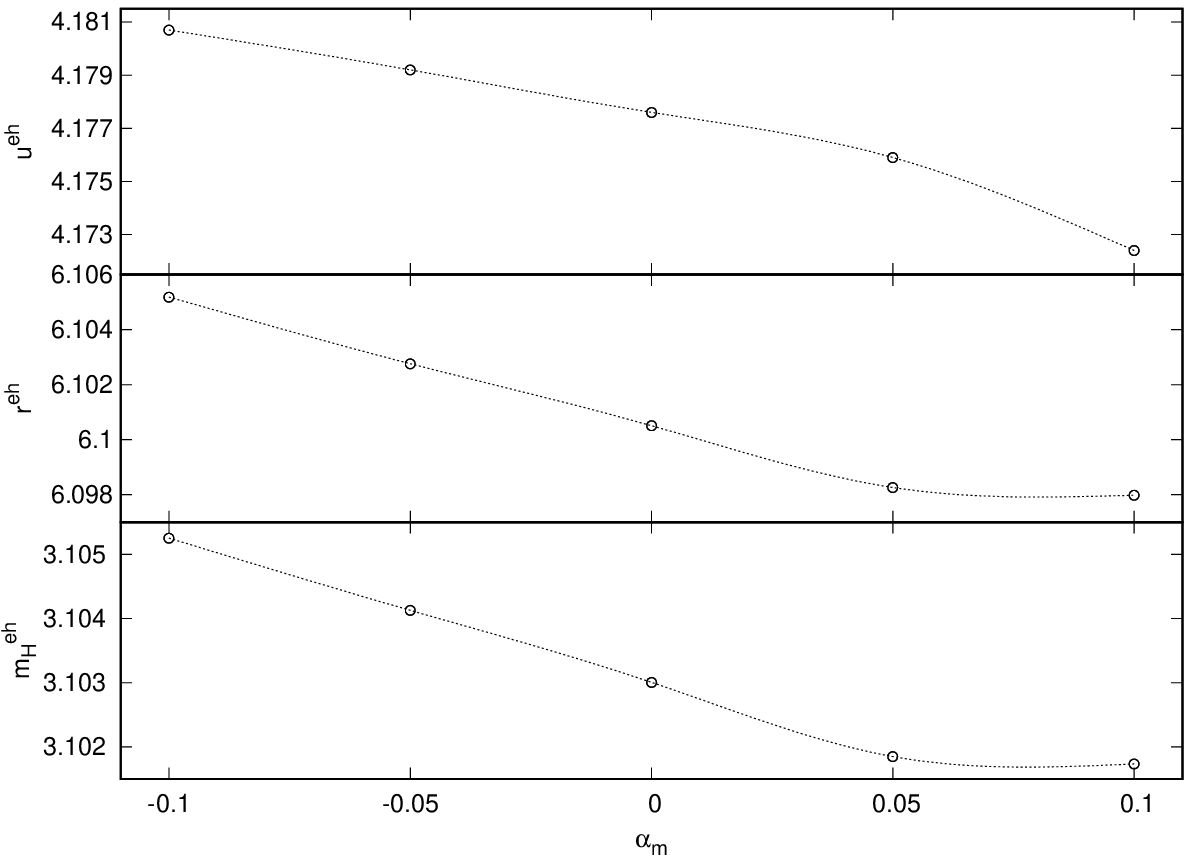}}\\
\subfigure[][]{\includegraphics[width=0.45\textwidth]{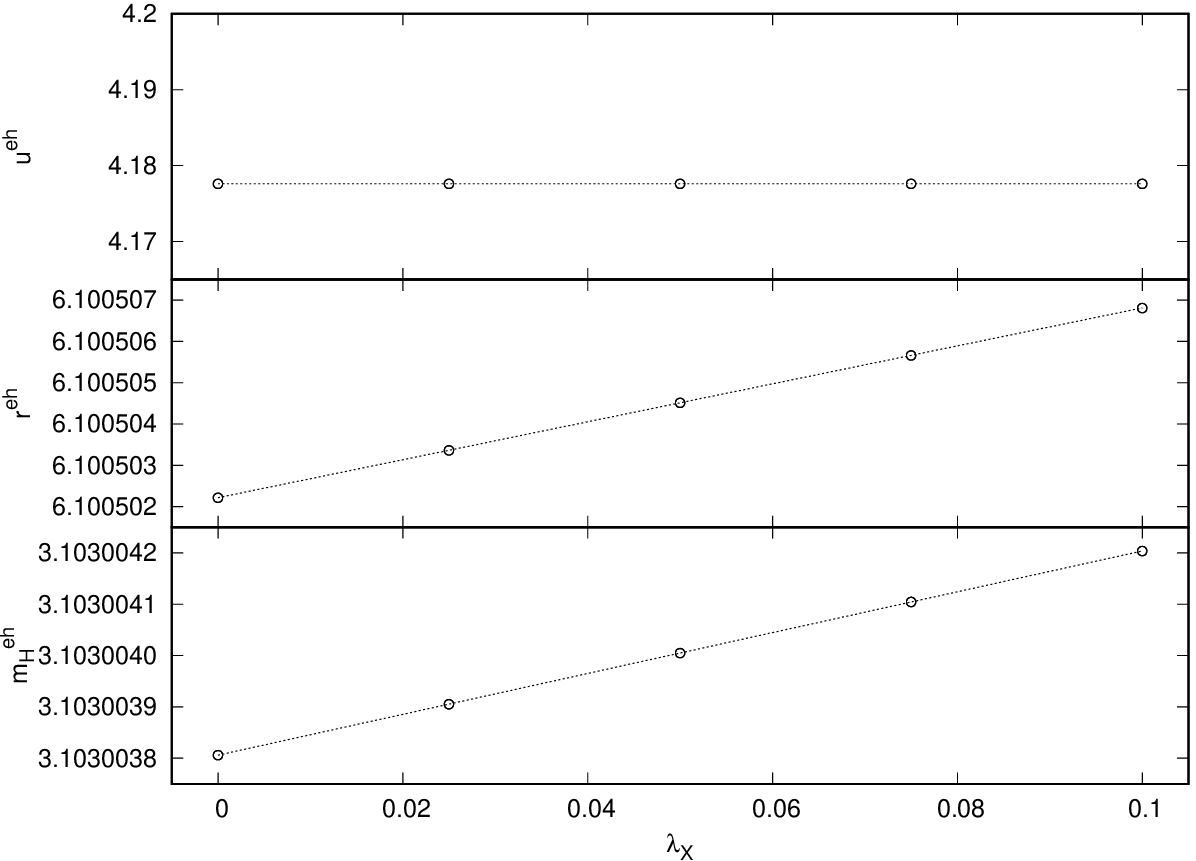}}
\hfill
\subfigure[][]{\includegraphics[width=0.45\textwidth]{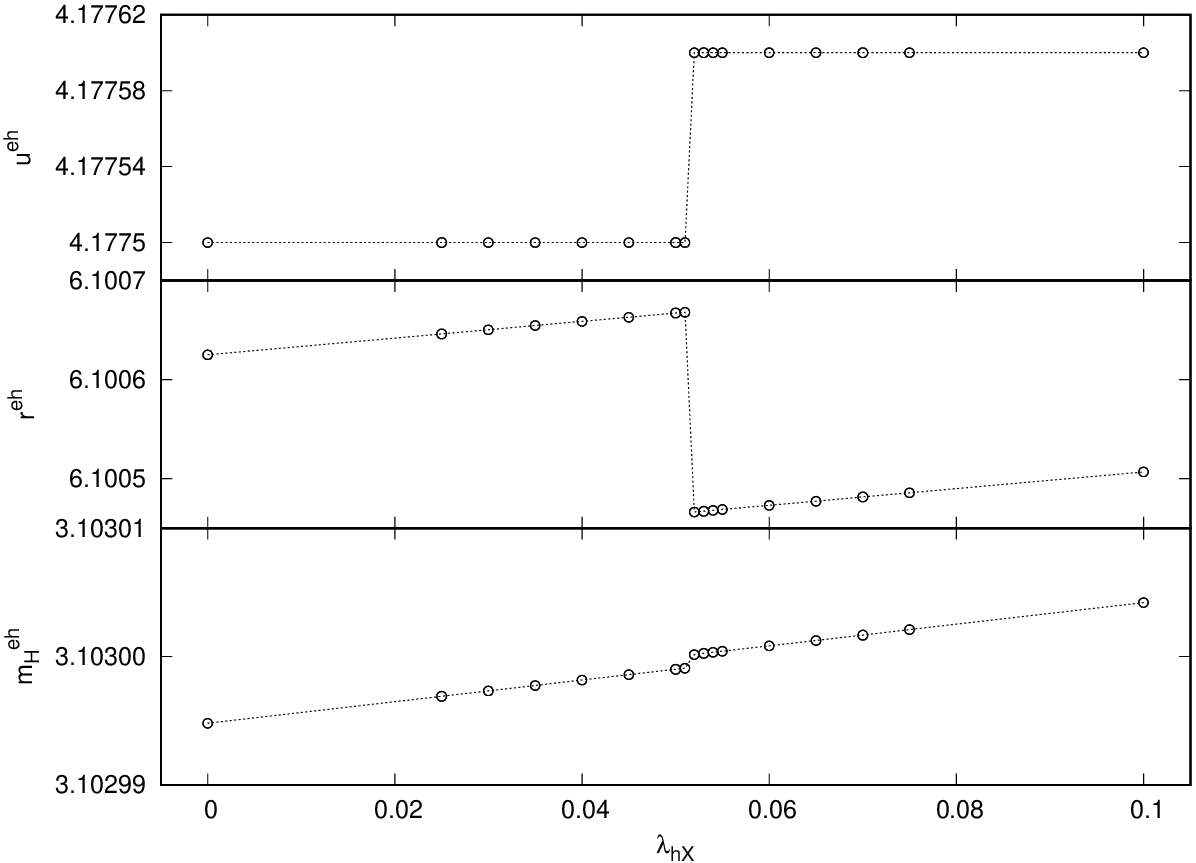}
\label{fig:noncoupl-char-urm-d}}
\caption{The $u$-locations of the event horizons,~$u^{eh}$, radii,~$r^{eh}$, and masses,~$m_H^{\ eh}$, of black holes formed during the gravitational collapse with scalars minimally coupled to gravity as functions of (a)~$m^2$, (b)~$\alpha_m$, (c)~$\lambda_X$ and (d)~$\lambda_{hX}$. The~non-varying parameters were $\lambda_X=\lambda_{hX}=0.1$, $m^2=0.25$, $\xi_X=\xi_h=0$, $\alpha_m=0$, $\aX=\ah=0.05$.}
\label{fig:noncoupl-char-urm}
\end{figure}

The dependence of the $u$-locations of the event horizons, radii and masses of black holes formed during the gravitational collapse within the model of interest as functions of $m^2$, $\alpha_m$, $\lambda_X$ and $\lambda_{hX}$ is presented in figure~\ref{fig:noncoupl-char-urm}. The~$u$-locations of the black hole event horizons increase as $m^2$ and $\lambda_{hX}$ increase, decreases with $\alpha_m$ and is not affected by a~value of $\lambda_X$. The~changes of the black hole radii and masses are the same qualitatively, that is they decrease with an increase of $m^2$ and $\alpha_m$ and increase with increasing quartic couplings. The~changes of $u^{eh}$, $r^{eh}$ and $m_H^{\ eh}$ against $m^2$ become much less dynamical as the mass parameter becomes bigger than $-0.15$ and the dependence on $\lambda_X$ is very weak within its whole range. Their changes with $\lambda_X$, $m^2$ and $\alpha_m$ are monotonic, while in the dependency on $\lambda_{hX}$ there is a~discontinuity for $\lambda_{hX}\approx 0.0515$.

\begin{figure}[tbp]
\begin{minipage}{0.5\textwidth}
\subfigure[][]{\includegraphics[width=0.83\textwidth]{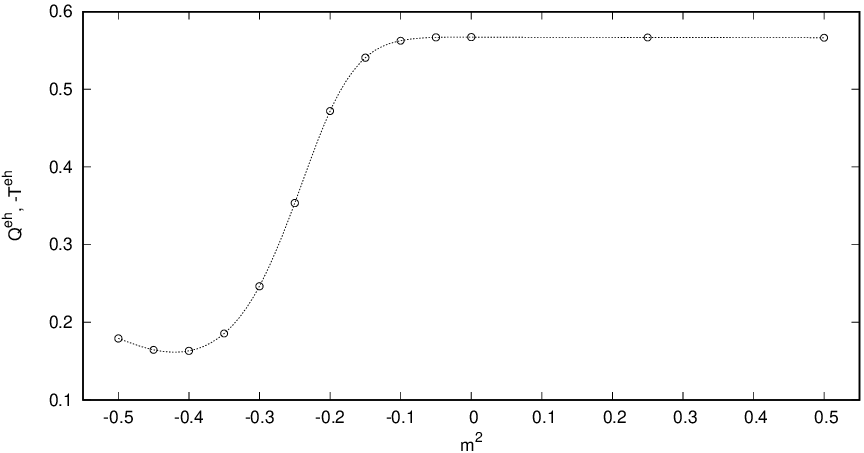}}
\end{minipage}
\begin{minipage}{0.5\textwidth}
\subfigure[][]{\includegraphics[width=0.83\textwidth]{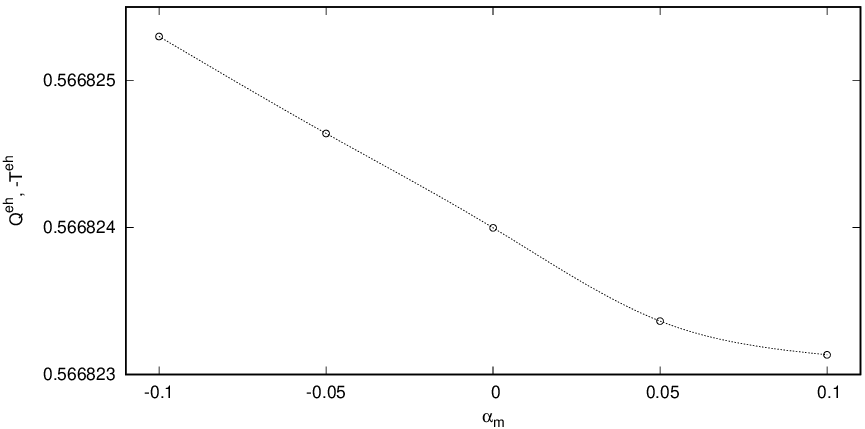}}
\end{minipage}
\begin{minipage}{0.5\textwidth}
\subfigure[][]{\includegraphics[width=0.83\textwidth]{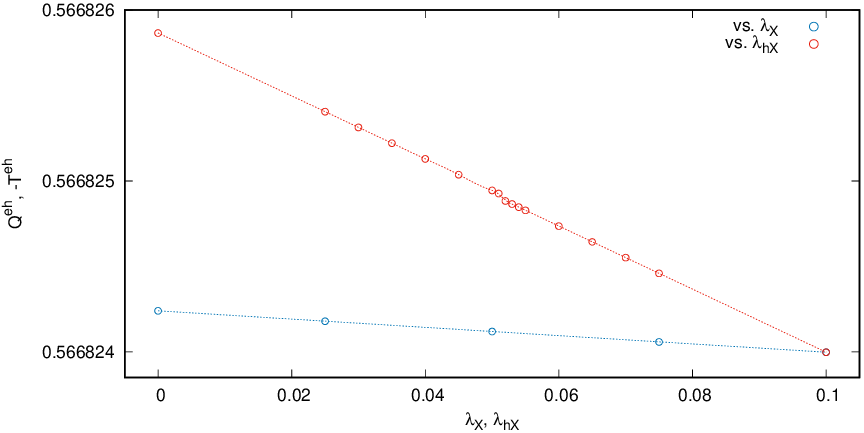}}
\end{minipage}
\begin{minipage}{0.475\textwidth}
\caption{(color online) The~black hole charges related to the $U(1)$ gauge fields, $Q^{eh}$ and $T^{eh}$, as functions of (a)~$m^2$, (b)~$\alpha_m$ and (c)~$\lambda_X,\lambda_{hX}$, for evolutions with scalars minimally coupled to gravity and non-varying parameters as in figure~\ref{fig:noncoupl-char-urm}.}
\label{fig:noncoupl-char-QT}
\end{minipage}
\end{figure}

The black hole charges related to the $U(1)$ gauge fields, $Q^{eh}$ and $T^{eh}$, versus $m^2$, $\alpha_m$ and $\lambda_X,\lambda_{hX}$ are depicted in figure~\ref{fig:noncoupl-char-QT}. An increase of $Q^{eh}$ with the mass parameter is small when the parameter exceeds $-0.05$. When its value is smaller, the dependence is not monotonic, the charge slightly decreases for small values of $m^2$ and after reaching a~minimum around $m^2=-0.4$, starts to increase. $Q^{eh}$ weakly decreases with $\alpha_m$ and both quartic couplings with a~characteristic discontinuity around $\lambda_{hX}=0.0515$, which was also observed for the black hole characteristics discussed above, as shown in figure~\ref{fig:noncoupl-char-urm-d}.

\begin{figure}[tbp]
\begin{minipage}{0.5\textwidth}
\subfigure[][]{\includegraphics[width=0.8\textwidth]{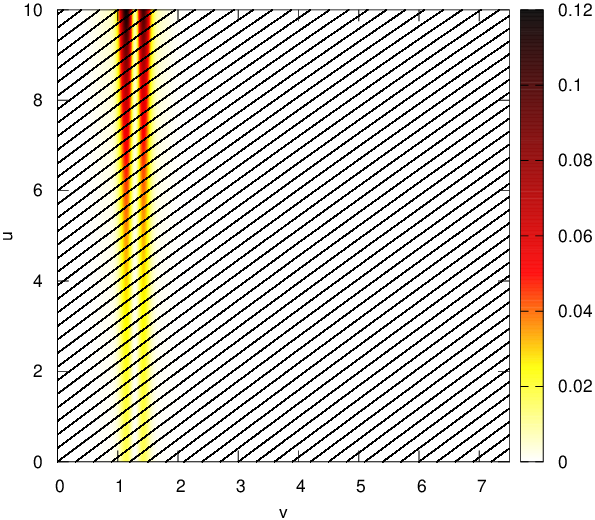}}
\end{minipage}
\begin{minipage}{0.5\textwidth}
\subfigure[][]{\includegraphics[width=0.8\textwidth]{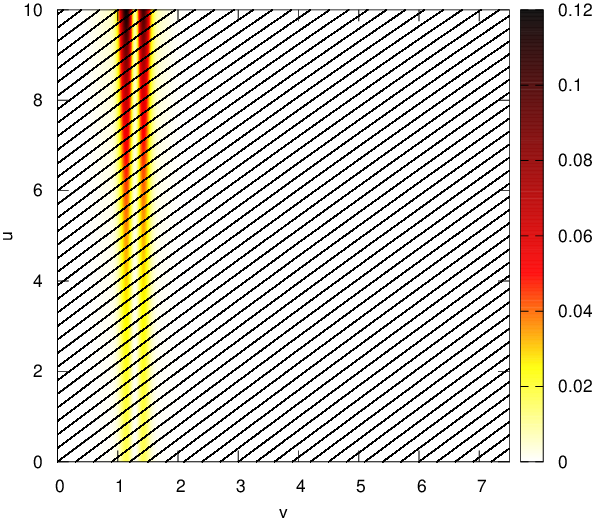}}
\end{minipage}
\begin{minipage}{0.5\textwidth}
\subfigure[][]{\includegraphics[width=0.8\textwidth]{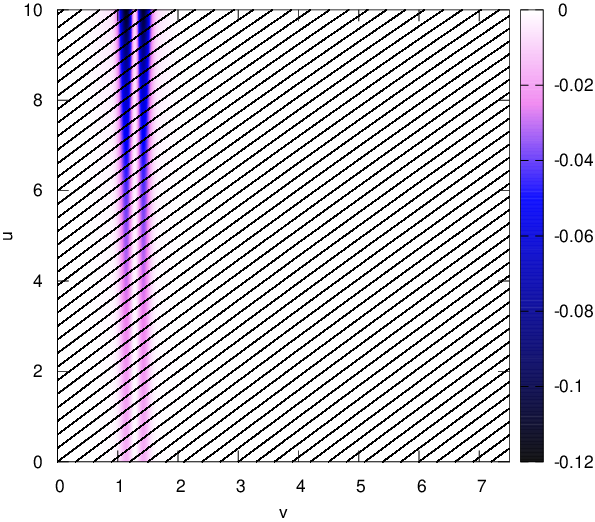}}
\end{minipage}
\begin{minipage}{0.475\textwidth}
\caption{(color online) The~$\left(vu\right)$-distribution of (a)~energy density, $\hat{\rho}$, (b)~radial pressure, $\hat{p}_{r}$, and (c)~pressure anisotropy, $\hat{p}_{a}$, for a~dynamical evolution characterized by parameters $\lambda_X=\lambda_{hX}=0.1$, $m^2=0.25$, $\xi_X=\xi_h=0$, $\alpha_m=0.1$ and the field amplitudes $\aX=\ah=0.01$.}
\label{fig:noncoupl-obs-nonsing}
\end{minipage}
\end{figure}

\begin{figure}[tbp]
\centering
\subfigure[][]{\includegraphics[width=0.38\textwidth]{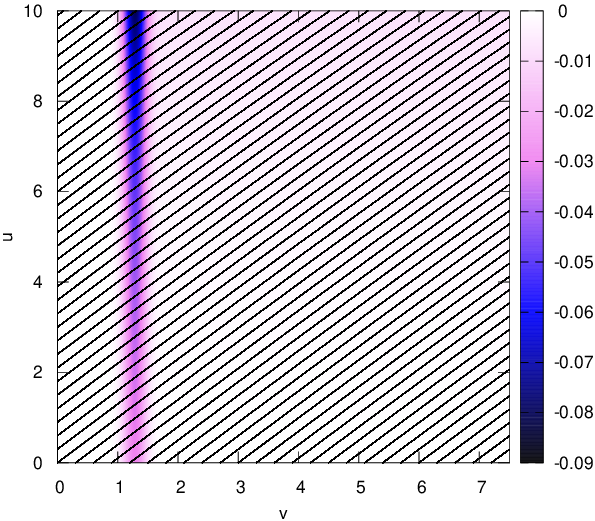}}
\hspace{1cm}
\subfigure[][]{\includegraphics[width=0.38\textwidth]{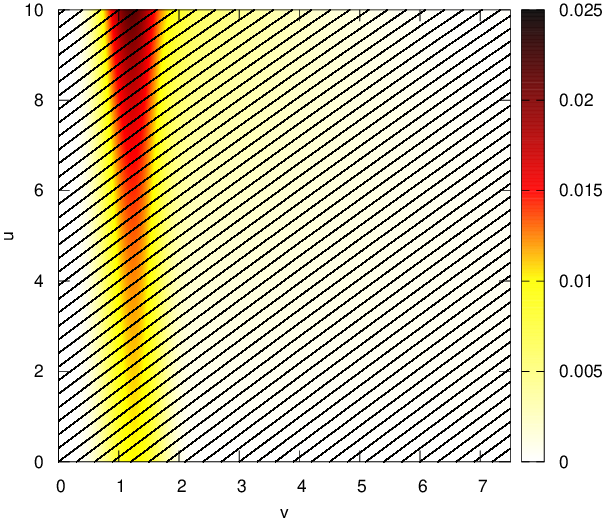}}
\caption{(color online) The~$\left(vu\right)$-distribution of (a)~the neutral scalar field, $h$, and (b)~the moduli of the complex scalar field, $|X|$, for the same parameters as in figure~\ref{fig:noncoupl-obs-nonsing}.}
\label{fig:noncoupl-fie-nonsing}
\end{figure}

\subsection{Observables and fields}
\label{sec:noncoupl-obs}

Figures~\ref{fig:noncoupl-obs-nonsing} and~\ref{fig:noncoupl-fie-nonsing} present the $\left(vu\right)$-distributions of observables defined in section~\ref{ssec:obs} and the evolving scalar fields, respectively, calculated for a~selected non-singular spacetime. Additionally, lines of constant $r$ were plotted on the graphs in order to visualize their behavior on a~Penrose diagram when the spacetime is flat. The~plots will serve as a~comparison for further analyses of singular spacetimes.

In a~spacetime that does not contain a~singularity, maximal absolute values of the energy density, radial pressure, pressure anisotropy, as well as the neutral scalar field and the moduli of the complex scalar field are distributed along a~null direction of constant advanced time. This is a~direction in which a~propagation of peaks of the field functions, whose profiles~\eqref{phi-prof} and~\eqref{psichi-prof} were imposed on the initial hypersurface $u=0$, advances as the dynamical collapse proceeds. 

\begin{figure}[tbp]
\centering
\subfigure[][]{\includegraphics[width=0.38\textwidth]{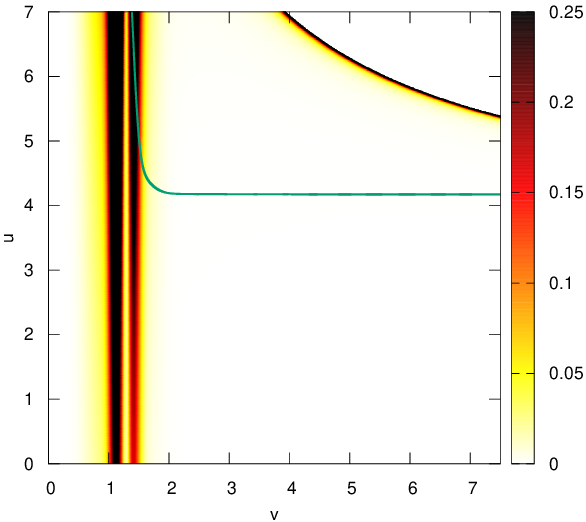}}
\hspace{1cm}
\subfigure[][]{\includegraphics[width=0.38\textwidth]{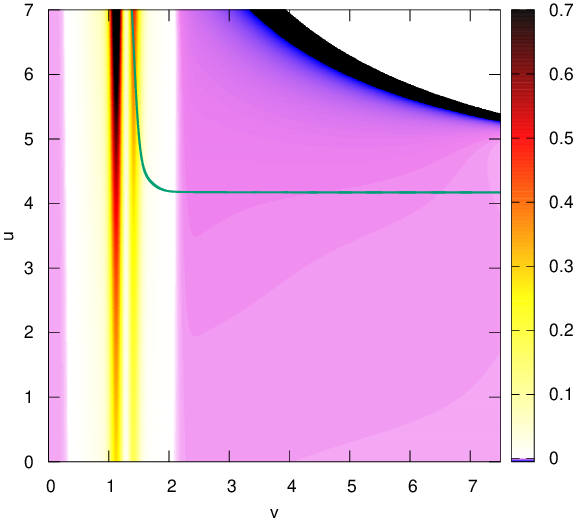}}\\
\subfigure[][]{\includegraphics[width=0.38\textwidth]{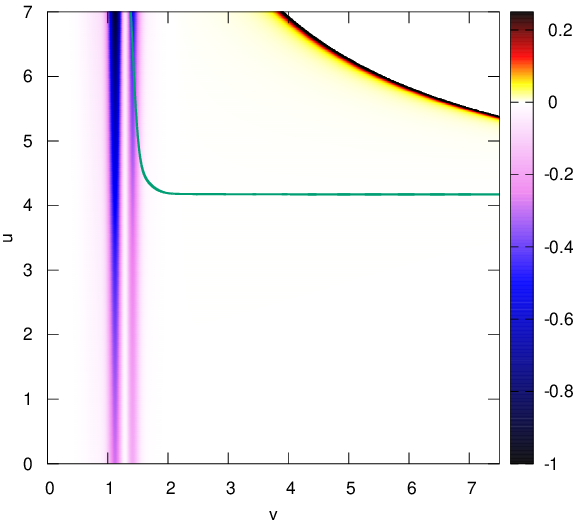}}
\hspace{1cm}
\subfigure[][]{\includegraphics[width=0.38\textwidth]{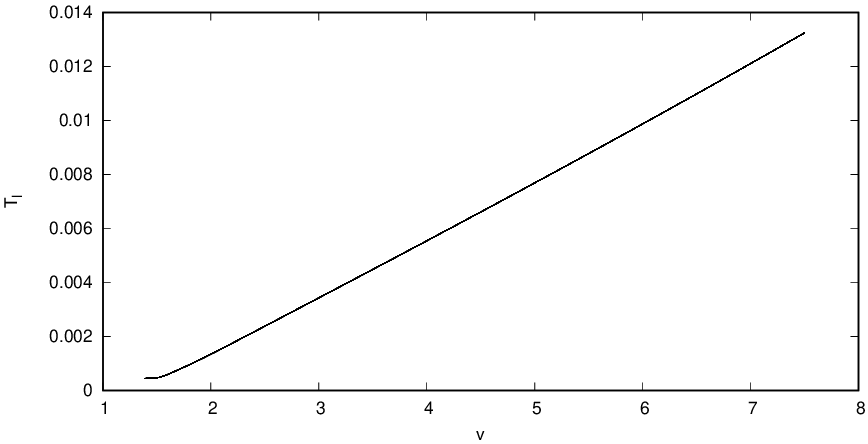}}
\caption{(color online) The~$\left(vu\right)$-distribution of (a)~energy density, $\hat{\rho}$, (b)~radial pressure, $\hat{p}_{r}$, and (c)~pressure anisotropy, $\hat{p}_{a}$, and (d)~local temperature along the black hole apparent horizon, $T_l$, as a~function of advanced time for a~dynamical evolution characterized by parameters $\lambda_X=\lambda_{hX}=0.1$, $m^2=0.25$, $\xi_X=\xi_h=0$, $\alpha_m=0.1$ (the same as in figure~\ref{fig:noncoupl-obs-nonsing}) and the field amplitudes $\aX=\ah=0.05$.}
\label{fig:noncoupl-obs-sing}
\end{figure}

\begin{figure}[tbp]
\centering
\subfigure[][]{\includegraphics[width=0.38\textwidth]{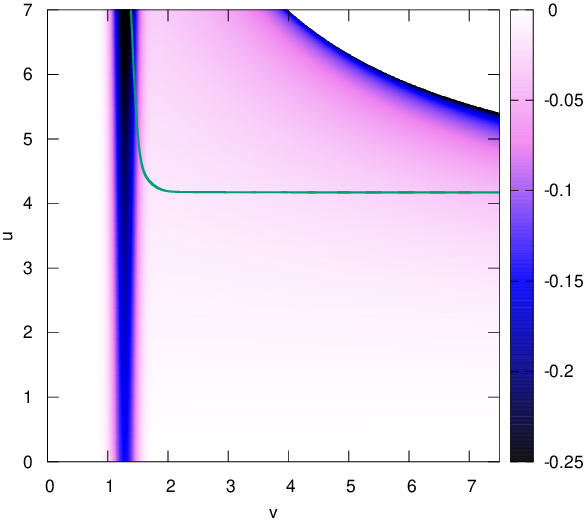}}
\hspace{1cm}
\subfigure[][]{\includegraphics[width=0.38\textwidth]{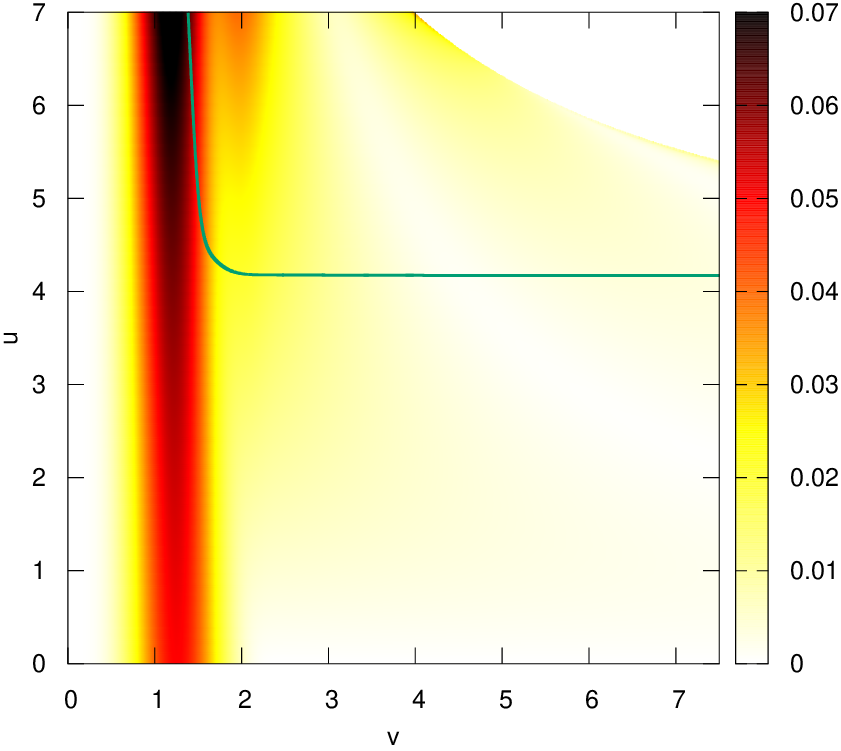}}
\caption{(color online) The~$\left(vu\right)$-distribution of (a)~the neutral scalar field, $h$, and (b)~the moduli of the complex scalar field, $|X|$, for the same parameters as in figure~\ref{fig:noncoupl-obs-sing}.}
\label{fig:noncoupl-fie-sing}
\end{figure}

Observables and fields distributions in a~spacetime containing a~black hole resulting from a~collapse with the same model parameters as the non-singular one described above and higher values of field amplitudes are presented in figures~\ref{fig:noncoupl-obs-sing} and~\ref{fig:noncoupl-fie-sing}. The~structure of the spacetime is presented in figure~\ref{fig:noncoupl-str-b}. As in the case of a~flat spacetime, peaks in absolute values of the energy density, radial pressure, pressure anisotropy and both neutral and complex scalar fields are observed along the null direction which indicates the field propagation in spacetime. Moreover, an increase of absolute values of the energy density, radial pressure and pressure anisotropy is visible in a~close vicinity of the central singularity. A~similar behavior is observed for the absolute values of the field $h$, while the complex scalar field with the modulus $|X|$ does not display such a~behavior as its values increase less considerably as the singularity is approached. All the quantities of interest remain finite within the whole spacetime, up to a~close vicinity of the central singularity, as far as it is reachable during numerical computations. The~energy density is positive within the whole spacetime. The~sign of the radial pressure and pressure anisotropy functions change as the singularity is approached, in comparison to the region of their high values located along a~constant value of advanced time. The~local temperature calculated along the black hole apparent horizon using the definition of surface gravity for dynamical spacetimes adopted in the current paper~\eqref{eqn:surfgrav} is positive and increases monotonically with advanced time. Such a~behavior as $v\to\infty$ is intuitive, as the black hole radius and mass also increase with advanced time.

\section{Spacetime and matter dynamics within the Higgs--dark matter toy model}
\label{sec:coupl}

The outcomes of simulations of gravitational dynamics within the non-truncated version of the model of interest, that includes the non-minimal scalars--gravity couplings, will be presented twofold. The~first set of results will refer to both non-zero coupling constatnts $\xi_X$ and $\xi_h$. The~second set will comprise the evolutions with one non-vanishing scalar--gravity coupling, either $\xi_X$ or $\xi_h$.

\subsection{Spacetime structures}
\label{sec:coupl-struc}

\begin{figure}[tbp]
\subfigure[][]{\includegraphics[width=0.4\textwidth]{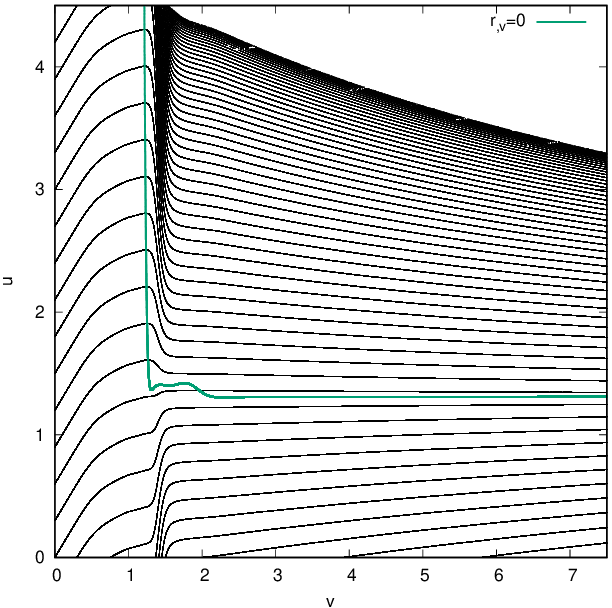}\label{fig:coupl-str-1a}}
\hfill
\subfigure[][]{\includegraphics[width=0.4\textwidth]{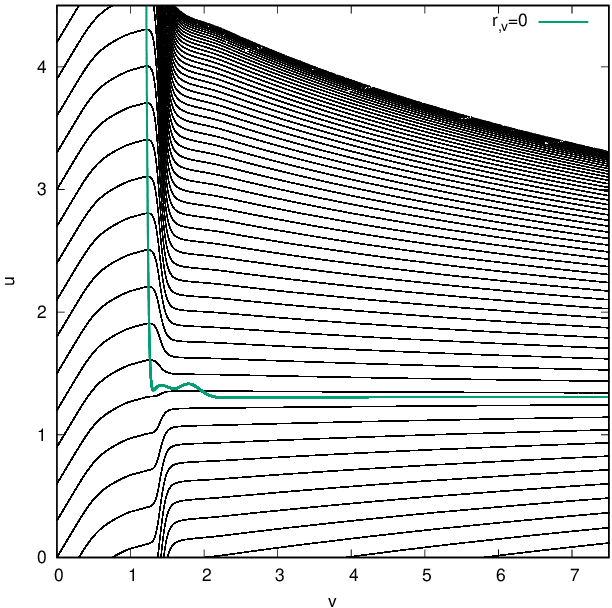}\label{fig:coupl-str-1b}}\\
\subfigure[][]{\includegraphics[width=0.4\textwidth]{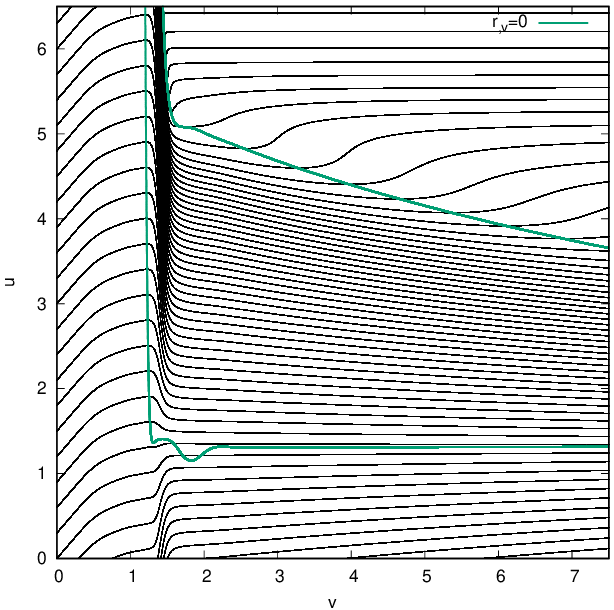}}
\hfill
\subfigure[][]{\includegraphics[width=0.4\textwidth]{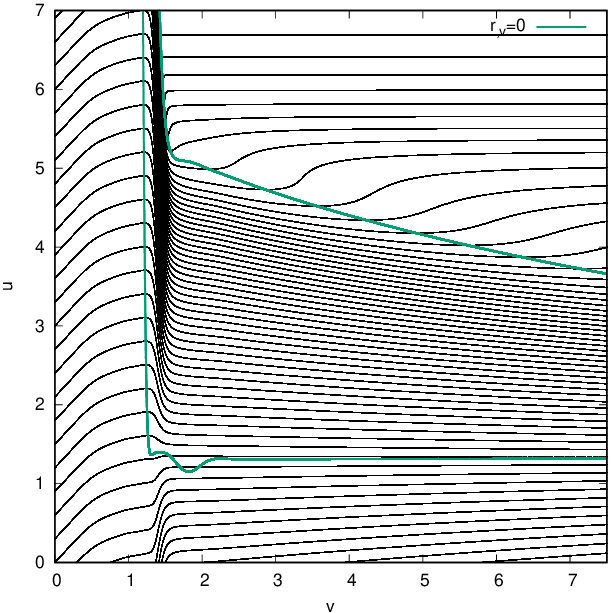}\label{fig:coupl-str-1d}}
\caption{(color online) Penrose diagrams of spacetimes emerging from evolutions with non-minimal scalar--gravity couplings. The~parameters $\lambda_{hX}$ and $\xi_X=\xi_h$ were equal to $0.1$ and $-0.5$, respectively. The~field amplitudes were $\aX=\ah=0.025$. The~remaining parameters were (a)~$\lambda_X=0.1$, $m^2=0.25$, $\alpha_m=0$, (b)~$\lambda_X=0.1$, $m^2=0.25$, $\alpha_m=0.1$, (c)~$\lambda_X=0$, $m^2=-0.25$, $\alpha_m=0$ and (d)~$\lambda_X=0$, $m^2=-0.25$, $\alpha_m=0.1$.}
\label{fig:coupl-str-1}
\end{figure}

Spacetime structures stemming from gravitational evolutions with $\xi_X\neq 0$ and $\xi_h\neq 0$ which lead to a~formation of a~black hole are presented in figure~\ref{fig:coupl-str-1}. For the cases with positive values of $m^2$, presented on the plots~\ref{fig:coupl-str-1a} and~\ref{fig:coupl-str-1b}, the spacetimes are dynamical Schwarzschild spacetimes with a~spacelike singular $r=0$ line surrounded by an apparent horizon that settles along a~$u=const.$ hypersurface as $v\to\infty$. The~non-vanishing coupling constants between scalar fields and gravity influence the behavior of the apparent horizon in the region where it changes its character from spacelike to null (between $v$ equal to $1.2$ and $2.3$). There are several timelike, spacelike and null portions of the horizon, in contrast to a~smooth behavior which was observed in the minimally coupled case, in figure~\ref{fig:noncoupl-str}.

When the parameter $m^2$ is negative, the collapse results in the dynamical Reissner-Nordstr\"{o}m--type spacetime. It contains a~spacelike central singularity along the vanishing $r$ line. The~singularity is surrounded by two apparent horizons. The~outer one is spacelike for small values of advanced time and becomes null as $v$ tends to infinity, indicating the location of the event horizon. Similarly to the case discussed above, there are several timelike, spacelike and null portions of the apparent horizon in the region where it transforms from spacelike to null. The~inner apparent horizon is spacelike within the whole domain of computations. An appearance of additional apparent horizons has been already observed during a~collapse involving exotic matter, precisely phantom scalar fields~\cite{NakoniecznaRogatkoModerski2012-044043}. There also exists a~Cauchy horizon in the emerging spacetime. It is located at an infinite-$v$ hypersurface. Its existence is manifested by a~collection of $r=const.$ lines that settle along constant-$u$ and are located between the apparent horizon and the singularity~\cite{OrenPiran2003-044013}.

\begin{figure}[tbp]
\subfigure[][]{\includegraphics[width=0.4\textwidth]{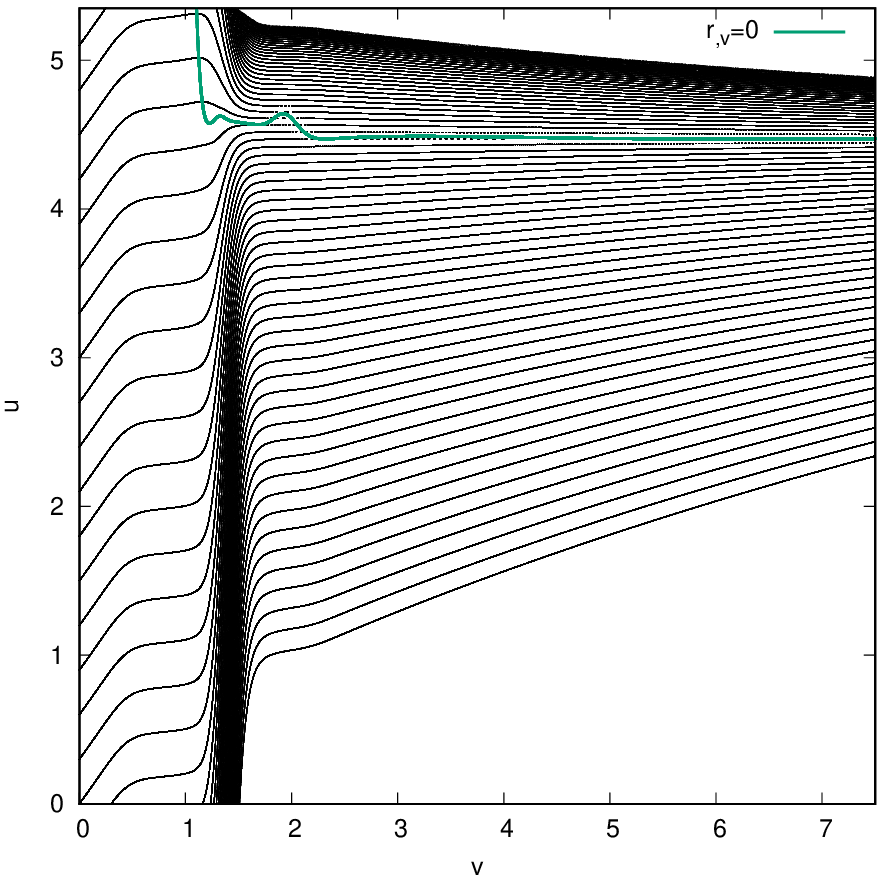}\label{fig:coupl-str-2a}}
\hfill
\subfigure[][]{\includegraphics[width=0.4\textwidth]{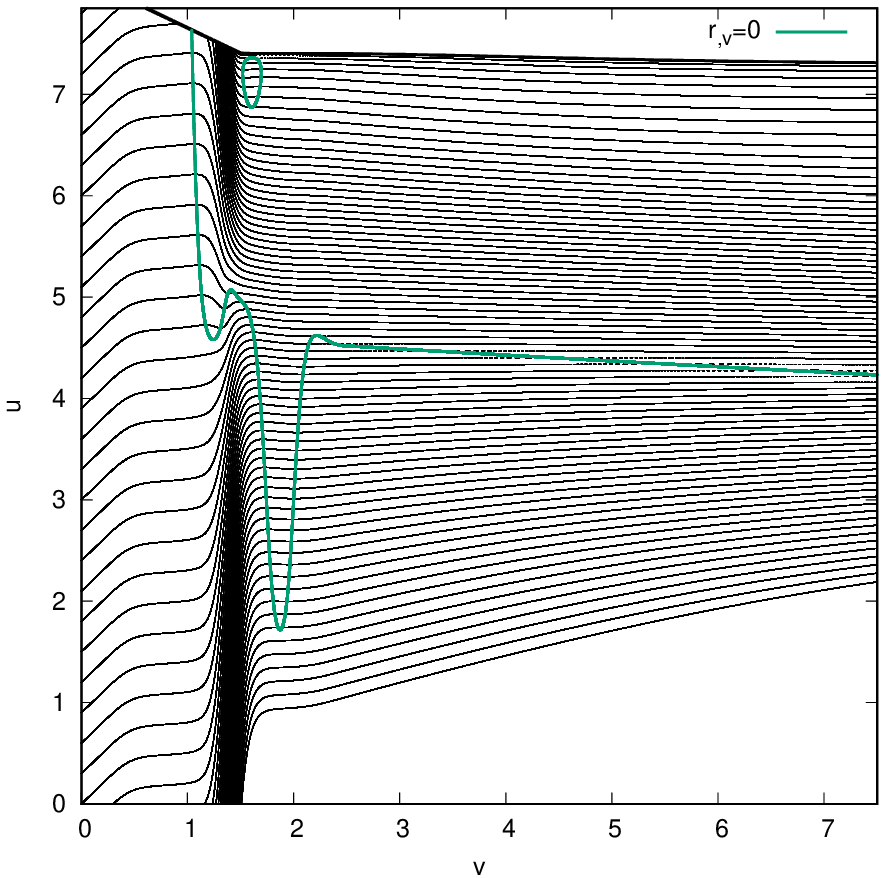}\label{fig:coupl-str-2b}}\\
\subfigure[][]{\includegraphics[width=0.4\textwidth]{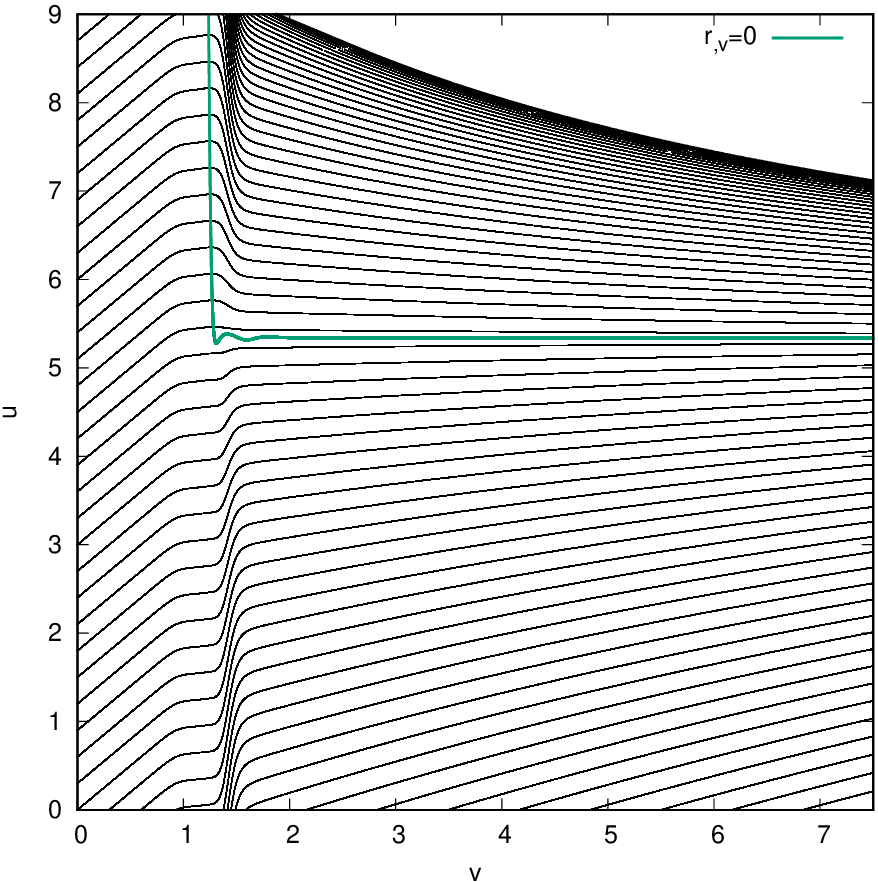}}
\hfill
\subfigure[][]{\includegraphics[width=0.4\textwidth]{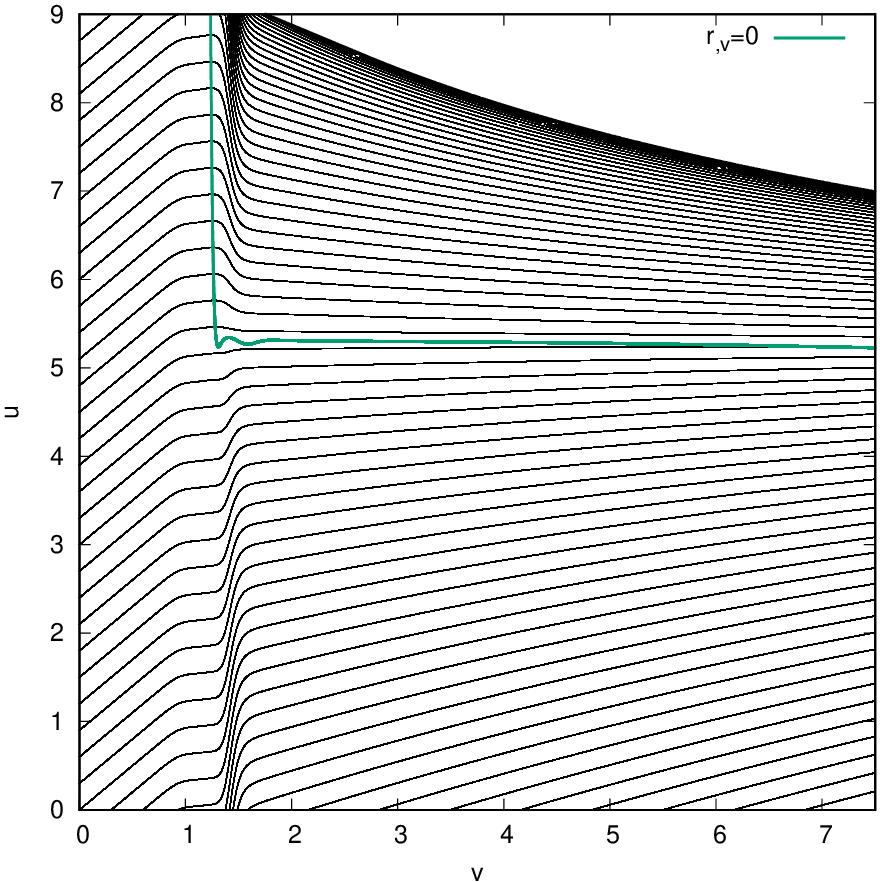}}
\caption{(color online) Penrose diagrams of spacetimes emerging from evolutions with one non-zero scalar--gravity coupling constant. The~parameters $\lambda_X$ and $\lambda_{hX}$ were equal to $0.1$ and $10^{-6}$, respectively. The~remaining parameters were (a)~$m^2=0.25$, $\xi_X=-1$, $\xi_h=0$, $\alpha_m=0$, $\aX=\ah=0.035$, (b)~$m^2=-0.25$, $\xi_X=-1$, $\xi_h=0$, $\alpha_m=0.1$, $\aX=\ah=0.035$, (c)~$m^2=0.25$, $\xi_X=0$, $\xi_h=-1$, $\alpha_m=0.1$, $\aX=\ah=0.015$, and (d)~$m^2=-0.25$, $\xi_X=0$, $\xi_h=-1$, $\alpha_m=0.1$, $\aX=\ah=0.015$.}
\label{fig:coupl-str-2}
\end{figure}

Penrose diagrams of spacetimes stemming from the gravitational collapse within the investigated model with one non-zero scalar--gravity coupling constant are shown in figure~\ref{fig:coupl-str-2}. In all cases the spacetimes are dynamical Schwarzschild spacetimes. Central singularities along $r=0$ are spacelike within them and surrounded by apparent horizons, which are spacelike for small values of $v$ and become null as $v$ increases. The~behavior of the apparent horizons in the regions where they transform from spacelike to null again involves several changes of their character between timelike, spacelike and null. These changes are more significant when $\xi_X\neq 0$, in comparison with evolutions with a~non-vanishing $\xi_h$.

\subsection{Black hole characteristics}
\label{sec:coupl-char}

In the case of both non-zero scalar--gravity coupling constants, the characteristics of forming black holes were examined for the evolution with the following parameters: $\lambda_X=0$, $\lambda_{hX}=0.1$, $m^2=-0.25$, $\xi_X=\xi_h=-0.1$, $\alpha_m=0.1$ and $\aX=\ah=0.025$. The~results for the case when only one scalar--gravity coupling constant $\xi_h$ does not vanish were obtained for the parameters $\lambda_X=0.1$, $\lambda_{hX}=10^{-6}$, $m^2=0.25$, $\xi_h=-1$, $\alpha_m=0.1$ and $\aX=\ah=0.015$. While the dependence on the particular parameter of the model is presented, the remaining ones are as above.

\begin{figure}[tbp]
\begin{minipage}{0.5\textwidth}
\subfigure[][]{\includegraphics[width=0.9\textwidth]{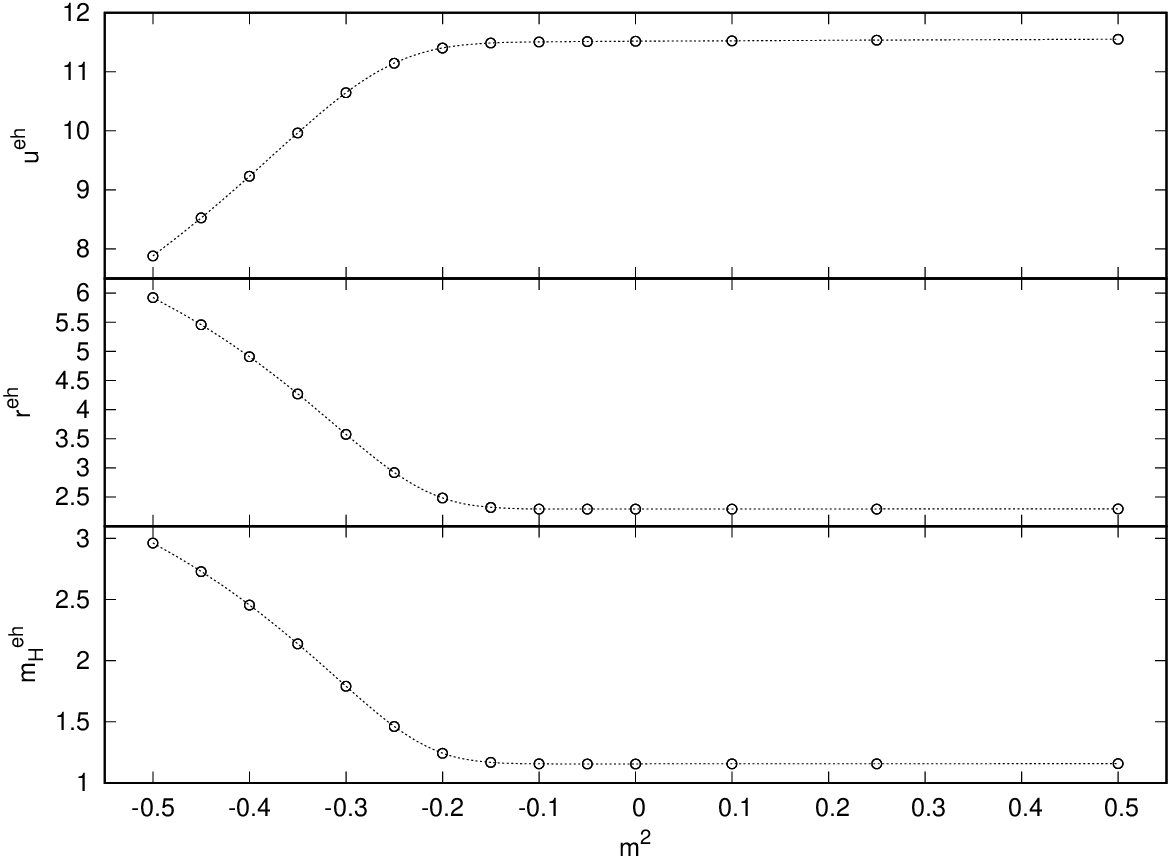}}
\end{minipage}
\hfill
\begin{minipage}{0.5\textwidth}
\subfigure[][]{\includegraphics[width=0.9\textwidth]{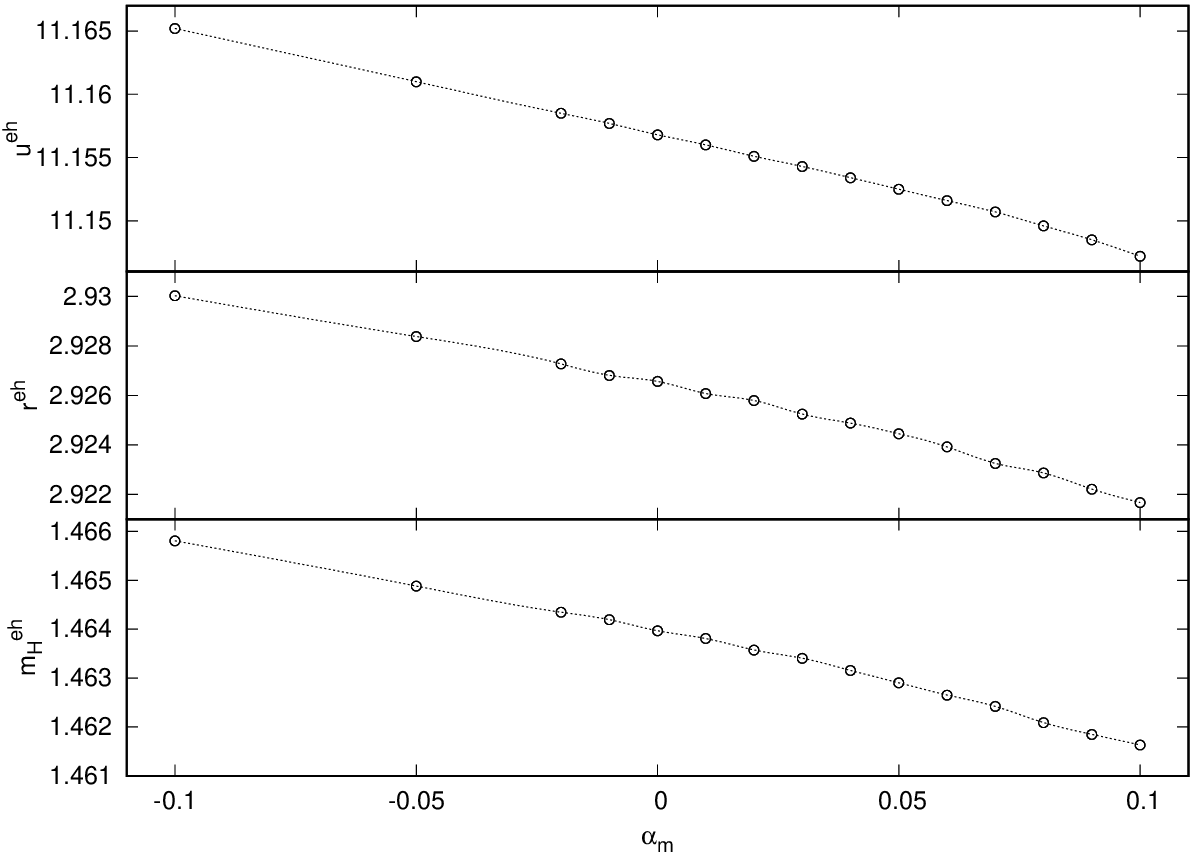}}
\end{minipage}
\begin{minipage}{0.5\textwidth}
\subfigure[][]{\includegraphics[width=0.9\textwidth]{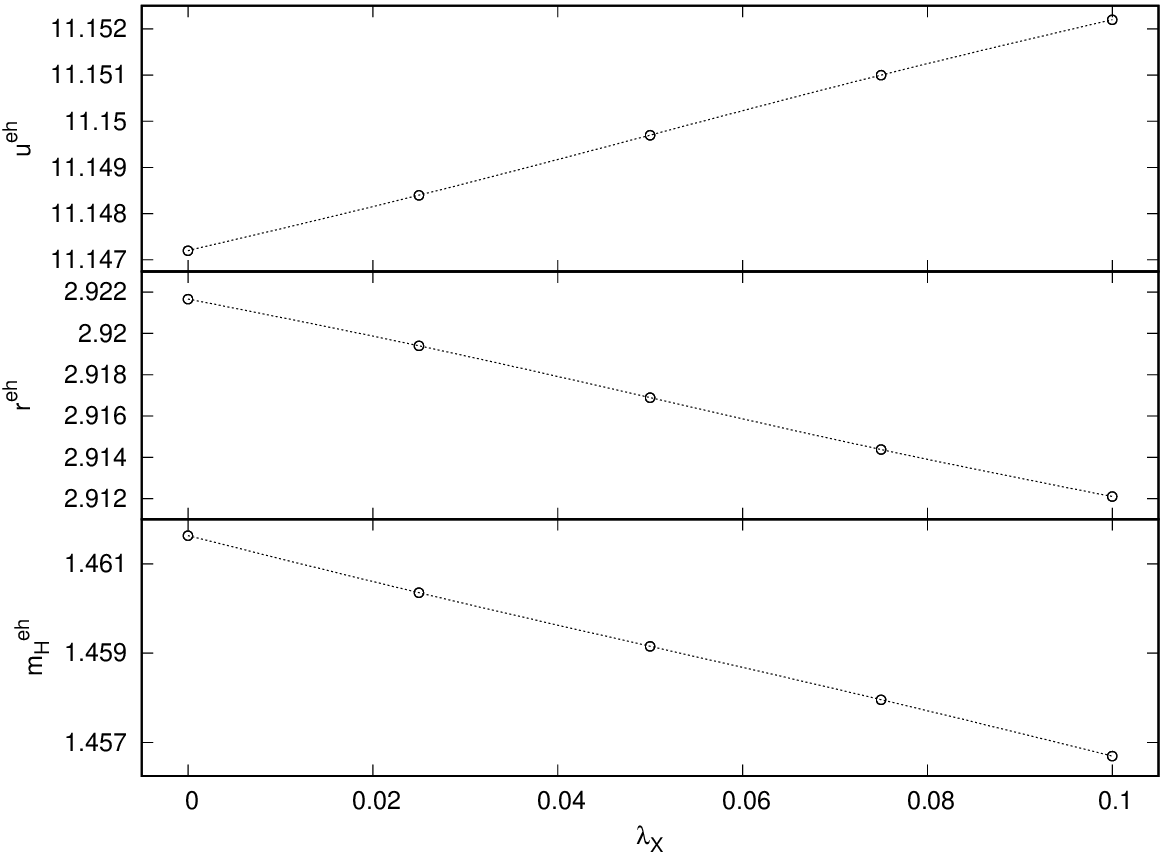}}
\end{minipage}
\hfill
\begin{minipage}{0.5\textwidth}
\subfigure[][]{\includegraphics[width=0.9\textwidth]{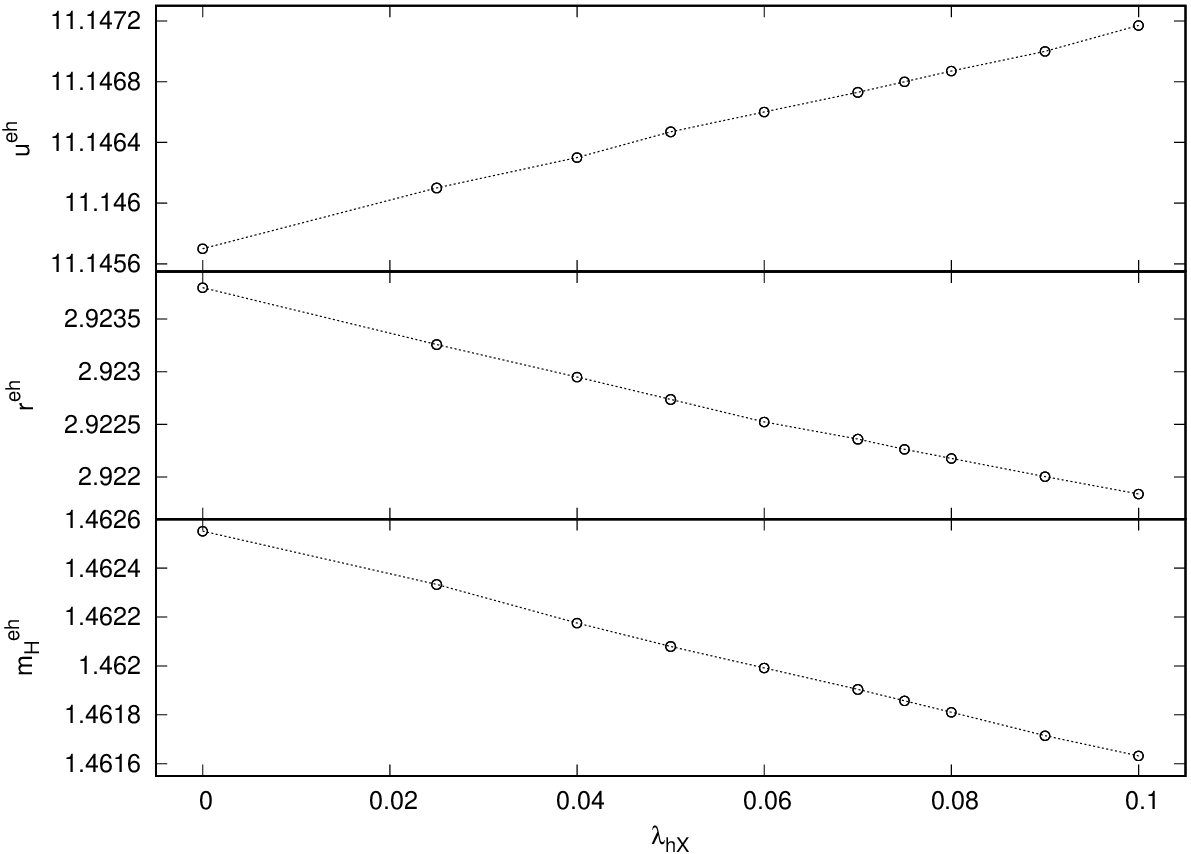}}
\end{minipage}
\begin{minipage}{0.5\textwidth}
\subfigure[][]{\includegraphics[width=0.9\textwidth]{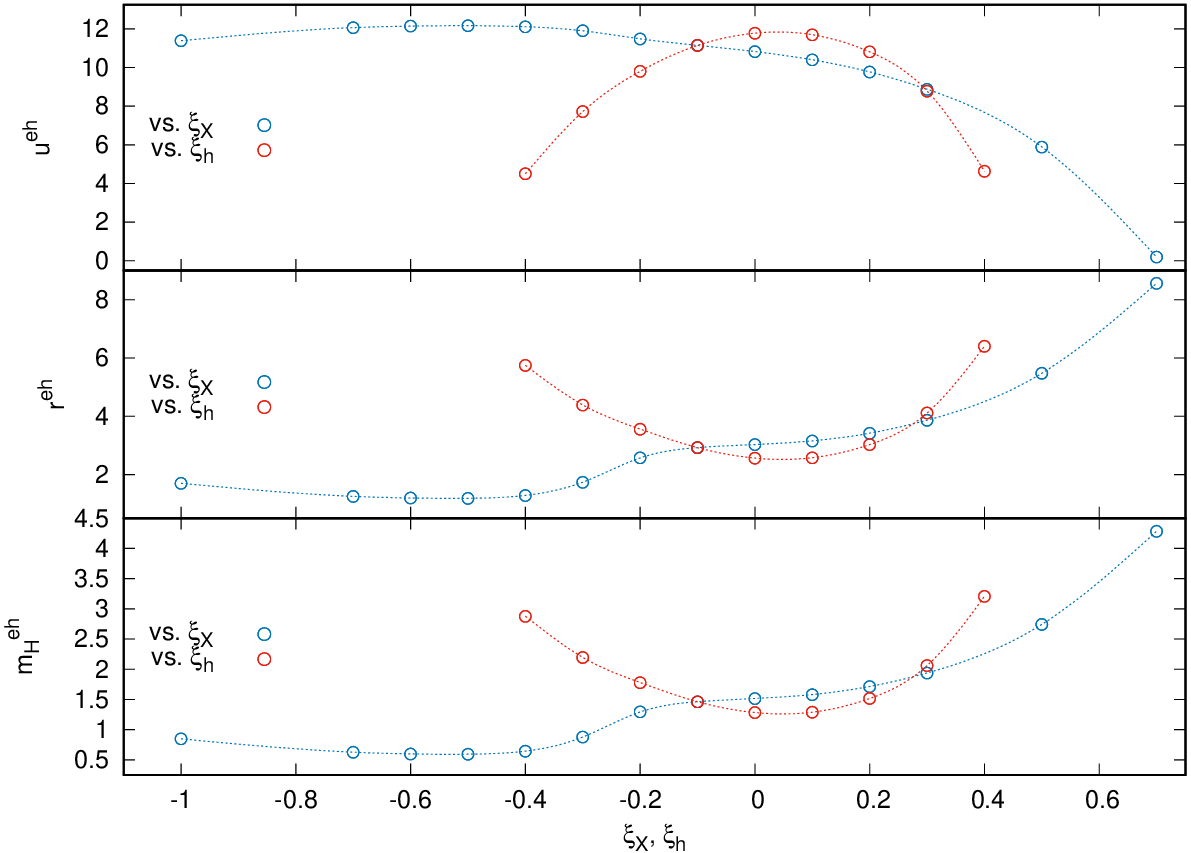}}
\end{minipage}
\hfill
\begin{minipage}{0.475\textwidth}
\caption{(color online) The~$u$-locations of the event horizons,~$u^{eh}$, radii,~$r^{eh}$, and masses,~$m_H^{\ eh}$, of black holes formed during the gravitational collapse with non-minimal scalar--gravity couplings as functions of (a)~$m^2$, (b)~$\alpha_m$, (c)~$\lambda_X$, (d)~$\lambda_{hX}$ and (e)~$\xi_X,\xi_h$. The~non-varying parameters were $\lambda_X=0$, $\lambda_{hX}=0.1$, $m^2=-0.25$, $\xi_X=\xi_h=-0.1$, $\alpha_m=0.1$, $\aX=\ah=0.025$.}
\label{fig:coupl2-char-urm}
\end{minipage}
\end{figure}

\begin{figure}[tbp]
\begin{minipage}{0.5\textwidth}
\subfigure[][]{\includegraphics[width=0.9\textwidth]{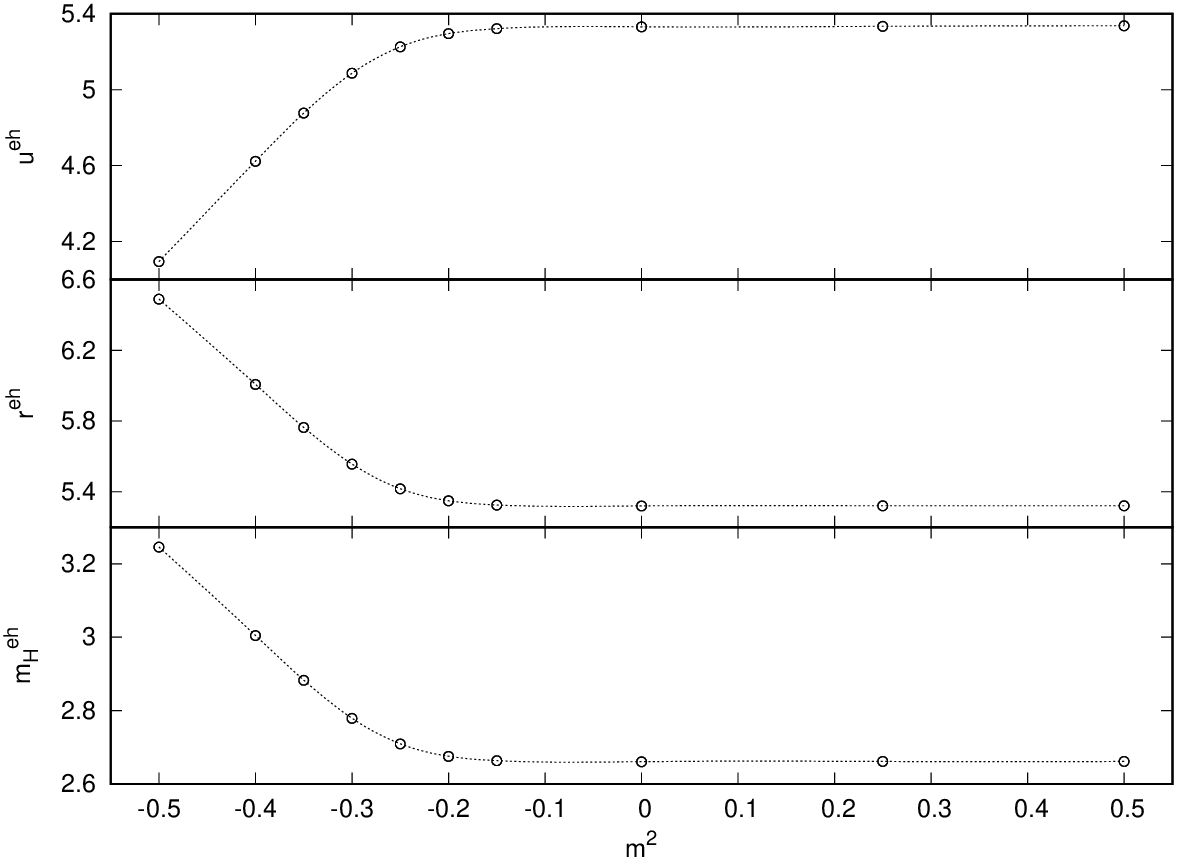}}
\end{minipage}
\hfill
\begin{minipage}{0.5\textwidth}
\subfigure[][]{\includegraphics[width=0.9\textwidth]{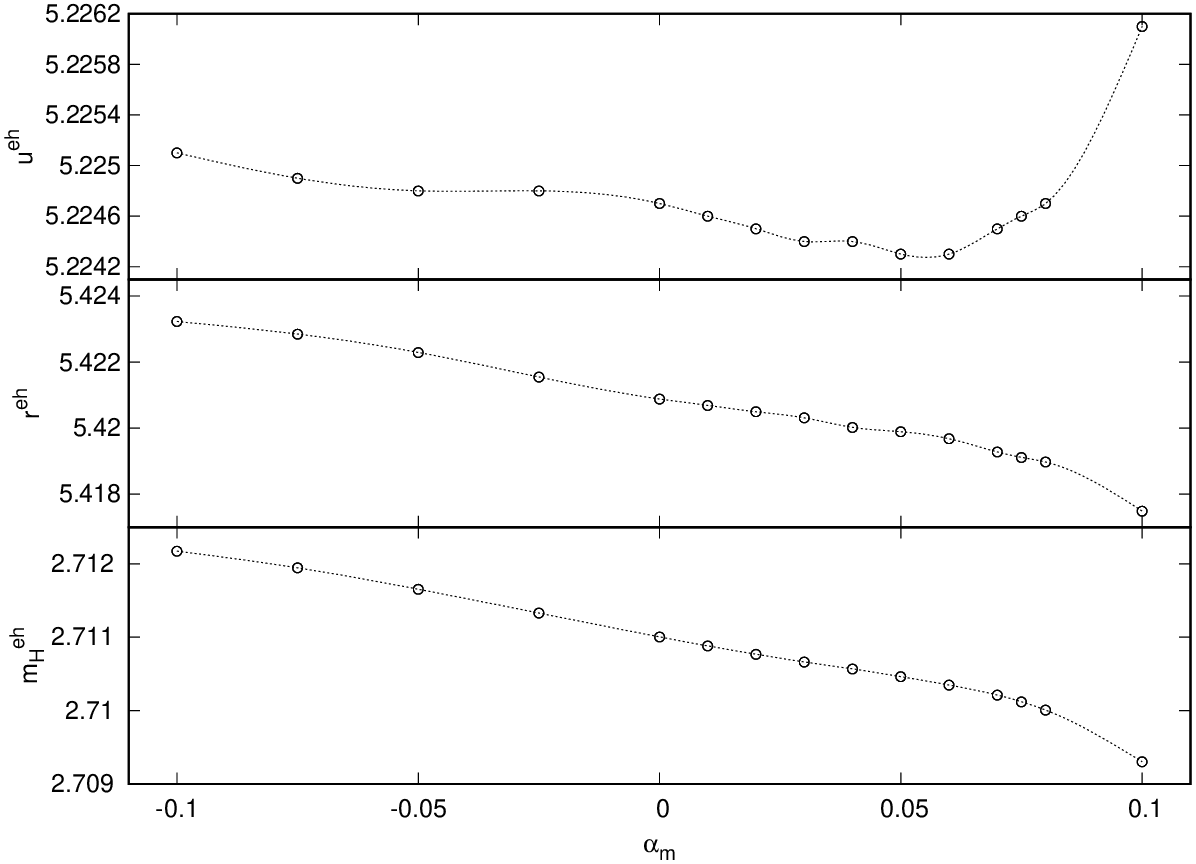}}
\end{minipage}
\begin{minipage}{0.5\textwidth}
\subfigure[][]{\includegraphics[width=0.9\textwidth]{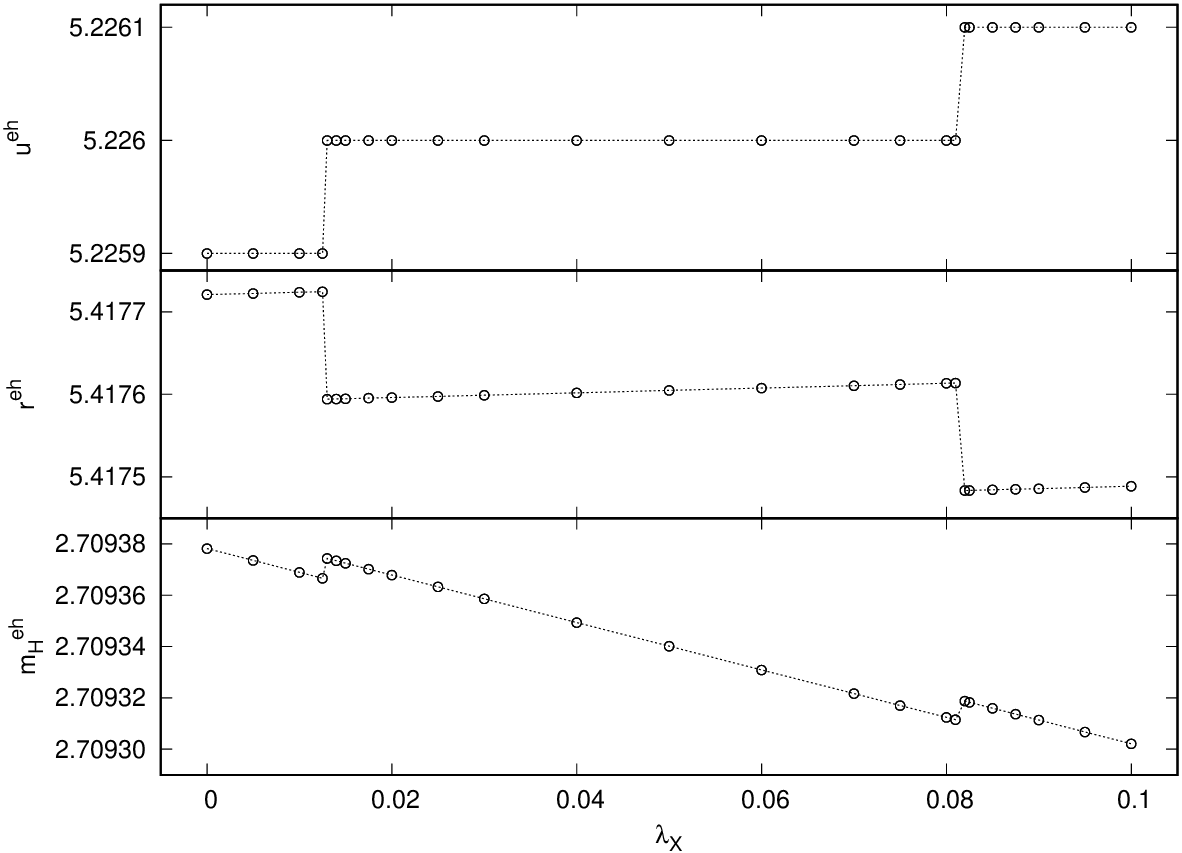}}
\end{minipage}
\hfill
\begin{minipage}{0.5\textwidth}
\subfigure[][]{\includegraphics[width=0.9\textwidth]{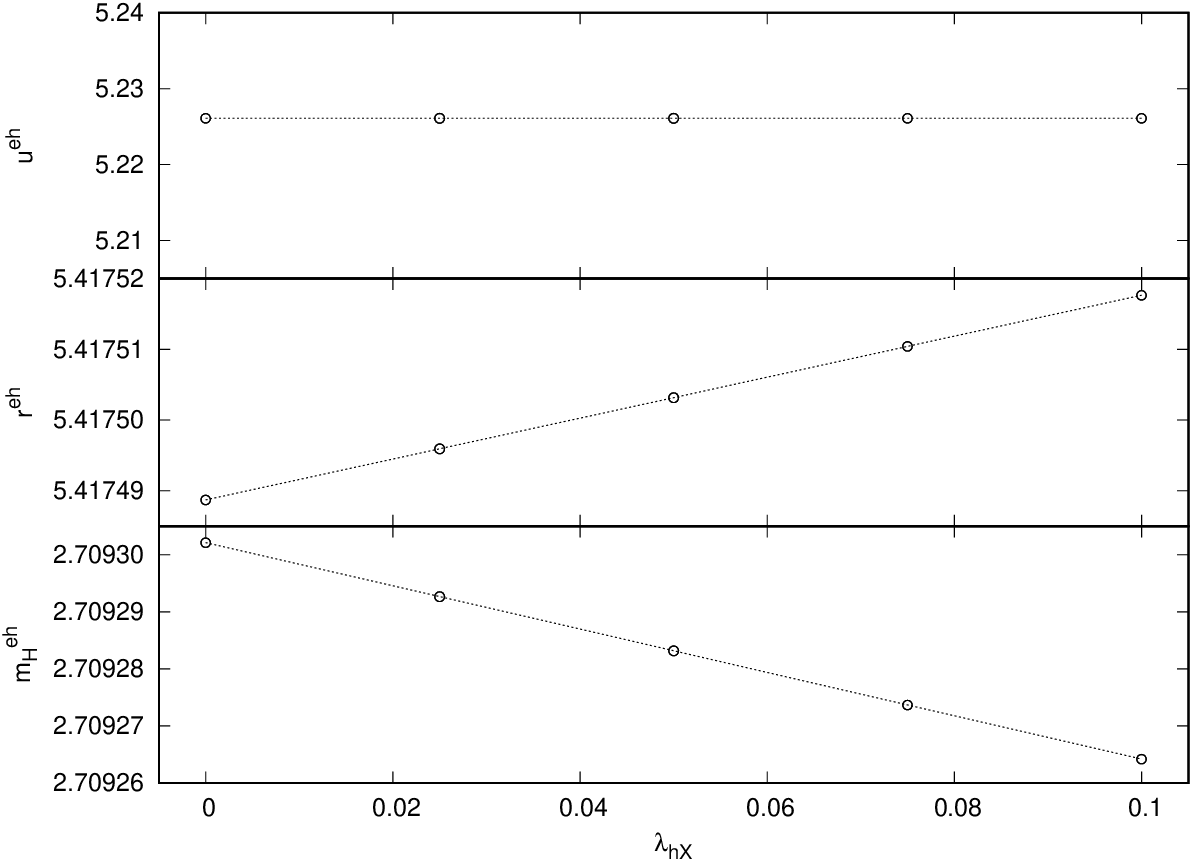}}
\end{minipage}
\begin{minipage}{0.5\textwidth}
\subfigure[][]{\includegraphics[width=0.9\textwidth]{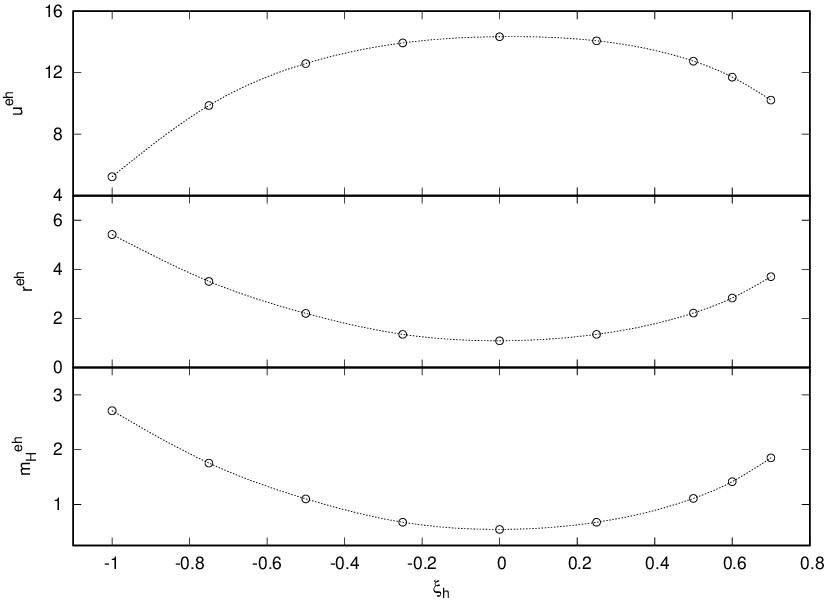}}
\end{minipage}
\hfill
\begin{minipage}{0.475\textwidth}
\caption{The $u$-locations of the event horizons,~$u^{eh}$, radii,~$r^{eh}$, and masses,~$m_H^{\ eh}$, of black holes formed during the gravitational collapse with one non-vanishing scalar--gravity coupling constant as functions of (a)~$m^2$, (b)~$\alpha_m$, (c)~$\lambda_X$, (d)~$\lambda_{hX}$ and (e)~$\xi_h$. The~non-varying parameters were $\lambda_X=0.1$, $\lambda_{hX}=10^{-6}$, $m^2=0.25$, $\xi_h=-1$, $\alpha_m=0$, $\aX=\ah=0.015$.}
\label{fig:coupl3-char-urm}
\end{minipage}
\end{figure}

The dependence of the $u$-locations of the event horizons, radii and masses of black holes formed during the gravitational collapse within the model of interest as functions of $m^2$, $\alpha_m$, $\lambda_X$, $\lambda_{hX}$ and $\xi_X,\xi_h$ is presented in figures~\ref{fig:coupl2-char-urm} and~\ref{fig:coupl3-char-urm} for the cases of two and one non-zero scalar--gravity couplings, respectively. The~dependencies of all these quantities on the mass parameter are qualitatively the same in both cases. After a~significant increase of $u^{eh}$ and decrease of $r^{eh}$ and $m_H^{\ eh}$ up to $m^2$ equal approximately $-0.15$ their values vary much less. This behavior is similar to the one observed in the minimally coupled case discussed in section~\ref{sec:noncoupl-char}. All the discussed characteristics decrease with $\alpha_m$ in both examined cases, except $u^{eh}$ when $\xi_X=0$, whose value, after a~decrease, begins to increase for large values of $\alpha_m$. The~decrease was also observed in the simplified version of the model with $\xi_X=\xi_h=0$. In the case of two non-vanishing scalar--gravity coupling constants, the relations between $u^{eh}$, $r^{eh}$, $m_H^{\ eh}$ and the quartic couplings are qualitatively the same, i.e., the first of the quantities increases and the remaining ones decrease as $\lambda_X$ and $\lambda_{hX}$ increase. In the dependencies on $\lambda_X$ in the case when $\xi_X=0$, which are qualitatively the same as the ones discussed above, two discontinuities around $\lambda_X$ equal to $0.013$ and $0.0815$ appear. The~$u$-locations of the event horizons do not change with $\lambda_{hX}$, while the radii and masses of nascent black holes increase and decrease with the parameter, respectively. The~changes of $u^{eh}$, $r^{eh}$ and $m_H^{\ eh}$ with the quartic self-interaction coupling constant of the Higgs field are qualitatively the same in both cases of two and one non-zero scalar--gravity couplings. There is a~maximal value of $u^{eh}$ and minima of $r^{eh}$ and $m_H^{\ eh}$, present at $\xi_h=0$. The~relations between the black hole features and the quartic self-interaction coupling constant of the complex scalar field in the case of $\xi_X\neq 0$ and $\xi_h\neq 0$ also display a~maximum in the case of the $u$-location of the event horizon and maxima in the cases of black hole radii and masses, but these shallow extrema are located at $\xi_X=-0.5$.

\begin{figure}[tbp]
\subfigure[][]{\includegraphics[width=0.47\textwidth]{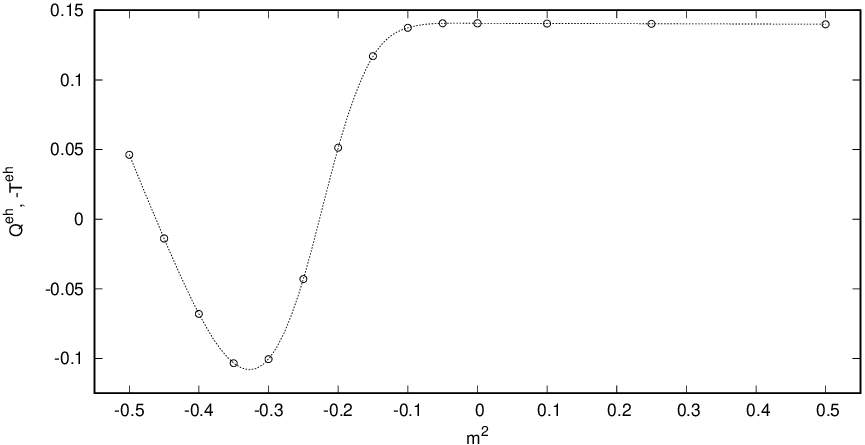}}
\hfill
\subfigure[][]{\includegraphics[width=0.47\textwidth]{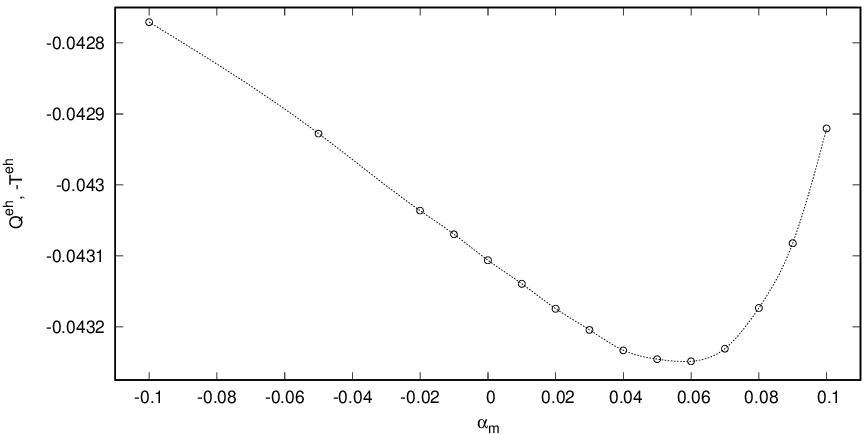}}\\
\subfigure[][]{\includegraphics[width=0.47\textwidth]{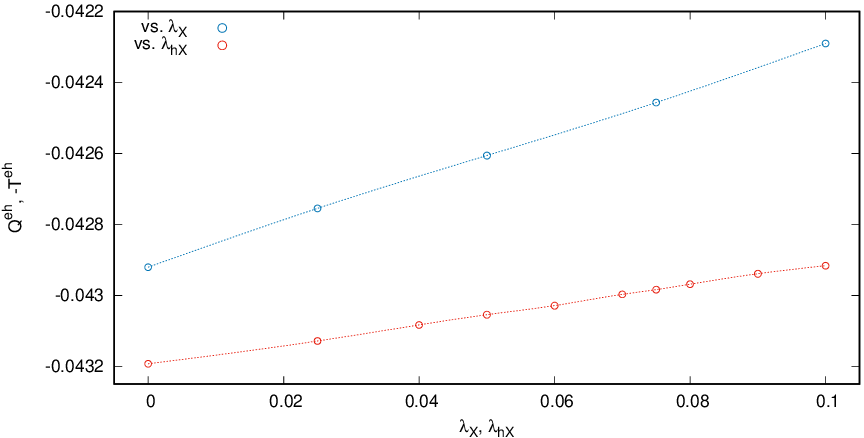}}
\hfill
\subfigure[][]{\includegraphics[width=0.47\textwidth]{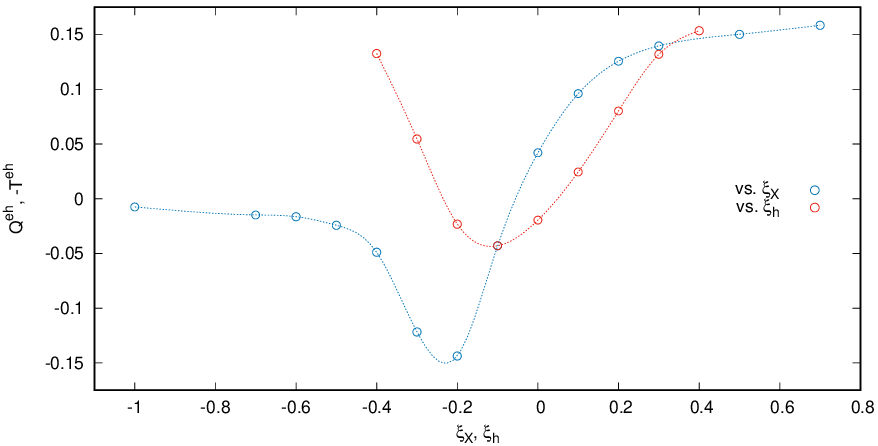}}
\caption{(color online) The~black hole charges related to the $U(1)$ gauge fields, $Q^{eh}$ and $T^{eh}$, as functions of (a)~$m^2$, (b)~$\alpha_m$, (c)~$\lambda_X,\lambda_{hX}$ and (d)~$\xi_X,\xi_h$, for evolutions with non-minimal scalar--gravity couplings and non-varying parameters as in figure~\ref{fig:coupl2-char-urm}.}
\label{fig:coupl2-char-QT}
\end{figure}

\begin{figure}[tbp]
\subfigure[][]{\includegraphics[width=0.47\textwidth]{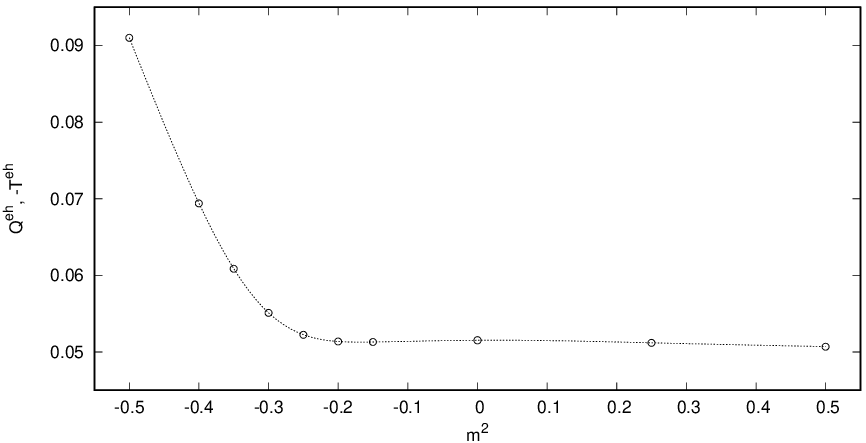}}
\hfill
\subfigure[][]{\includegraphics[width=0.47\textwidth]{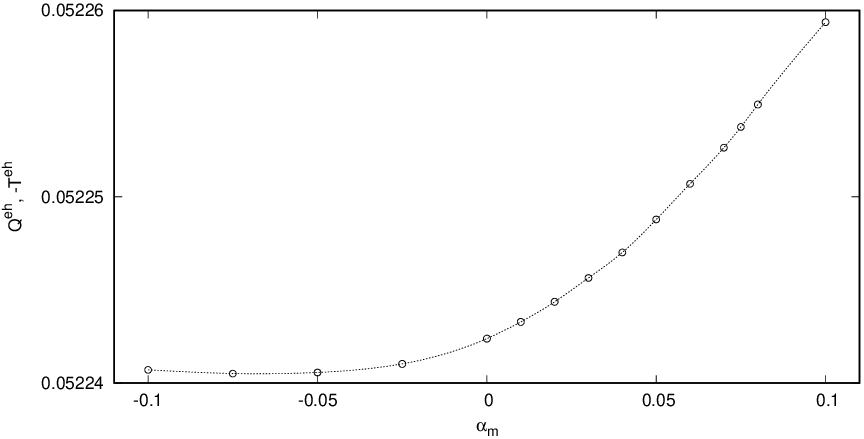}}\\
\subfigure[][]{\includegraphics[width=0.47\textwidth]{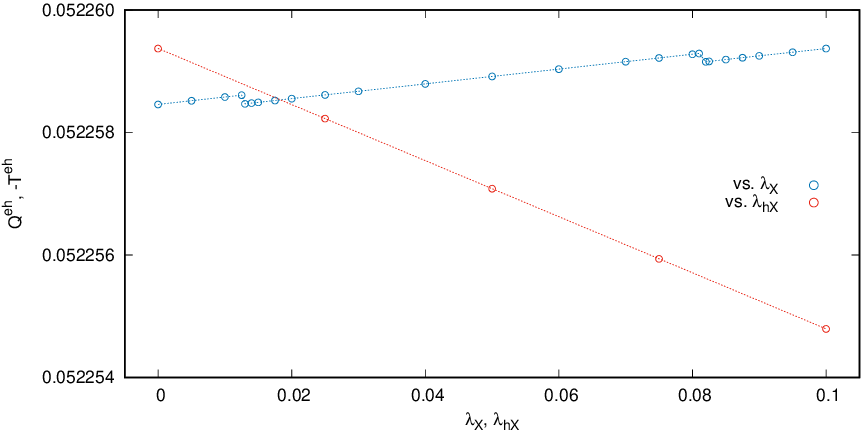}}
\hfill
\subfigure[][]{\includegraphics[width=0.47\textwidth]{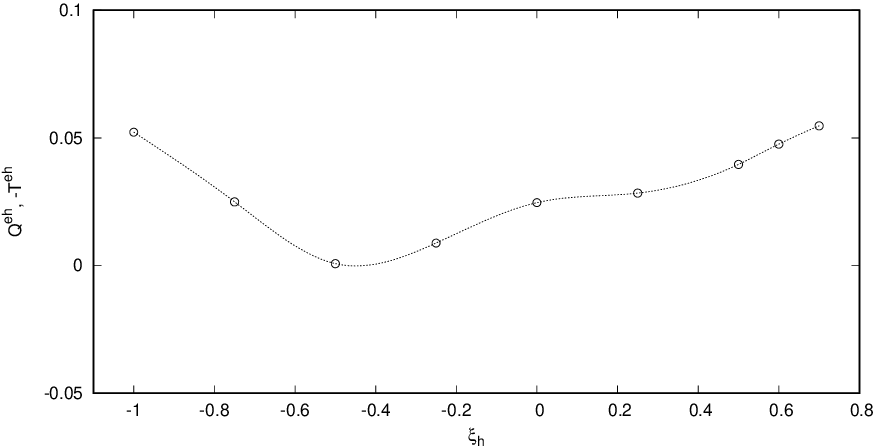}}
\caption{(color online) The~black hole charges related to the $U(1)$ gauge fields, $Q^{eh}$ and $T^{eh}$, as functions of (a)~$m^2$, (b)~$\alpha_m$, (c)~$\lambda_X,\lambda_{hX}$ and (d)~$\xi_h$, for evolutions with one non-minimal scalar--gravity coupling and non-varying parameters as in figure~\ref{fig:coupl3-char-urm}.}
\label{fig:coupl3-char-QT}
\end{figure}

The black hole charges related to the $U(1)$ gauge fields, $Q^{eh}$ and $T^{eh}$, as functions of $m^2$, $\alpha_m$, $\lambda_X,\lambda_{hX}$ and $\xi_X,\xi_h$ are depicted in figures~\ref{fig:coupl2-char-QT} and~\ref{fig:coupl3-char-QT}. When both non-minimal scalar--gravity couplings do not vanish, the relation between the charges and $m^2$ is similar to the one observed for the minimally coupled case, described in section~\ref{sec:noncoupl-char}. It possesses a~minimum around $m^2=-0.35$ and its variations become small when the parameter becomes bigger than $-0.05$. For the case with $\xi_X=0$, $Q^{eh}$ decreases within the whole $m^2$ range and the changes become less significant when the mass parameter exceeds $0.2$. The~dependency of the charge on $\alpha_m$ displays a~minimum at $0.06$ when both non-minimal couplings are non-zero and is monotonically increasing when only $\xi_h\neq 0$. The~charge increases as $\lambda_X$ and $\lambda_{hX}$ increase when both scalar--gravity couplings do not vanish. It decreases with $\lambda_{hX}$ and increases with two discontinuities at $0.013$ and $0.0815$ as $\lambda_X$ increases in the case with $\xi_X=0$. The~dependencies against $\xi_X$ and $\xi_h$ possess local minima, situated at $\xi_X=-0.25$ and $\xi_h$ equal to $-0.1$ and $-0.45$ for the cases of two and one non-minimal scalar--gravity couplings, respectively.

\subsection{Observables and fields}
\label{sec:coupl-obs}

The $\left(vu\right)$-distributions of observables defined in section~\ref{ssec:obs} and the evolving scalar fields along with the relation between local temerature along the apparent horizon and $v$ will be depicted and discussed for selected spacetimes whose structures were shown in section~\ref{sec:coupl-struc}. Two representative cases with spacetimes resulting from gravitational evolutions with two non-zero scalar--gravity coupling constants will be presented. The~behavior of the quantities of interest within spacetimes, which stem from collapses when one of the non-minimal couplings vanishes are qualitatively the same.

\begin{figure}[tbp]
\centering
\subfigure[][]{\includegraphics[width=0.38\textwidth]{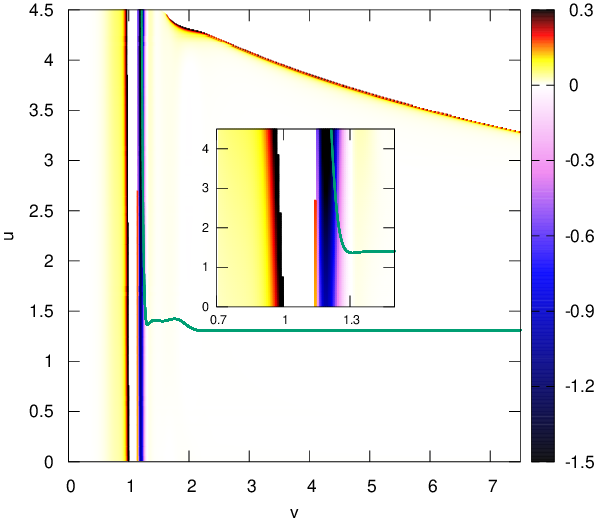}}
\hspace{1cm}
\subfigure[][]{\includegraphics[width=0.38\textwidth]{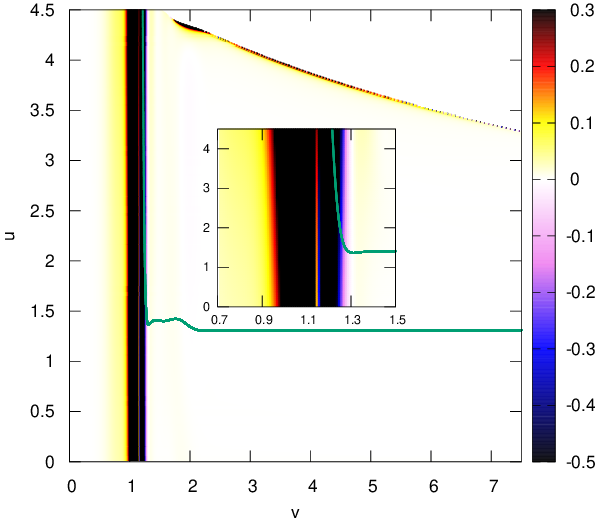}}\\
\subfigure[][]{\includegraphics[width=0.38\textwidth]{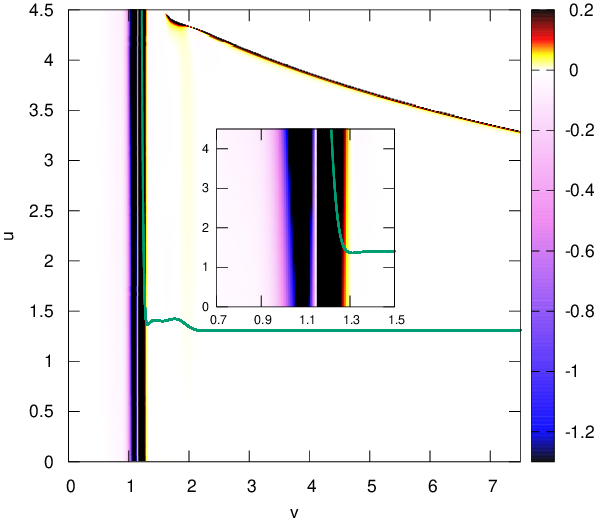}}
\hspace{1cm}
\subfigure[][]{\includegraphics[width=0.38\textwidth]{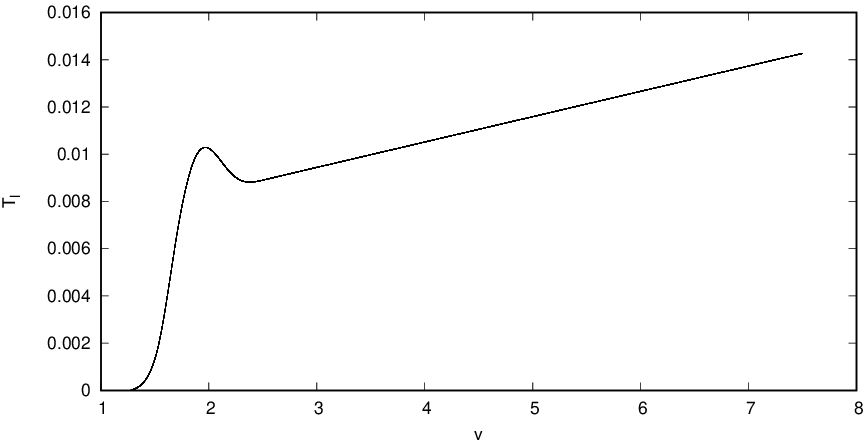}}
\caption{(color online) The~$\left(vu\right)$-distribution of (a)~energy density, $\hat{\rho}$, (b)~radial pressure, $\hat{p}_{r}$, and (c)~pressure anisotropy, $\hat{p}_{a}$, and (d)~local temperature along the black hole apparent horizon, $T_l$, as a~function of advanced time for a~dynamical evolution characterized by parameters $\lambda_X=\lambda_{hX}=0.1$, $m^2=0.25$, $\xi_X=\xi_h=-0.5$, $\alpha_m=0$ and $\aX=\ah=0.025$ (the same as in figure~\ref{fig:coupl-str-1a}).}
\label{fig:coupl-obs-1}
\end{figure}

\begin{figure}[tbp]
\centering
\subfigure[][]{\includegraphics[width=0.38\textwidth]{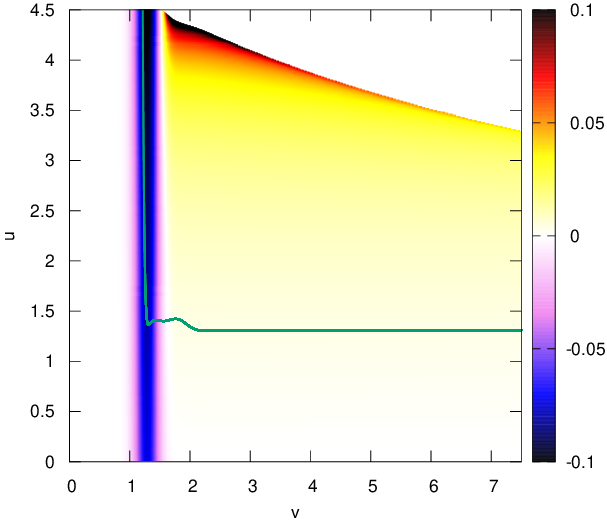}}
\hspace{1cm}
\subfigure[][]{\includegraphics[width=0.38\textwidth]{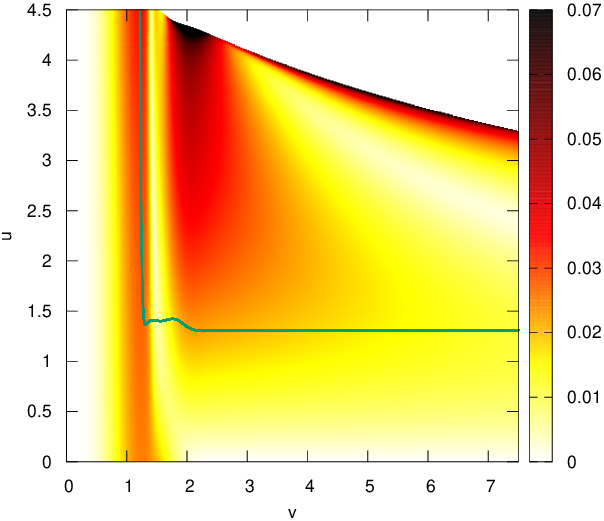}}
\caption{(color online) The~$\left(vu\right)$-distribution of (a)~the neutral scalar field, $h$, and (b)~the moduli of the complex scalar field, $|X|$, for the same parameters and field amplitudes as in figure~\ref{fig:coupl-obs-1}.}
\label{fig:coupl-fie-1}
\end{figure}

Figures~\ref{fig:coupl-obs-1} and~\ref{fig:coupl-fie-1} present the distributions for a~Schwarzschild-type spacetime with a~structure shown in figure~\ref{fig:coupl-str-1a}. Similarly to the case of a~singular spacetime stemming from the collapse in which the non-minimal scalar--gravity couplings are excluded discussed in section~\ref{sec:noncoupl-obs}, the highest absolute values of the observables, as well as the neutral scalar field function and the moduli of the complex scalar field are distributed along a~null direction of constant advanced time being a~direction of propagation of peaks of initially imposed field functions. Moreover, an increase of their values is observed in the vicinity of the central singularity. This increase is most intense nearby the part of the singularity situated for large values of retarded time. The~energy density, radial pressure, pressure anisotropy and both field functions are positive in the vicinity of the singularity along $r=0$ and remain finite there. Their signs in the spacetime region further away from the singularity vary, which is a~result of non-minimal couplings of scalars to gravity. It can be deduced when compared with the behavior of observables in spacetimes in the minimally coupled case, which was discussed in section~\ref{sec:noncoupl-obs}. The~changes of the black hole local temperature, which is positive, along the apparent horizon as $v$ increases are not monotonic. It increases for small values of advanced time, reaches a~maximum in the $v$-range, within which the apparent horizon changes its character from spacelike to null and then, after a~slight decrease. the temperature increases along the null segment of the horizon.

\begin{figure}[tbp]
\centering
\subfigure[][]{\includegraphics[width=0.38\textwidth]{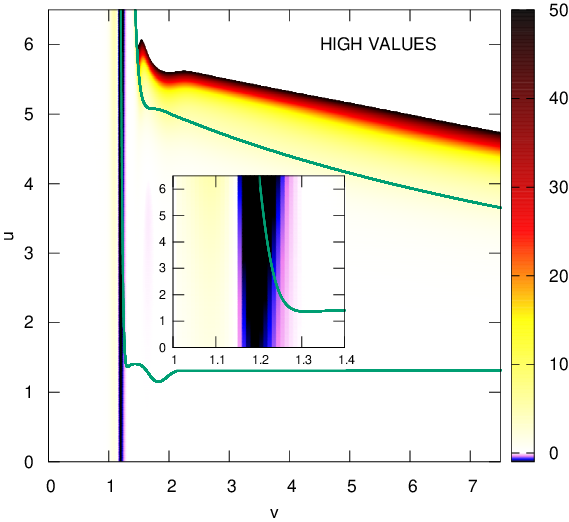}}
\hspace{1cm}
\subfigure[][]{\includegraphics[width=0.38\textwidth]{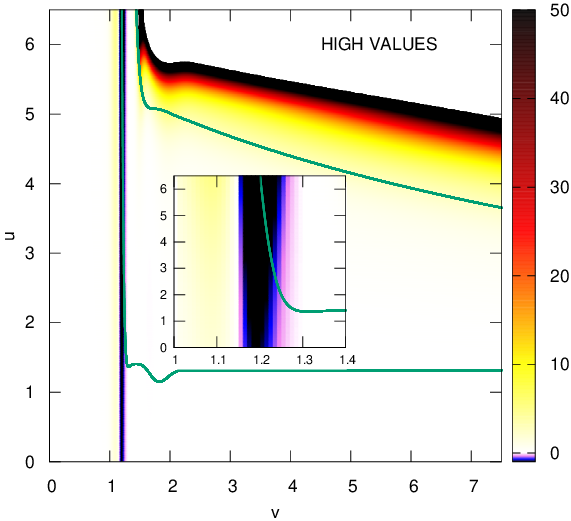}}\\
\subfigure[][]{\includegraphics[width=0.38\textwidth]{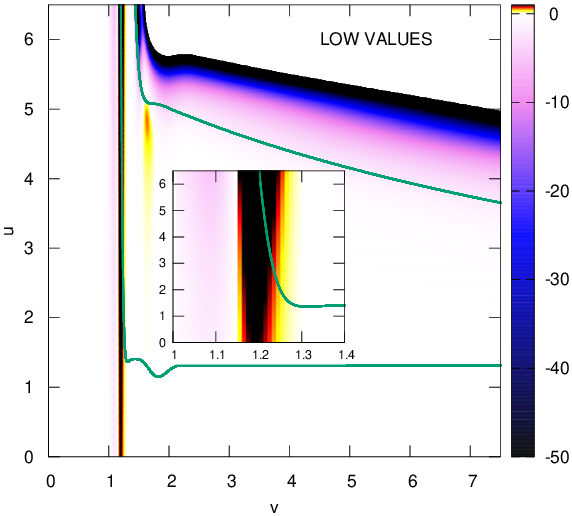}}
\hspace{1cm}
\subfigure[][]{\includegraphics[width=0.38\textwidth]{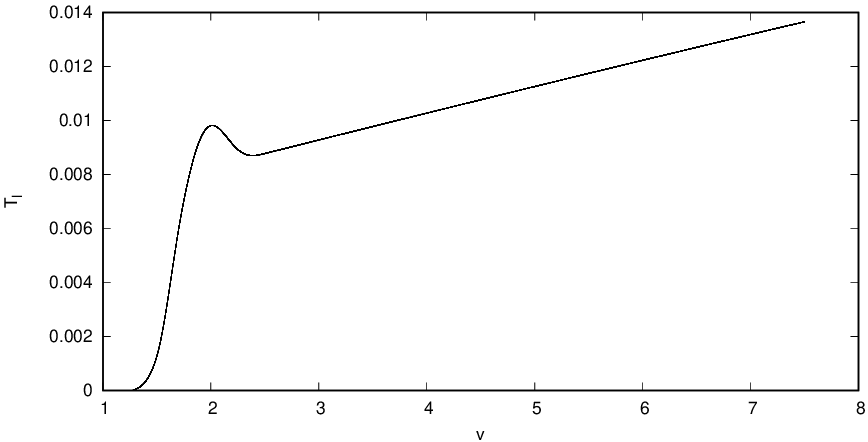}}
\caption{(color online) The~$\left(vu\right)$-distribution of (a)~energy density, $\hat{\rho}$, (b)~radial pressure, $\hat{p}_{r}$, and (c)~pressure anisotropy, $\hat{p}_{a}$, and (d)~local temperature along the black hole outer apparent horizon, $T_l$, as a~function of advanced time for a~dynamical evolution characterized by parameters $\lambda_X=0$, $\lambda_{hX}=0.1$, $m^2=-0.25$, $\xi_X=\xi_h=-0.5$, $\alpha_m=0.1$, $\aX=\ah=0.025$ (the same as in figure~\ref{fig:coupl-str-1d}).}
\label{fig:coupl-obs-2}
\end{figure}

\begin{figure}[tbp]
\centering
\subfigure[][]{\includegraphics[width=0.38\textwidth]{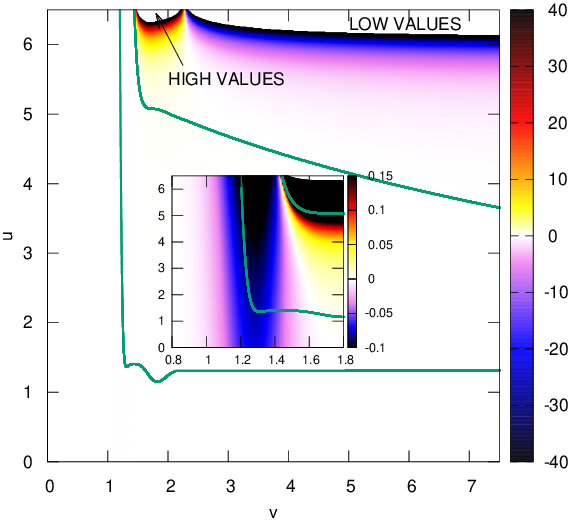}}
\hspace{1cm}
\subfigure[][]{\includegraphics[width=0.38\textwidth]{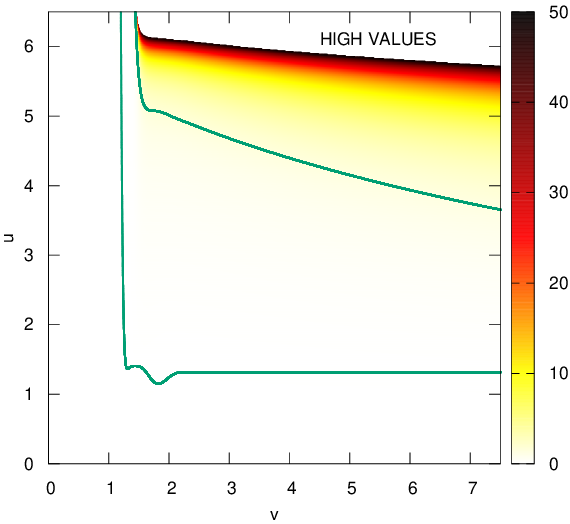}}
\caption{(color online) The~$\left(vu\right)$-distribution of (a)~the neutral scalar field, $h$, and (b)~the moduli of the complex scalar field, $|X|$, for the same parameters and field amplitudes as in figure~\ref{fig:coupl-obs-2}.}
\label{fig:coupl-fie-2}
\end{figure}

The spacetime distributions of observables and field functions stemming from a~dynamical evolution, which results in a~Reissner-Nordstr\"{o}m-type spacetime are shown in figures~\ref{fig:coupl-obs-2} and~\ref{fig:coupl-fie-2}, respectively. The~structure of the spacetime was depicted in figure~\ref{fig:coupl-str-1d}. As in the cases discussed above, a~region of high absolute values of the energy density, radial pressure, pressure anisotropy, the neutral scalar field and the moduli of the complex scalar field is situated along the direction of initial peaks propagation along constant $v$. The~values of all discussed quantities increase considerably beyond the inner apparent horizon, where the $r=const.$ lines settle down along null hypersurfaces of constant retarded time indicating the location of the Cauchy horizon at infinite advanced time, and diverge as the central singularity is approached. The~local temperature in the studied case is positive and behaves in the same manner as in both cases discussed above. It increases for small values of advanced time, where the apparent horizon is spacelike. Then, it reaches an extremum in the region, where the character of the apparent horizon transforms from spacelike to null. Afterwards, after a~slight decrease, it begins to increase along the null part of the horizon.

\section{Conclusions}
\label{sec:conclusions}

Gravitational dynamics within a~toy model which involves a~Higgs field and two dark matter candidates was investigated. The~theoretical setup consisted of two scalar fields and two $U(1)$ gauge fields. The~gauge fields described electromagnetism and the dark photon, which was one of the dark matter candidates. One of the scalars was not charged under any of the $U(1)$ gauge fields and represented a~real part of the Higgs doublet written in the unitary gauge. The~second scalar was complex and charged under a~$U(1)$ gauge field. It~represented a~second stable dark matter candidate. A~possibility that both scalars possess non-minimal couplings to gravity was also included. The~coupling channels among the ordinary matter sector, which consisted of the real scalar and electromagnetic field, and the dark sector, composed of the complex scalar and an additional $U(1)$ field, were given by a~kinetic mixing between the $U(1)$ gauge fields and the Higgs portal coupling among the scalars.

The course and results of gravitational collapse within the setup described above was simulated numerically for two cases with different characters of couplings between the scalars and gravity. One of them was a~minimal coupling with vanishing scalar--gravity coupling constants $\xi_X=\xi_h=0$. The~other one was the full version of the model which involves the non-minimal couplings $\xi_X\neq 0$ and $\xi_h\neq 0$.

The evolutions led to either non-singular spacetimes or spacetimes containing black holes. In the case of a~model with scalars minimally coupled to gravity the outcome of a~singular collapse is a~dynamical Schwarzschild spacetime. There is a~central spacelike singularity within it, surrounded by an apparent horizon, which is spacelike in the dynamical spacetime region, that is for small values of advanced time. For positive values of the parameter $m^2$ the horizon settles down along a~null $u=const.$ hypersurface as $v$ tends to infinity and indicates the location of the event horizon in the spacetime. When the mass parameter is negative, there is a~null part of the apparent horizon, which is followed by its second spacelike part visible for large values of advanced time.

Inclusion of the non-minimal couplings between scalars and gravity results in a~formation of more complex spacetime structures. In all cases, the spacetimes contain central spacelike singularities. When both of the scalar--gravity coupling constants are non-zero and the mass parameter is positive and when only one of the couplings does not vanish, irrespectively of the sign of $m^2$, the emerging spacetimes are dynamical Schwarzschild spacetimes. The~apparent horizon surrounding the singularity is spacelike and null for small and large values of advanced time, respectively. Between these two parts there exist also timelike portions of the horizon. It seems characteristic for the non-minimality of the scalar--gravity couplings, as was not observed when only minimally coupled scalars were involved in the process, neither in the current paper nor in previous works on the subject (see, e.g.,~\cite{NakoniecznaNakoniecznyYeom2019-1930006} and references therein). 

For both non-zero couplings $\xi_X$ and $\xi_h$ and negative values of the parameter $m^2$ dynamical Reissner-Nordstr\"{o}m spacetimes form. The~behavior of the outer apparent horizon is qualitatively similar to the one observed in the cases of non-minimally coupled scalars decribed in the previous paragraph. The~second, inner apparent horizon is spacelike within the whole spacetime. There also exists a~Cauchy horizon at infinite advanced time. The~type of the spacetime which forms during the gravitational process involving non-minimal couplings of scalars to gravity depends on the mass parameter of the complex scalar field. A~similar dependency was observed during a~collapse with both dark energy and dark matter present in spacetime~\cite{NakoniecznaRogatkoNakonieczny2015-012}.

The dependencies of the $u$-locations of the event horizons, radii and masses of black holes emerging from the investigated process on $m^2$ in both minimally and non-minimally coupled cases are qualitatively the same. As the mass parameter increases, the black holes form later in terms of retarded time and both their radii and masses become smaller. Qualitatively similar relations were observed within the outcomes of gravitational collapse of an electrically charged scalar field accompanied by dark matter and, possibly, additionally by dark energy~\cite{NakoniecznaRogatkoNakonieczny2015-012}. The~changes are significant up to about $m^2=-0.15$ and become much less visible for its larger values. The~dependence of the absolute values of black hole $U(1)$ charges on the mass parameter posesses a~minimum for its negative values for the cases of both couplings $\xi_X$, $\xi_h$ either vanishing or not equal to zero. When $m^2$ becomes bigger than approximately $-0.05$, the charges increase become less significant. When only one scalar--gravity coupling constant is non-zero, $Q^{eh}$ and $|T^{eh}|$ decrease with $m^2$ and the changes become less dynamical as the mass parameter increases.

The black holes form earlier in terms of retarded time and their radii, masses and charges become smaller as $\alpha_m$ increases. Exceptions from this rule are observed in the case of only one non-vanishing scalar--gravity coupling, for which an increase of the $u$-locations of the event horizons and absolute values of black hole charges is observed for large values of the parameter $\alpha_m$.

Black holes form later in terms of $u$, their radii and masses decrease while their charges increase with both $\lambda_X$ and $\lambda_{hX}$ model parameters in the case when $\xi_X\neq 0$ and $\xi_h\neq 0$. When either both or only one scalar--gravity coupling vanishes, the quartic couplings do not influence the time of the event horizon formation. With their increase, it turns out that the black holes possess bigger radii and masses and smaller charges. The~dependencies of the $u$-locations of the event horizons, the radii, masses and charges of forming black holes on $\lambda_{hX}$ in the minimally coupled case and on $\lambda_X$ when only one scalar--gravity coupling does not vanish are complicated as they exhibit discontinuities.

The dependencies of the black hole features on the scalar--gravity coupling $\xi_h$ possess extrema. There exists a~value of $\xi_h$ for which the nascent black hole forms the latest in terms of retarded time and simultaneously possesses the smallest radius and mass. What is interesting, there is also a~minimum in the relation between the black hole charges and~$\xi_h$, but it does not overlap with the one for $u^{eh}$, $r^{eh}$ and $m_H^{\ eh}$. The~dependencies on the scalar--gravity coupling $\xi_X$ are qualitatively the same with the difference that the observed extrema are shallower.

In all the investigated cases, both non-singular and singular, an increase of the energy density, radial pressure, pressure anisotropy and values of the evolving scalar fields was observed along a~null direction of the propagation of the maxima of initially imposed field profiles in spacetime. When dynamical Schwarzschild black holes formed, another increase in values of the quantities measured by an observer moving with the collapsing matter was visible in a~close vicinity of the emerging singularity. For dynamical Reissner-Nordstr\"{o}m spacetimes the increase was also considerable in the region, in which the $r=const.$ lines settled at null hypersurfaces of constant retarded time, indicating the existence of the Cauchy horizon at infinite advanced time. It seems that calculating the proposed observables within the dynamical spacetimes may allow to distinguish between their types. The~increase of these quantities is much bigger as the spacelike singularity is approached in dynamical Reissner-Nordstr\"{o}m spacetimes containing the Cauchy horizon, when compared to dynamical spacetimes of the Schwarzschild type.

The local temperature calculated along the apparent horizon of the nascent black holes using the definition for dynamical black holes adopted in the current paper increases for large values of advanced time for all investigated cases. This is a~region, in which the apparent horizons are situated along null hypersurfaces. In dynamical regions of spacetimes, where the horizons are either spacelike or timelike, the changes of the values of local temperature are monotonic and increasing in the minimally coupled case and non-monotonic when at least one non-minimal coupling is involved.

\appendix
\section{Numerical computations}
\label{sec:appendix}

\subsection{Algorithm setup}

The dynamics of the physical system of interest described by equations~\eqref{eqn:P1-2}--\eqref{eqn:C2} was resolved numerically. The~set of equations of motion involves the following functions: $d$,~$z_1$,~$z_2$, $y$, $X_1$, $X_2$, $h$, $a$, $w_1$, $w_2$, $x$, $r$, $f$, $g$, $Q$, $\beta$, $T$, $\gamma$, each of which is a~function of two null coordinates, i.e., advanced and retarded times. The~dynamics of functions $d$, $z_1$, $z_2$ and $y$ was followed along the $u$-coordinate according to equations $E4$, $X_{_{\left(Re\right)}}$, $X_{_{\left(Im\right)}}$ and $H$, respectively. The~remaining functions, $X_1$, $X_2$, $h$, $a$, $w_1$, $w_2$, $x$, $r$, $f$, $g$, $Q$, $\beta$, $T$ and $\gamma$, evolved along the $v$-coordinate in line with the respective equations $P6$, $P7$, $P2$, $X_{_{\left(Re\right)}}$, $X_{_{\left(Im\right)}}$, $H$, $P4$, $E3$, $E2$, $D2$, $D1$, $G2$ and $G1$.

The system was solved within a~bounded region of the $\left(vu\right)$-plane, which is presented in figure~\ref{fig:domain} in section~\ref{sec:particulars}. A~null hypersurface~$u=const.$ was taken as an initial data surface. The~boundary conditions were posed on a~hypersurface of constant $v$. For purposes of numerics, the two surfaces were marked as $u=0$ and $v=0$, respectively. 

Initial conditions involve arbitrary profiles of the evolving fields functions, $X_1$, $X_2$ and~$h$, which were assigned according to~\eqref{psichi-prof} and~\eqref{phi-prof}. The~initial values of~$z_1$, $z_2$ and~$y$ were calculated analytically using the~relations $P6$ and~$P7$. Since in the employed setup the distribution of matter is shell--shaped, the boundary is not affected by it and the field functions $X_1$, $X_2$ and $h$ vanish there. The~boundary values of $z_1$, $z_2$ and $y$ were obtained via integration of equations $X_{_{\left(Re\right)}}$, $X_{_{\left(Im\right)}}$ and $H$, respectively. 

Within the investigated setup there remains gauge freedom to choose the initial and boundary profiles of the $r$ function. We chose $r\left(0,0\right)$ to be equal to $7.5$ for numerical purposes. The~initial and boundary values of $g$ and $f$, respectively, which determine the distances between the null lines were chosen to be constant, namely $g\left(0,v\right)=\frac{1}{2}$ and $f\left(v,0\right)=-\frac{1}{2}$. Their precise values are justified by the fact that mass~\eqref{eqn:mH} should vanish at the central point $\left(0,0\right)$. The~$r$ values along the initial null segment were obtained with the use of the equation $P4$ and along the boundary using the equation $P3$. The~initial and boundary profiles of $f$ and $g$, respectively, were calculated via the integration of the equation $E3$.

The initial values of function $d$ were calculated out of the equation $E2$, while its boundary values were obtained with the use of the equation $E4$. The~abovementioned spherical shell shape of matter distribution justifies setting the following boundary values: $a\left(u,0\right)=1$, $Q\left(u,0\right)=\beta\left(u,0\right)=T\left(u,0\right)=\gamma\left(u,0\right)=0$ and $w_1\left(u,0\right)=w_2\left(u,0\right)=x\left(u,0\right)=0$. The~initial profiles of these functions were obtained using the equations $P2$, $D2$, $D1$, $G2$, $G1$, $X_{_{\left(Re\right)}}$, $X_{_{\left(Im\right)}}$ and $H$, respectively.

\subsection{Employed schemes}

The numerical code was written from scratch in Fortran. The~integration along the retarded time coordinate was conducted with the use of the 2$^{nd}$ order accurate Runge--Kutta method. The~integration of the partial differential equations along the $v$-coordinate was performed with the 2$^{nd}$ order accurate Adams--Bashforth--Moulton method, apart from the first point, where the trapezoidal rule was applied. Adequate $v$-derivatives of functions $z_1$, $z_2$ and $y$ whose calculation was indispensable to perform the integration of~\eqref{eqn:E2} were obtained with 2$^{nd}$ order accurate rules, symmetrical everywhere apart from the points at boundaries of the computational region. The~function $c$ is not explicitly involved in the integration of the set of evolution equations~\eqref{eqn:P1-2}--\eqref{eqn:C2} described above, however, it is needed to compute the observables~\eqref{eqn:rho}--\eqref{eqn:pa}. For this purpose, the values of c were calculated according to its definition $c=\frac{a\POu}{a}$.

The double null coordinates ensure regular behaviour of all the evolving quantities within the domain of integration. However, considerable numerical difficulties arise as the event horizon, where function $f$ diverges, is approached.  A~relatively dense numerical grid is necessary to determine its location and to examine the behaviour of fields beyond it, especially for large values of advanced time.  The~efficiency of calculations requires using an adaptive grid and performing integration with a~smaller step in particular regions. For the gravitational collapse investigations, the refinement algorithm making the grid denser solely in the $u$-direction, is sufficient~\cite{BorkowskaRogatkoModerski2011-084007}. In order to determine the area of the integration grid, where it should be denser, a~local error indicator is needed.  This quantity should be bounded with the evolving quantities and change its value significantly in the adequate region. The~function $\frac{\Delta r}{r}$ along the $u$-coordinate thus indicates the surrounding of the event horizon in spacetime and hence meets our requirements~\cite{OrenPiran2003-044013}.

\subsection{Tests of the code}

The accuracy of the numerical code was checked indirectly, as no analytical solutions exist for the investigated process. The~tests were performed for two evolutions initiated with the following parameters. Evolution 1 was conducted with $\lambda_X=\lambda_{hX}=0.1$, $m^2=0.25$, $\xi_X=\xi_h=0$, $\alpha_m=0$, $\aX=\ah=0.05$ and Evolution 2 with $\lambda_X=\lambda_{hX}=0.1$, $m^2=0.25$, $\xi_X=\xi_h=-0.5$, $\alpha_m=0$, $\aX=\ah=0.025$. The~spacetime structures are presented in figures~\ref{fig:noncoupl-str-a} and~\ref{fig:coupl-str-1a}, respectively.

The first test was based on checking the convergence of the code. In order to monitor the convergence, the computations for Evolutions 1 and 2 were conducted on four grids with integration steps being multiples of the basic value $\delta=10^{-4}$. A~step of a~particular grid was twice the size of a~denser one. The~convergence was examined on a~$u=const.$ hypersurface chosen arbitrarily with the value of $u$ equal to $1$. The~selected hypersurface was situated close to the emerging event horizon, but in the region where the adaptive mesh on neither of the grids was active, which was necessary for performing a~proper comparison of the results.

\begin{figure}[tbp]
 \subfigure[][]{\includegraphics[width=0.47\textwidth]{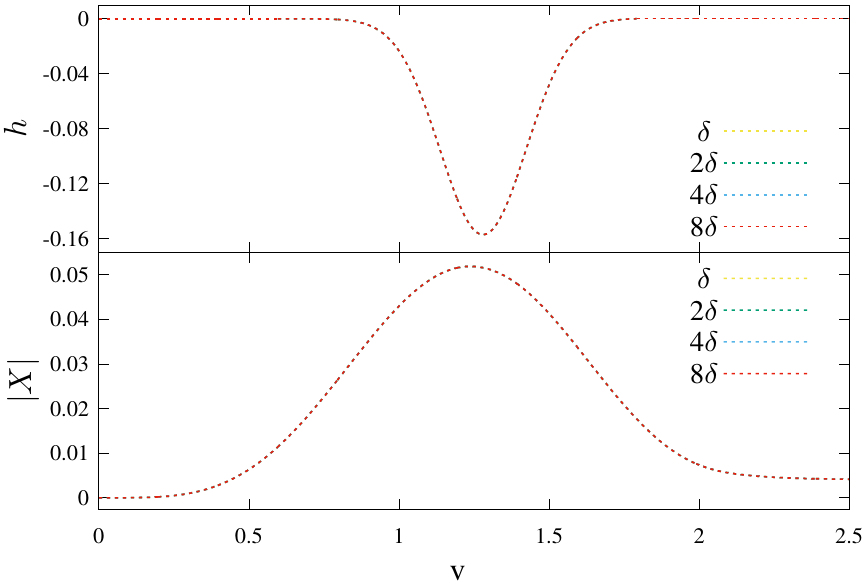}}
\hfill
\subfigure[][]{\includegraphics[width=0.47\textwidth]{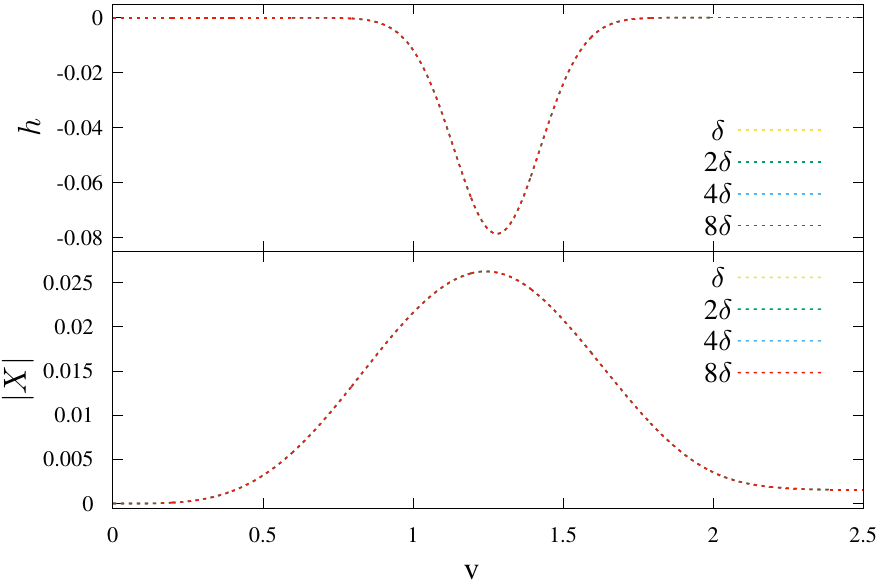}}
\caption{(color online) The~convergence of field functions. The~scalar field, $h$, and the moduli of complex scalar field, $|X|$, were plotted versus $v$ for evolutions conducted with integration steps, which were multiples of \mbox{$\delta=10^{-4}$}, along hypersurfaces of constant $u$ equal to $1$ for (a)~Evolution 1 and (b)~Evolution 2.}
\label{fig:Conv1}
\end{figure}

\begin{figure}[tbp]
\subfigure[][]{\includegraphics[width=0.47\textwidth]{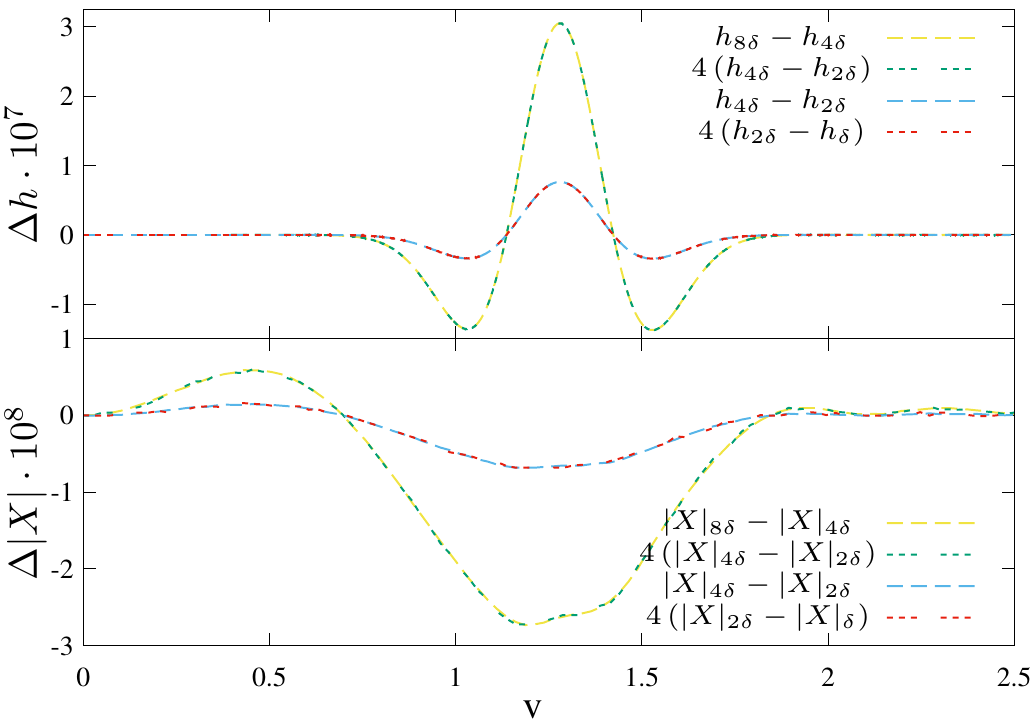}}
\hfill
\subfigure[][]{\includegraphics[width=0.47\textwidth]{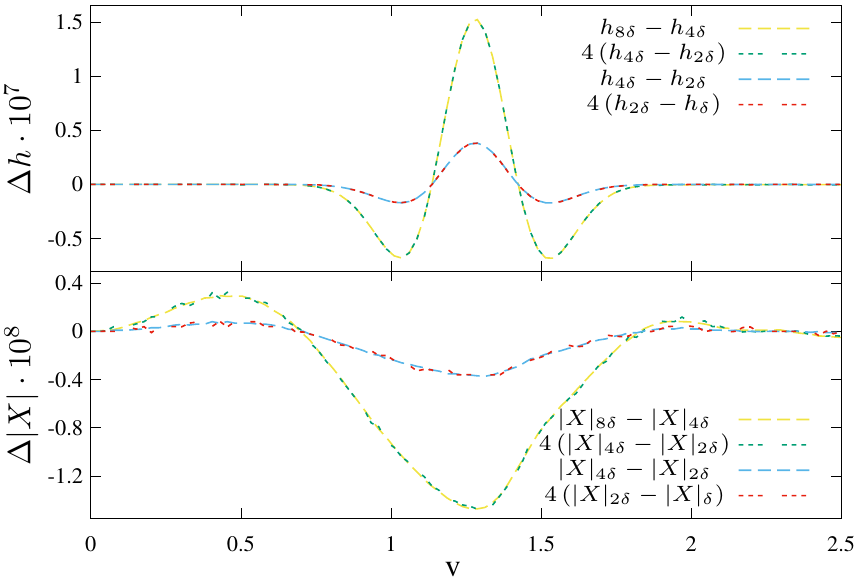}}
\caption{(color online) The~convergence of the code. The~differences between the scalar field functions, $\Delta h$, and the moduli of complex scalar field, $\Delta|X|$, calculated on grids with different integration steps (multiples of \mbox{$\delta=10^{-4}$}) and their multiples were obtained along the same hypersurfaces of constant $u$ as in figure~\ref{fig:Conv1} for (a)~Evolution 1 and (b)~Evolution 2.}
\label{fig:Conv2}
\end{figure}

The field functions along the selected hypersurface of constant $u$ from within the $v$-range in which the functions are initially non-zero for the examined integration steps are shown in figure~\ref{fig:Conv1}. The~maximal observed discrepancy between the functions calculated on the finest and coarsest grids was equal to $3\cdot 10^{-4}\%$. Figure~\ref{fig:Conv2} presents the 2$^{nd}$ order convergence of the numerical code. The~maximal divergence between the field profiles obtained on two grids with a~quotient of integration steps equal to $2$ and their respective quadruples was $10^{-1}\%$. As expected, the errors decreased with an increase of the grid density. The~overall error analysis revealed that the expected convergence was achieved and both the algorithm and the numerical code were appropriate for solving the system of equations~\eqref{eqn:P1-2}--\eqref{eqn:C2}.

\begin{figure}[tbp]
\centering
\includegraphics[width=0.47\textwidth]{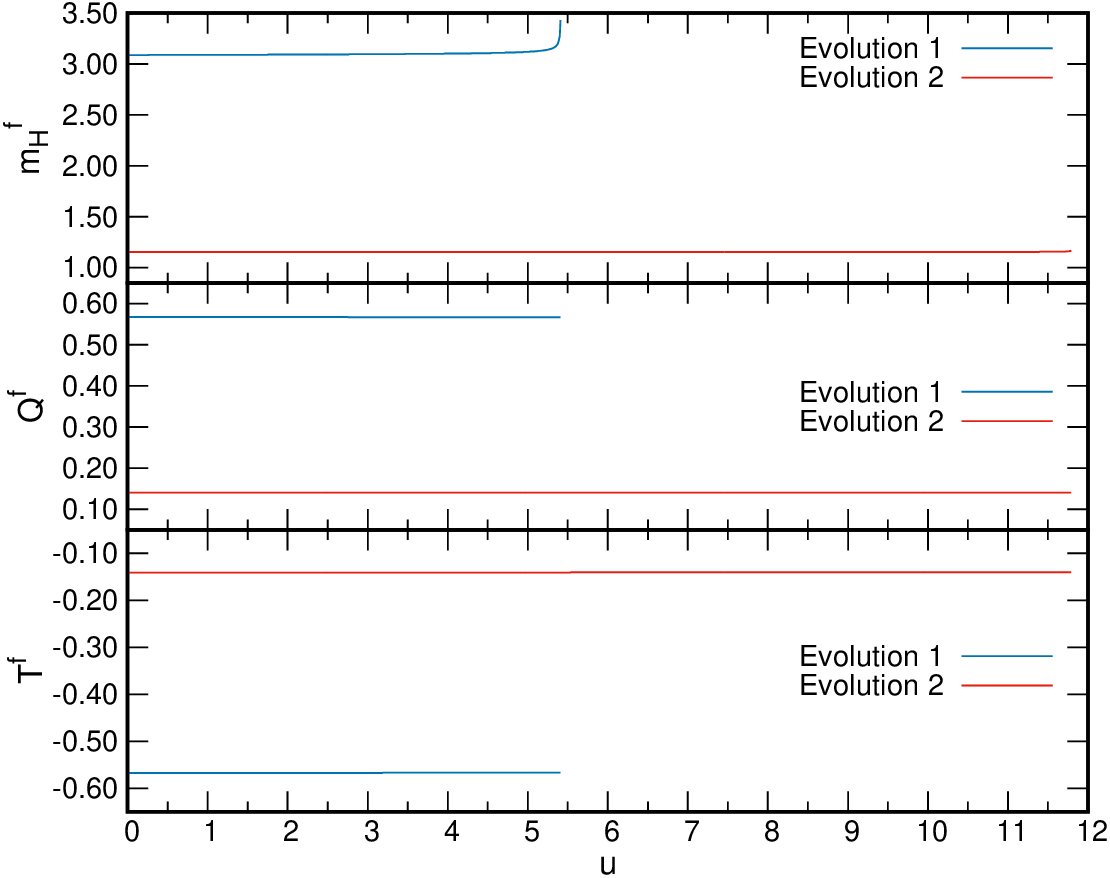}
\caption{(color online) Hawking masses, $m_H^{\ \ f}$, and~charges related to~the~$U(1)$ gauge fields, $Q^f$ and~$T^f$, calculated along $v=7.5$ as~functions of~retarded time, $u$, for~both tested Evolutions.}
\label{fig:Cons}
\end{figure}

The second test of the numerical code was based on checking the mass and charge conservation in spacetime. The~Hawking mass~\eqref{eqn:mH} and charges related with the $U(1)$ gauge fields defined by relations~\eqref{eqn:A-2} and~\eqref{eqn:C-2} as functions of retarded time along the line $v=7.5$, which was a~maximal value of advanced time achieved numerically, are presented in figure~\ref{fig:Cons} for the investigated Evolutions. Since during the course of gravitational collapse the matter was scattered by the gravitational potential barrier when the collapsing shell approached its gravitational radius, the plotted physical quantities were not conserved during the whole evolution. The~effect of the outgoing flux was negligible everywhere except for the vicinity of the event horizon. The~deviation from the constancy increased with advanced time, as the horizon was approached. The~maximal percentage deviations from the particular quantity conservation up to the value of $u$ corresponding to the location of the event horizon were equal to $0.49\%$, $0.07\%$ and $0.07\%$ for the mass and charges $Q$ and $T$, respectively. The~analysis of mass and charge conservation in spacetime led to the conclusion that the behavior of matter investigated numerically was correct within the computational domain.

\acknowledgments

AN was supported by the National Science Centre, Poland, under a~postdoctoral scholarship DEC-2016/20/S/ST2/00368. 
LN was supported by the National Science Centre, Poland, under a~grant DEC-2017/26/D/ST2/00193.





\bibliographystyle{stylebib}
\bibliography{higgscollapse.bib}

\end{document}